\documentclass[rmp,aps,twocolumn,nofootinbib,floats]{revtex4}

\pdfoutput=1
\usepackage{graphics}
\usepackage{epsfig}

\newcommand\beq{\begin{equation}}
\newcommand\eeq{\end{equation}}
\newcommand\beqar{\begin{eqnarray}}
\newcommand\eeqar{\end{eqnarray}}

\def\lsim{\mathrel{\rlap{\lower4pt\hbox{\hskip1pt$\sim$}}
    \raise1pt\hbox{$<$}}}         %less than or approx. symbol
\def\gsim{\mathrel{\rlap{\lower4pt\hbox{\hskip1pt$\sim$}}
    \raise1pt\hbox{$>$}}}         %greater than or approx. symbol

\begin{document}

\title{Solar fusion cross sections II: the pp chain and CNO cycles}

\author{E. G. Adelberger}
\affiliation{Department of Physics and Center for Experimental Nuclear Physics and Astrophysics, \\
University of Washington, Seattle, WA 98195 USA}
\author{A. B. Balantekin}
\affiliation{Department of Physics, University of Wisconsin, Madison, WI 53706 USA}
\author{D. Bemmerer}
\affiliation{Forschungszentrum Dresden-Rossendorf, D-01314 Dresden, Germany}
\author{C. A. Bertulani}
\affiliation{Department of Physics and Astronomy, Texas A\&M University, Commerce, TX 75429 USA}
\author{J.-W. Chen}
\affiliation{Department of Physics and Center for Theoretical Sciences, National Taiwan University, Taipei 10617, Taiwan}
\author{H. Costantini}
\affiliation{Universit\`{a} di Genova and INFN Sezione di Genova, Genova, Italy}
\author{M. Couder}
\affiliation{Department of Physics and JINA, University of Notre Dame, 
Notre Dame, IN 46556 USA}
\author{R. Cyburt}
\affiliation{JINA and National Superconducting Cyclotron Laboratory,
Michigan State University, East Lansing, MI 48824 USA}
\author{B. Davids}
\affiliation{TRIUMF, 4004 Wesbrook Mall, Vancouver, BC, Canada V6T 2A3}
\author{S. J. Freedman}
\affiliation{Department of Physics, University of California, Berkeley, and \\
Lawrence Berkeley National Laboratory, 
Berkeley, CA 94720 USA }
\author{M.Gai}
\affiliation{Laboratory for Nuclear Sciences at Avery Point, University of Connecticut, CT 06340-6097 and \\
Department of Physics, Yale University, New Haven, CT 06520-8124 USA}
\author{A. Garc\'{\i}a}
\affiliation{Department of Physics and Center for Experimental Nuclear Physics and Astrophysics, \\
University of Washington, Seattle, WA 98195 USA}
\author{D. Gazit}
\affiliation{Institute for Nuclear Theory, University of Washington, Seattle, WA 98195 USA and \\
Racah Institute of Physics, The Hebrew University, Jerusalem, 91904, Israel}
\author{L. Gialanella}
\affiliation{Dipartimento di Scienze Fisiche, Universit\`{a} di Napoli,  and 
INFN Sezione di Napoli, Napoli, Italy}
\author{U. Greife}
\affiliation{Department of Physics, Colorado School of Mines, Golden, CO 80401 USA}
\author{M. Hass}
\affiliation{Department of Particle Physics and Astrophysics, The Weizmann Institute, Rehovot, Israel}
\author{K. Heeger}
\affiliation{Department of Physics, University of Wisconsin, Madison, WI 53706 USA}
\author{W. C. Haxton}
\affiliation{Department of Physics, University of California, Berkeley, and Lawrence Berkeley National Laboratory,\\
Berkeley, CA 94720 and 
Institute for Nuclear Theory, University of Washington, Seattle, WA 98195 USA}
\author{G. Imbriani}
\affiliation{Dipartimento di Scienze Fisiche, Universit\`{a} di Napoli, and 
INFN Sezione di Napoli, Napoli, Italy}
\author{T. Itahashi}
\affiliation{Research Center for Nuclear Physics, Osaka University, Ibaraki, Osaka 567-0047 Japan}
\author{A. Junghans}
\affiliation{Forschungszentrum
Dresden-Rossendorf, D-01314 Dresden, Germany}
\author{K. Kubodera}
\affiliation{Department of Physics and Astronomy, University of South Carolina, Columbia, SC 29208 USA}
\author{K. Langanke}
\affiliation{GSI Helmholtzzentrum f\"{u}r Schwerionenforschung, D-64291 Darmstadt, Germany, Institut  f\"{u}r Kernphysik, Universit\"{a}t Darmstadt, Germany and
Frankfurt Institute for Advanced Studies, Frankfurt, Germany}
\author{D. Leitner}
\affiliation{Lawrence Berkeley National Laboratory, 
Berkeley, CA 94720 USA}
\author{M. Leitner}
\affiliation{Lawrence Berkeley National Laboratory, 
Berkeley, CA 94720 USA}
\author{L. E. Marcucci}
\affiliation{Department of Physics "E. Fermi", University of Pisa, and \\
INFN Sezione di Pisa, Largo B. Pontecorvo 3, I-56127, Pisa, Italy}
\author{T. Motobayashi}
\affiliation{The Institute of Physical and Chemical Research (RIKEN), 2-1 Hirosawa, Wako, Saitama 351-0198, Japan}
\author{A. Mukhamedzhanov}
\affiliation{Cyclotron Institute, Texas A\&M University, College Station, TX 77843 USA}
\author{Kenneth M. Nollett}
\affiliation{Physics Division, Argonne National Laboratory, 9700 S. Cass Ave., Argonne, IL 60439 USA}
\author{F. M. Nunes}
\affiliation{National Superconducting Cyclotron Laboratory and Department of Physics and Astronomy,\\
Michigan State University, East Lansing, MI 48824 USA}
\author{T.-S. Park}
\affiliation{Department of Physics and BAERI, Sungkyunkwan University, Suwon 440-746 Korea}
\author{P. D. Parker}
\affiliation{Wright Nuclear Structure Laboratory, Yale University, New Haven, CT 06520 USA}
\author{P. Prati}
\affiliation{Universit\`{a} di Genova and INFN Sezione di Genova, Genova, Italy}
\author{M. J. Ramsey-Musolf}
\affiliation{Department of Physics, University of Wisconsin, Madison, WI 53706 USA}
\author{R. G. Hamish Robertson}
\affiliation{Department of Physics and Center for Experimental Nuclear Physics and Astrophysics, \\
University of Washington, Seattle, WA 98195 USA}
\author{R. Schiavilla}
\affiliation{Department of Physics, Old Dominion University, Norfolk, VA 23529 and \\
Jefferson Laboratory, Newport News, VA 23606 USA}
\author{E. C. Simpson}
\affiliation{Department of Physics, University of Surrey, Guildford, Surrey GU2 7XH, United Kingdom}
\author{K. A. Snover}
\affiliation{Department of Physics and Center for Experimental Nuclear Physics and Astrophysics, \\
University of Washington, Seattle, WA 98195 USA}
\author{C. Spitaleri}
\affiliation{INFN Laboratori Nazionali del Sud \& DMFCI, Universit\`{a} di Catania, Catania, Italy}
\author{F. Strieder}
\affiliation{Institut f\"{u}r Experimentalphysik III, Ruhr-Universit\"{a}t Bochum, Bochum, Germany}
\author{K. Suemmerer}
\affiliation{GSI Helmholtzzentrum f\"{u}r Schwerionenforschung GmbH,
Planckstra\ss{}e 1, D-64291 Darmstadt, Germany}
\author{H.-P. Trautvetter}
\affiliation{Institut f\"{u}r Experimentalphysik III, Ruhr-Universit\"{a}t Bochum, Bochum, Germany}
\author{R. E. Tribble}
\affiliation{Cyclotron Institute, Texas A\&M University, College Station, TX 77843 USA}
\author{S. Typel}
\affiliation{Excellence Cluster Universe, 
Technische Universit\"{a}t M\"{u}nchen,
Boltzmannstra\ss{}e 2, D-85748 Garching and \\
GSI Helmholtzzentrum f\"{u}r Schwerionenforschung GmbH, 
Planckstra\ss{}e 1, D-64291 Darmstadt, Germany}
\author{E. Uberseder}
\affiliation{Department of Physics and JINA, University of Notre Dame, 
Notre Dame, IN 46556 USA}
\author{P. Vetter}
\affiliation{Lawrence Berkeley National Laboratory, 
Berkeley, CA 94720 USA}
\author{M. Wiescher}
\affiliation{Department of Physics and JINA, University of Notre Dame, 
Notre Dame, IN 46556 USA}
\author{L. Winslow}
\affiliation{Lawrence Berkeley National Laboratory, 
Berkeley, CA 94720 USA}

\begin{abstract}  
We summarize and critically evaluate the available data on nuclear fusion
cross sections important to energy generation in the Sun and other hydrogen-burning stars and to 
solar neutrino production.   Recommended values and uncertainties are provided
for key cross sections, and a recommended spectrum is given for $^8$B solar neutrinos.
We also discuss opportunities for further increasing the precision
of key rates, including new facilities, new experimental techniques, and improvements
in theory.  This review, which summarizes the conclusions of a workshop held at the
Institute for Nuclear Theory, Seattle, in January 2009, is intended as a 10-year update
and supplement to Reviews of Modern Physics {\bf 70} (1998) 1265.

\end{abstract}                                                                 

%\date{May 2001}
\maketitle
\tableofcontents

\section{INTRODUCTION}
\label{sec:intro}
In 1998 the Reviews of Modern Physics published a summary and critical analysis of 
the nuclear reaction cross sections important to solar burning.  That effort,
\citet{Adel98} and denoted here as Solar Fusion I, began with a meeting hosted
by the Institute for Nuclear Theory, University of Washington, 17-20 February 1997.
A group of international experts in the nuclear physics and astrophysics of hydrogen-burning
stars met to begin critical discussions of the existing data on relevant nuclear reactions, with
the aim of determining ``best values" and uncertainties for the contributing
low-energy S-factors.  The group also considered
opportunities for further improvements in both measurements and theory.  

Such data and related nuclear theory have been crucial to the
standard solar model (SSM) and the neutrino fluxes it predicts.  Indeed, measurements
of nuclear reactions gave the field its start.  In 1958
\citet{HJ58,HJ59} showed that the rate for
$^3$He+$^4$He $\rightarrow$ $^7$Be +$\gamma$ was $\sim$ 1000 times larger than
expected, and thus that the pp chain for $^4$He synthesis would have additional terminations
beyond $^3$He+$^3$He $\rightarrow$ $^4$He + 2p.  This result led Davis
to recognize that his chlorine detector
might be able to see the higher energy neutrinos from these other terminations,
and spurred Bahcall and others to develop a quantitative model
of the Sun capable of predicting those fluxes \cite{bd}.

At the time of the 1997
meeting, three decades of effort in solar neutrino physics
had produced four measurements that were at variance with
the SSM and the standard model of electroweak interactions.  The 
measurements came from the pioneering work of Ray Davis, Jr.
\cite{Davis68, Davis94}; the observation
of $^8$B neutrinos in the Kamiokande water Cerenkov detector \cite{Kamiokande96};
and the GALLEX \cite{Kirsten03} and SAGE \cite{Gavrin03} radiochemical detectors 
sensitive primarily  to pp and $^7$Be neutrinos.  The resulting pattern of fluxes that emerged
from these experiments was difficult to reconcile with any plausible variation in the
SSM, requiring a much sharper reduction in the $^7$Be neutrino flux than in the
$^8$B flux, despite the greater sensitivity of the latter to changes in the solar core
temperature.

For this reason it was argued in Solar Fusion I that the measurements provided
evidence for new physics beyond the standard model.  New solar neutrino experiments that
promised much more precise data -- the 50-kiloton successor to Kamiokande, Super-Kamiokande,
and the heavy-water-based Sudbury Neutrino Observatory (SNO), with sensitivity
to both electron and heavy-flavor neutrinos -- were then underway.   The authors
of Solar Fusion I, recognizing that the impact of these new experiments would depend
in part on the quality of the nuclear microphysics input to the SSM, thus undertook 
an extended study of the key reaction rates for the pp chain and CNO bi-cycle.
The effort appears to have been of some value to the community, as Solar Fusion I
has become one of the most heavily cited papers in nuclear astrophysics.

\subsection{Solar Fusion II: the 2009/10 effort}
Ten years after publication of Solar Fusion I a proposal was made to the INT to
revisit this process, in order to produce a new evaluation that would reflect the
considerable progress made in the past decade, as well as new motivations
for further constraining the SSM.  Examples of advances in the
nuclear physics include the
LUNA II program at Gran Sasso \cite{LUNAII}, which has provided remarkable 
low-energy measurements of key reactions such as $^3$He($\alpha$,$\gamma$)$^7$Be
and $^{14}$N(p,$\gamma$)$^{15}$O; several high-precision measurements
addressing the key pp-chain uncertainty identified in Solar Fusion I, 
$^7$Be(p,$\gamma$)$^8$B; the application of new theoretical techniques
to the p+p and hep neutrino reactions; and the resolution of several unresolved
questions about screening corrections in plasmas.

The context for these measurements has also changed.  In 1997 the field's
central concern was, in some sense, a qualitative one, the origin of the solar
neutrino problem.  This question was answered in spectacular fashion by
the dual discoveries of Super-Kamiokande \cite{SK} and SNO \cite{SNO} -- 
two distinct neutrino oscillations  responsible for the missing 
atmospheric and solar neutrinos, largely determining the
pattern of the light neutrino masses.   But issues remain, and most of these 
require precision.  There is intense interest in extending direct measurements
to the low-energy portion of the solar neutrino spectrum ($\lsim$ 2 MeV), 
where experiments with good energy resolution can determine the separate
contributions of pep, CNO, 
$^7$Be, and pp neutrinos.  There is the potential to further constrain the solar
neutrino mixing angle $\theta_{12}$:  the solar luminosity determines the
pp flux to high accuracy, and the low-energy spectrum lies in the vacuum 
region of the MSW triangle, in contrast to the high-energy $^8$B neutrinos, where
matter effects are significant.  Thus precise low-energy measurements have
considerable ``leverage" to test $\theta_{12}$ and the consistency of the 
conclusions we have drawn from SNO, Super-Kamiokande, and the KamLAND
reactor neutrino experiment.  Borexino, now entering its calibration phase, is
the first effort in this program of high-precision spectroscopy of low-energy
solar neutrinos.

But the resolution of the solar neutrino problem has also returned the field
to its roots:  Davis built the chlorine detector to probe the interior of the Sun and
thereby test directly the theory of stellar evolution and nuclear energy generation \cite{bd}.
Davis was diverted from that goal by the missing solar neutrinos.  But as the
weak interaction effects responsible for that anomaly are now reasonably well understood,
solar neutrinos again have become a quantitative tool for astronomy.  
Indeed, the program carried out by SNO and
Super-Kamiokande has already yielded one remarkable constraint
on the Sun, a direct determination
of the core temperature to high precision, through
measurement of the $^8$B neutrino flux
($\phi(^8$B) $\propto T_c^{22}$).  The 8.6\%  precision of the SNO NCD-phase results \cite{NCD}, 
$\phi(^8$B) = $(5.54 {}^{+0.33}_{-0.31} {}^{+0.36}_{-0.34}) \times 10^6$/cm$^2$/s,
implies a sensitivity to core temperature of $\sim$ 0.5\%.

New questions have arisen about the Sun that neutrinos could potentially
address, provided the associated laboratory astrophysics has been done.
One important success of the SSM in the 1990s
was in predicting the local sound
speed $c(r)$.  Comparisons between $c(r)$ deduced from
helioseismology and the predictions of the SSM yielded agreement
at $\sim$ 0.2\% throughout much of the Sun.  Bahcall and
others argued \cite{Bahcall2001} that helioseismology is a more severe and detailed
test of the SSM than neutrino production, so that SSM success in
reproducing $c(r)$ made
a particle-physics resolution of the solar neutrino problem more likely.

The sound speed is a function of the Sun's interior pressure and density profiles,
which in turn reflect thermal transport properties that depend on the Sun's metal
content, through the opacity.
Thus the comparison between
helioseismology and the SSM tests a key assumption of the SSM, 
that the metals are distributed uniformly throughout the Sun, apart from small
corrections due to diffusion.  This assumption allows one to equate
SSM interior metal abundances to convective-zone abundances deduced
from analyses of photospheric absorption lines.  Such analyses had been 
based on 1D models of the photosphere.  Recently {\it ab initio} 3D analyses
have been developed, yielding significant improvements in
predicted line shapes and in the consistency
of metal abundance determinations from various atomic and molecular lines.  However,
this work also reduced metallicity estimates from Z $\sim$  0.0169 to $\sim$ 0.0122 \cite{Asplund},
destroying the once excellent agreement between helioseismology 
and the SSM.

It has been suggested that this difficulty may reflect, contrary to the
SSM, differences in solar core and
convective-zone metallicities that could have arisen from the late-stage evolution of the solar disk:
as a great deal of metal was scoured out of the disk by the formation of the giant planets,
the last few percent of gas deposited onto the Sun could have been
depleted of metals \cite{hax08}.  Indeed, recent studies of ``solar twins" show abundance
trends that correlate with the existence of planets \cite{asplanet,liplanet}.  \citet{hax08}
argued that a
direct measurement of solar core metallicity could be made by observing CNO solar
neutrinos.

In both of the above examples -- using neutrinos to determine the solar core
temperature and metallicity -- nuclear physics uncertainties remain one of the limiting
factors in the analyses.

The proposal to revisit in 2009 the deliberations of 1997 thus had several motivations:
\begin{itemize}
\item providing a set of standard S-factors and uncertainties that reflect the progress
made in laboratory and theoretical nuclear astrophysics over the last decade; 
\item enabling more precise analyses of solar neutrino experiments
designed to constrain neutrino oscillations and other new physics, e.g., future pp and
pep neutrino experiments that exploit these well understood fluxes;  and
\item enabling analyses in which solar neutrinos are used as a probe of the solar core. 
\end{itemize}
The 2009 INT workshop\footnote[1]{The workshop was proposed in a letter to the
Institute for Nuclear Theory's National Advisory Committee (NAC) and
approved by the NAC and INT Director at the time of the NAC's
August 2008 annual meeting.  Wick Haxton (lead), Eric Adelberger, Heide Costantini, 
Peter Parker, R. G. Hamish Robertson, Kurt Snover, Frank Strieder, and Michael
Wiescher formed the organizing committee and served as co-editors of this paper.
Additional community members joined this group to act as working group heads: Jiunn-Wei Chen, 
Barry Davids, Stuart Freedman, Alejandro Garcia, Uwe Greife, Michael Hass, 
Gianluca Imbriani, Kuniharu Kubodera, Daniela Leitner,
Laura Marcucci, Filomena Nunes, Tae-Sun Park,
Paolo Prati, Hanns-Peter Trautvetter, and Stefan Typel.  The working group heads
were responsible for organizing discussions, creating section drafts, and 
responding to subsequent criticisms of the drafts.  Organizing committee members, in their
capacity as co-editors, were responsible for creating from the drafts a coherent
document, and for addressing any issues unresolved by the working groups. 
Workshop presentations are archived
on the INT's web site, http://www.int.washington.edu/PROGRAMS/solar\_fusion.html.}
was modeled after that of 1997, with invitations
extended to and accepted by representatives
from most of the experimental groups active in the nuclear physics of hydrogen burning
stars.  There was also active involvement of theorists, reflecting the progress that has
been made in {\it ab initio} calculations. 
The workshop participants are the authors of this
manuscript.  As in 1997, early organizing included the selection of working group leaders
who identified key papers, which were then entered in a database for review, prior
to the start of the workshop.   These materials were then summarized and discussed
during the workshop, as the various working groups considered the state of the data
and outlined any additional work that would be needed for this review.
The process of critically analyzing both new and older data and working toward a consensus
on best-value cross sections and uncertainties continued throughout 2009.  
A few new topics not considered in 1997 but now recognized to be quite important, such as the
shape of the $^8$B neutrino spectrum, were addressed.
(The $^8$B neutrino spectrum is one of the inputs to SNO and 
Super-Kamiokande
analyses.) The workshop included working groups on indirect techniques for
constraining cross sections, to summarize the progress that has been made in validating
such approaches, and on new facilities and instrumentation, in view of the facility
 investments that are being considered
in laboratory nuclear astrophysics (above and below ground).

\subsection{Contents of this review}
The review begins in Section II with a description of
hydrogen burning by the pp chain and CNO bi-cycle, and the
neutrino byproducts of these reaction chains.  The role of S-factors and
the associated questions of screening and of extrapolating data to the solar Gamow
peak are discussed.  We provide a fairly complete overview of 
progress in theory, which in some cases provides our only estimate of
S-factors, and in other cases determines the forms of the functions
that are needed for data extrapolations.

Discussions of individual reactions are organized by chapter: Secs.
III-IX discuss the
pp chain reactions p+p $\rightarrow$ d+e$^+$+$\nu_e$; d+p $\rightarrow$ $^3$He+$\gamma$;
$^3$He+$^3$He $\rightarrow$ $^4$He+p+p; $^3$He+$^4$He $\rightarrow$ $^7$Be+$\gamma$; $^3$He+p $\rightarrow$ $^4$He+e$^+$+$\nu_e$; $^7$Be, pp, and
CNO nuclei electron capture; and $^7$Be+p $\rightarrow$ $^8$B+$\gamma$.
Sec. X discusses the spectrum of $^8$B neutrinos produced in the $\beta$ decay
to a broad resonance in $^8$Be. Sec. XI discusses $^{14}$N+p $\rightarrow$ $^{15}$O+$\gamma$ and other reactions contributing to the CNO cycles.  Sec. XII
describes the progress that has been made in developing and validating indirect methods,
while Sec. XIII describes future facilities and instrumentation that could further advance the
field. 

The conclusions of this review, in some cases, required the working groups to
make some judgments.  There are discrepant data sets, and there are cases where
data extrapolations have some dependence on models.   We have tried to 
treat such questions as consistently as possible,
aware that excessively optimistic treatments of uncertainties could be misleading,
while excessively conservative treatments would degrade
the value of the best experiments done in the field. 
In most cases our working groups
were able to reach consensus.  In cases where significant differences remained
among the experts, we have tried to identify the source of the disagreement, so that
``consumers" will be aware that full consensus may have to await future
measurements.

Table \ref{tab:summary} summarizes the conclusions of this review.

\begin{table*}
\caption{The Solar Fusion II recommended values for S(0), its derivatives, and related quantities, and for the 
resulting uncertainties on S($E$) in the region of the solar Gamow peak -- the most 
probable reaction energy -- defined for a temperature
of 1.55 $\times$ 10$^7$K characteristic of the Sun's center.  See the text for detailed discussions
of the range of validity for each S($E$).
Also see Sec. \ref{sec:ec} for 
recommended values of CNO electron
capture rates, Sec. \ref{sec:N114other} for other CNO S-factors, and Sec. \ref{sec:spectrum} for 
the $^8$B neutrino spectral shape.
Quoted uncertainties are 1$\sigma$.}
\label{tab:summary}
\begin{tabular}{lccccc}
\hline\hline
Reaction~~~~~~ & ~~~~~Section~~~~~&~~~~~S(0)~~~~~ & ~~~~~S$^\prime$(0)~~~~~ & ~~~~~S$^{\prime \prime}$(0)~~~~~&
~~~~~~Gamow peak~~~~~~ \\
 &  & (keV-b) & (b) & (b/keV) & uncertainty (\%) \\
\hline
 p(p,e$^+\nu_e$)d & \ref{sec:s11} & (4.01 $\pm$ 0.04)$\times$10$^{-22}$ &
  (4.49 $\pm$ 0.05)$\times$10$^{-24}$ & $-$ & $\pm$ 0.7 \\
 d(p,$\gamma$)$^3$He & \ref{sec:dp} & (2.14$^{+0.17}_{-0.16}$)$\times$10$^{-4}$& 
 $(5.56^{+0.18}_{-0.20})\times$10$^{-6}$& $(9.3^{+3.9}_{-3.4})\times$10$^{-9}$ & ~~$\pm$ 7.1~\footnote{Error
 from phenomenological quadratic fit.  See text.} \\
 ${}^3$He(${}^3$He,2p)${}^4$He & \ref{sec:s33}& (5.21 $\pm$ 0.27) $\times$ 10$^3$ & $-$4.9 $\pm$ 3.2 & (2.2 $\pm$ 1.7) $\times$ 10$^{-2}$ & ~~$\pm$ 4.3~$^a$ \\
 ${}^3$He(${}^4$He,$\gamma$)${}^7$Be &\ref{sec:s34} &0.56 $\pm$ 0.03 & ($-$3.6 $\pm$ 0.2)$\times$10$^{-4}$~\footnote{S$^\prime$(0)/S(0) taken from theory; error is that due to S(0).  See text.} & (0.151 $\pm$ 0.008)$\times$10$^{-6}$~\footnote{S$^{\prime \prime}$(0)/S(0) taken from theory; error is that due to S(0).  See text.} &  $\pm$ 5.1\\
 ${}^3$He(p,e$^+\nu_e$)${}^4$He &\ref{sec:hep} & (8.6 $\pm$ 2.6)$\times$10$^{-20}$ &$-$ &$-$ & $\pm$ 30~ \\
 ${}^7$Be(e$^-,\nu_e$)${}^7$Li &\ref{sec:ec} & See Eq.~(\ref{eq:bec})& $-$& $-$ & $\pm$ 2.0 \\
  p(pe$^-$,$\nu_e$)d & \ref{sec:ec} & See  Eq.~(\ref{eq:pep3}) &$-$  &$-$ & ~~$\pm$ 1.0~\footnote{Estimated error in the pep/pp rate ratio. See Eq.~(\ref{eq:pep3})} \\
 ${}^7$Be(p,$\gamma$)${}^8$B & \ref{sec:s17} & (2.08 $\pm$ 0.16)$\times$10$^{-2}$~\footnote{Error dominated by theory.} &($-$3.1 $\pm$ 0.3)$\times$10$^{-5}$ & (2.3 $\pm$ 0.8)$\times$10$^{-7}$ & $\pm$ 7.5 \\
 ${}^{14}$N(p,$\gamma$)${}^{15}$O & \ref{sec:N114} & 1.66 $\pm$ 0.12 & ($-$3.3 $\pm$ 0.2)$\times$10$^{-3}$~$^b$ & (4.4 $\pm$ 0.3)$\times$10$^{-5}$~$^c$ & $\pm$ 7.2\\
\hline\hline
\end{tabular}
\end{table*}

\section{NUCLEAR REACTIONS IN HYDROGEN-BURNING STARS}
\label{sec:theory}
Observations of stars reveal a wide variety of stellar conditions,
with luminosities relative to solar spanning a range $L \sim$
10$^{-4}$ to 10$^6$ $L_\odot$ and surface temperatures $T_s
\sim$2000--50000 K. The simplest relation one could propose
between luminosity $L$ and $T_{s}$ is 
\begin{equation}
    L ~ = ~4 \pi R^2 \sigma_\mathrm{SB} ~ T_s^4 \Rightarrow L/L_{\odot} ~=~
    (R/R_\odot)^2 ~ (T_s/T_\odot)^4,
 \label{eq1}
\end{equation}
where $\sigma_\mathrm{SB}$ is the Stefan-Boltzmann constant, and
$L_\odot$, $T_\odot$, and $R_\odot$ are the solar values. This relation
suggests that stars of a similar structure might
lie along a one--parameter path (in this simplified example,
defined by a function of the blackbody radii, $(R/R_\odot)^2$)
in the luminosity (or magnitude) vs.~temperature (or color)
plane. In fact, there is a dominant path in the
Hertzsprung--Russell color--magnitude diagram along which roughly
80$\%$ of the stars reside. This is the main sequence, those stars
supporting themselves by hydrogen burning through the pp chain,
\begin{equation}
4\rm{p} \to {}^4\rm{He} + 2e^+ + 2\nu_e,
\label{eq:burn}
\end{equation}
or CNO cycles.
The laboratory nuclear astrophysics of hydrostatic hydrogen burning
is the focus of this review.

\begin{figure}
\begin{center}
\includegraphics[width=8.5cm]{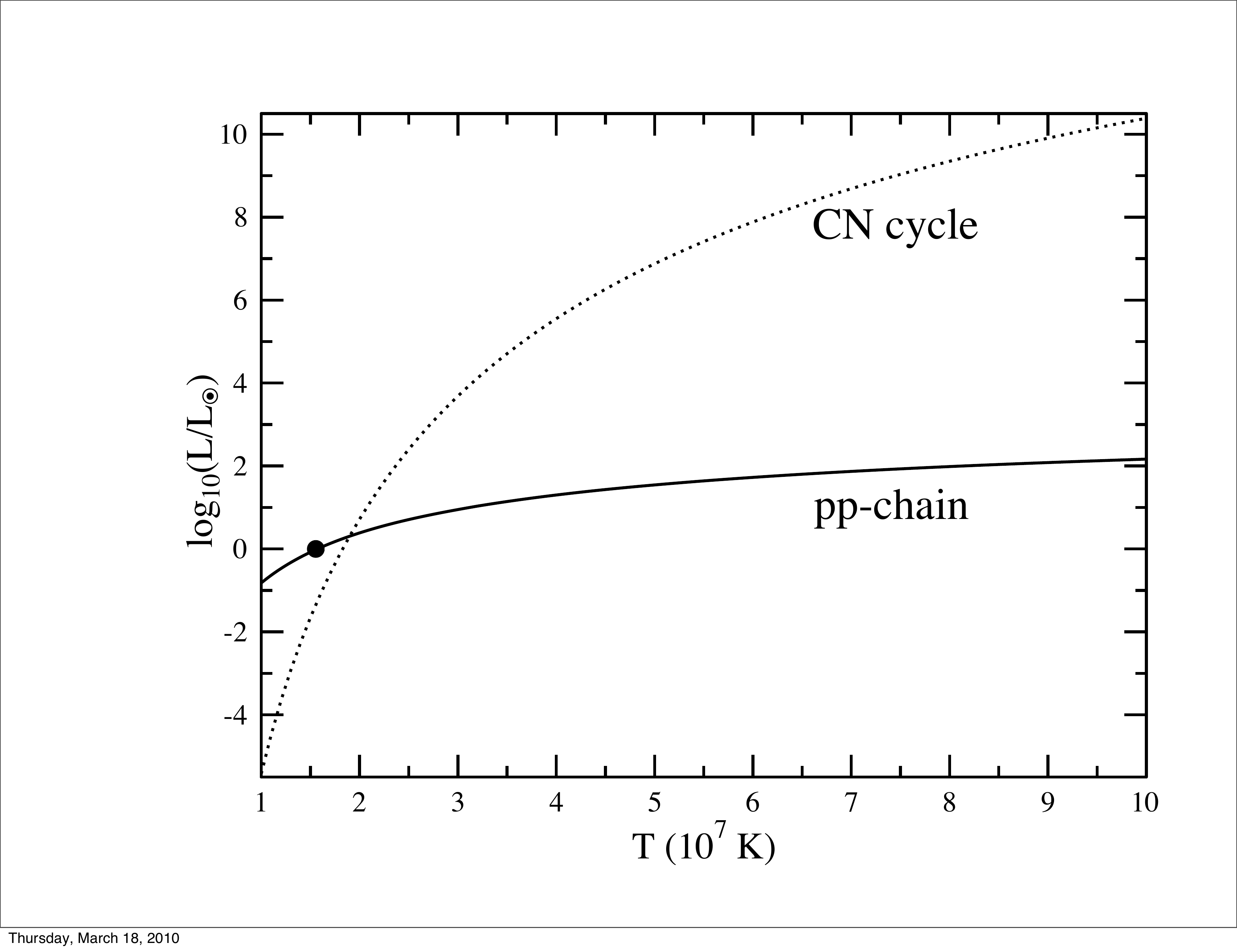}
\caption{The stellar energy production as a function of temperature for the pp chain
and CN cycle, showing the dominance of the former at solar temperatures.  Solar
metallicity has been assumed.  The dot
denotes conditions in the solar core: the Sun is powered dominantly by the pp chain.} 
\label{fig:cnopp}
\end{center}
\end{figure}

\begin{figure*}
\begin{center}
\includegraphics[width=18cm]{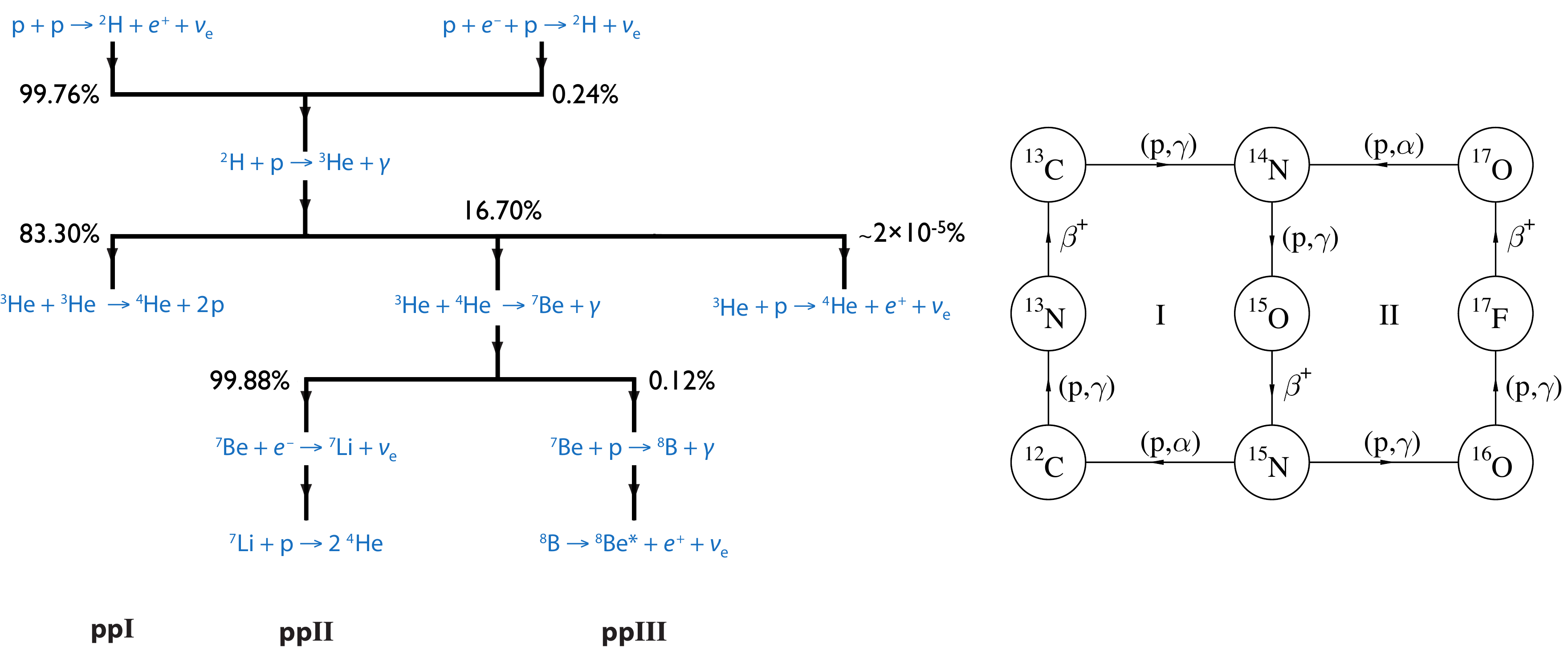}
\caption{The left frame shows the three principal cycles comprising the pp chain (ppI, ppII, and ppIII),
with branching  percentages indicated, each of which is ``tagged" by a distinctive neutrino.
Also shown is the minor branch ${}^3$He+p $\rightarrow$
${}^4$He+e$^+$+$\nu_e$, which burns only $\sim$ 10$^{-7}$ of ${}^3$He, but produces the most
energetic neutrinos.  The right frame shows the CNO bi-cycle.
The CN cycle, marked I, produces about 1\% of solar energy and 
significant fluxes of solar neutrinos.} 
\label{fig:cycles}
\end{center}
\end{figure*}

As one such star, the Sun is an important test of our theory of
main sequence stellar evolution:~its properties~--~age, mass,
surface composition, luminosity, and helioseismology~--~are by far
the most accurately known among the stars. The SSM
traces the evolution of the Sun over the past 4.6 Gyr
of main sequence burning, thereby predicting the
present--day temperature and composition profiles, the relative
strengths of competing nuclear reaction chains, and the neutrino
fluxes resulting from those chains. The SSM makes
four basic assumptions:

\begin{itemize}

\item The Sun evolves in hydrostatic equilibrium, maintaining a
local balance between the gravitational force and the pressure
gradient.  Knowledge of the equation of state as a function of 
temperature, density, and composition allows one to implement
this condition in the SSM.

\item Energy is transported by radiation and convection. The solar
envelope, about 2.6\% of the Sun by mass,  is convective.  Radiative
transport dominates in the interior, $r \lsim 0.72 R_\odot$, and thus
in the core region where thermonuclear reactions take place.  
The opacity is sensitive to composition.

\item The Sun generates energy through hydrogen burning, Eq.~(\ref{eq:burn}). 
Figure \ref{fig:cnopp} shows the competition between the pp chain and CNO
cycles as a function of temperature: the relatively cool temperatures of
the solar core favor the pp chain, which in the SSM produces
$\sim$ 99$\%$ of the Sun's energy.  The reactions contributing to the
pp chain and CNO bi-cycle  are shown in Fig. \ref{fig:cycles}.   The SSM 
requires as input rates for each of the contributing reactions, which are
customarily provided as S-factors, defined below.  Typically cross
sections are measured at somewhat higher energies, where rates are
larger, then extrapolated to the solar energies of interest.  Corrections also
must be made for the differences in the screening environments of 
terrestrial targets and the solar plasma.

\item The model is constrained to produce today's solar radius,
mass, and luminosity.  The primordial Sun's metal abundances
are generally determined from a combination of photospheric and meteoritic
abundances, while the initial $^4$He/H ratio is adjusted to reproduce, after
4.6 Gyr of evolution, the modern Sun's luminosity.

\end{itemize}

The SSM predicts that, as the Sun evolves, the core He abundance increases,
the opacity and core temperature rise, and the
luminosity increases (by a total of $\sim$ 44\% over 4.6 Gyr).  The details of this
evolution depend on a variety of model input parameters and their
uncertainties:  the photon luminosity $L_\odot$, the mean radiative
opacity, the solar age, the diffusion coefficients describing the gravitational
settling of He and metals, the abundances of the key metals, and the
rates of the nuclear reactions.

If the various nuclear rates are precisely known, the competition 
between burning paths can be used as a sensitive diagnostic of the
central temperature of the Sun. Neutrinos probe this competition,
as the relative rates of the ppI, ppII, and ppIII
cycles comprising the pp chain can be determined from the fluxes of
the pp/pep, $^7$Be, and $^8$B neutrinos.  This is one of the
reasons that laboratory astrophysics efforts to provide precise nuclear
cross section data have been so closely connected with solar neutrino
detection.

Helioseismology provides a second way to probe the solar interior,
and thus the physics of the radiative zone that the SSM was designed to describe.
The sound speed profile $c(r)$ has been determined rather precisely over
the outer 90$\%$ of the Sun and, as previously discussed, is now in
conflict with the SSM, when recent abundance determinations from
3D photospheric absorption line analyses are used.

\subsection{Rates and S-factors}
The SSM requires a quantitative description of relevant nuclear reactions.
Both careful laboratory measurements constraining rates
at near-solar energies and a supporting theory of sub-barrier 
fusion reactions are needed.

At the temperatures and densities in the solar interior
(e.g., $T_c \sim$ 15.5 $\times$ 10$^6$ K and $\rho_c
\sim$ 153 g/cm$^3$ at the Sun's center), interacting nuclei reach a
Maxwellian equilibrium distribution in a time that is
infinitesimal compared to nuclear reaction time scales. Therefore,
the reaction rate between two nuclei can be written
 \cite{Burbidge,Clayton}
\begin{equation} \label{eq2}
r_{12} \, = \, \frac{n_1 \, n_2}{1 + \delta_{12}} \, \langle
{\sigma v} \rangle_{12}.
\end{equation}
Here the Kronecker delta prevents double counting
in the case of identical particles, 
$n_1$ and $n_2$ are the number densities of nuclei of type 1
and type 2 (with atomic numbers $Z_1$ and $Z_2$, and mass numbers
$A_1$ and $A_2)$, and $\left\langle {\sigma v} \right\rangle
_{12}$ denotes the product of the reaction cross section $\sigma$ and the
relative velocity $v$ of the interacting nuclei, averaged over the
collisions in the stellar gas,
\begin{equation}
\langle \sigma v  \rangle_{12} ~=~ \int_0^\infty \sigma
(v)~ v ~ \Phi(v)\, \mbox{d} v.
    \label{eq3}
\end{equation}

Under solar conditions nuclear velocities are
very well approximated by a Maxwell--Boltzmann distribution. It
follows that the relative velocity distribution is also a
Maxwell--Boltzmann, governed by the reduced mass $\mu$ of the
colliding nuclei,
\begin{equation}
\Phi(v) \mbox{ d} v = \left( \frac{\mu}{2 \pi k T} \right)^{3/2} \mbox{
exp} \left(-\frac{\mu v^2}{2kT}\right) 4 \pi v^2 \mbox{ d} v.
       \label{eq4}
\end{equation}
Therefore,
\begin{equation}
\langle{\sigma  v}\rangle_{12} \, = \sqrt{ \frac{8}{\pi \mu (kT)^3}} \, \int_0^\infty E ~ \sigma(E) ~
\mathrm{exp}\left(-\frac{E}{kT} \right) \mbox{ d} E,
        \label{eq:sigmav}
\end{equation}
where $E$ is the relative kinetic energy and $k$ is the Boltzmann constant.
In order to evaluate $\langle {\sigma v} \rangle_{12}$
the energy dependence of the reaction cross section
must be determined. 

Almost all of the nuclear reactions relevant to solar energy generation
are nonresonant and charged--particle induced.  For such reactions
it is helpful to remove much of the
rapid energy dependence associated with the Coulomb barrier, 
by evaluating the probability of s-wave scattering off a 
point charge.  The nuclear physics (including effects of finite nuclear size,
higher partial waves, antisymmetrization, and any atomic screening
effects not otherwise explicitly treated) is then isolated in
the S-factor, defined by
\begin{equation}
\sigma \left( E \right) =\frac{\mathrm{S} \left( E
\right)}{E}\;\;\mathrm{exp}\left[ {-2\pi \eta(E) } \right],
\label{eq:S}
\end{equation}
with the Sommerfeld parameter 
$\eta(E)$ = $Z_1 Z_2$ $\alpha$/$v$, where $v=\sqrt{2E/\mu}$ is the relative velocity and
$\alpha$ the fine structure constant
($\hbar$ = c = 1).  Because the S-factor is slowly varying, one can extrapolate
S$(E)$ more reliably from
the range of energies spanned by data to the lower energies characterizing the
Gamow peak.

A substitution of Eq.~(\ref{eq:S}) into Eq.~(\ref{eq:sigmav}) followed by a
Taylor expansion of the argument of the exponentials then yields \cite{Bahcallbook}
\begin{eqnarray}
\langle \sigma v \rangle_{12}~&=&~\sqrt{2 \over \mu k T} {\Delta E_0 \over kT}
f_0~\mathrm{S_{eff}} \exp \left[-3E_0/ (kT) \right] \nonumber \\
&=& 1.301 \times 10^{-14}~\mathrm{cm^3/s}
 ~ \left( Z_1 Z_2 \over A \right)^{1/3}  \nonumber \\
 &\times& f_0~
{\mathrm{S_{eff}} \over \mathrm{~MeV~b~}}~
T_9^{-2/3} \exp \left[-3 E_0/ (kT) \right],
\label{eq:taylor}
\end{eqnarray}
where
\[ {E_0 \over kT} = (\pi Z_1 Z_2 \alpha/ \sqrt{2})^{2/3} \left[\mu /(kT)\right]^{1/3} ,\]
\[ {\Delta E_0 \over kT} = 4 \sqrt{{E_0 \over 3 kT}},~~~~~ A={A_1 A_2 \over A_1+A_2}, \]
and 
\begin{eqnarray}
\mathrm{S_{eff}} &=& \mathrm{S}(0)~ \left(1 + \frac{5kT}{36E_0} \right)+
\mathrm{S}^{\prime} (0) E_0 ~ \left(1 + \frac{35kT}{36E_0} \right) \nonumber \\
&+& \frac{1}{2}
\mathrm{S}^{\prime \prime} (0) E_0^2 ~ \left(1 + \frac{89kT}{36E_0}\right). \nonumber
\end{eqnarray}
$E_0$, the Gamow peak energy where the integrand
of Eq.~(\ref{eq:sigmav}) takes on its maximum value,
is the most probable energy of reacting
nuclei. $\Delta E_{0}$ corresponds to the full width
of the integrand at 1/$e$ of its maximum value, when approximated as a Gaussian. 
Equation~(\ref{eq:taylor}) includes a factor $f_0$, discussed below,
to correct for the effects of electronic screening on nuclear reactions
occurring in the solar plasma.

Rates in an astrophysical plasma can be calculated given
S($E$) which by virtue of its slow energy dependence, in the case
of non-resonant reactions, can be approximated
by its zero-energy value S($0$) and possible corrections determined
by its first and second derivatives, S$^{\prime}(0)$ and
S$^{\prime \prime}(0)$.  It is these quantities that we need to
determine by fitting laboratory data, or in cases
where such data cannot be obtained, through theory. 
For most of the reactions contributing to the pp chain and
CNO bi-cycle, data have been obtained only for energies
in regions above the Gamow peak, e.g., typically $E \gsim 100$ keV, so
that extrapolations to lower energies depend on the quality
of the fit to higher energy data.  Ideally one desires a fitting function
that is well motivated theoretically and tightly constrained by the 
existing, higher-energy data.  The purpose of this
review is to provide current best values and uncertainties for S$(0)$ and,
if feasible, its derivatives.

S-factor uncertainties, when folded into SSM calculations, then limit 
the extent to which that model can predict observables, such as the 
depth of the convective zone, the sound speed profile, and the neutrino fluxes.
It has become customary in the SSM to parameterize the consequences
of input uncertainties on observables  through logarithmic partial derivatives,
determined by calculating the SSM response to  variations
in individual input parameters.  SSM compilations of the logarithmic 
partial derivatives provide, for example, a way to
assess the importance of each S-factor uncertainty on neutrino
flux predictions.

The partial
derivatives $\alpha(i,j)$  for each neutrino  flux $\phi_i$ and 
SSM input parameter $\beta_j$ are defined by
\begin{equation}
\alpha(i,j)  \equiv  {\partial   \ln{\left[  \phi_i/\phi_i(0)  \right]}  \over
\partial \ln{\left[ \beta_j / \beta_j(0)\right]}}
\end{equation}
where  $\phi_i(0)$  and  $\beta_j(0)$   denote  the  SSM  best  values.    The $\alpha(i,j)$  
for 19 SSM input parameters $\beta_j$ are given by
\citet{PGS2008} in their 2008 SSM update.
The $\beta_j$ include parameters such as the
Sun's age and luminosity, the abundances of important metals,
and S-factors.  

The partial derivatives
define  the power-law  dependencies of
neutrino fluxes with respect to the SSM best-value prediction $\phi_i(0)$,
\begin{equation}
\phi_i   =   \phi_i(0)  \prod_{j=1}^N   \left[   {\beta_j  \over   \beta_j(0)}
\right]^{\alpha(i,j)}=\phi_i(0)  \prod_{j=1}^N   \left[ 1+  \delta \beta_j
\right]^{\alpha(i,j)},
\label{eq:prod}
\end{equation}
where the product extends over  $N$ SSM input parameters,
and where $\delta \beta_j \equiv \Delta \beta_j/\beta_j(0)$ is the
fractional uncertainty of input parameter $\beta_j$ with respect to its
SSM best value. This expression 
separates the impact of SSM parameter variations on $\phi_i$ into a
solar piece -- the infinitesimal SSM response described by $\alpha(i,j)$ --
and a laboratory or theory piece -- the estimated uncertainty $\delta \beta_j$ of 
an input parameter (in our case, that of an S-factor).   From SSM tabulations
of the $\alpha(i,j)$, one can estimate the change in a 
SSM flux prediction $\phi_i$, when a given
SSM parameter $\beta_j$ is perturbed away from its SSM best value
by an amount $\delta \beta_j$, without redoing the SSM calculation.
For example, to assess the impact of an improved
nuclear cross section measurement on $\phi_i$, 
one sets $\delta \beta_j$ to the estimated uncertainty of the corresponding S-factor,
to obtain the corresponding variation in $\phi_i$.
In this way one can identify nuclear physics improvements that will
have the most impact on reducing flux uncertainties.
Alternatively, the process can be
inverted:  a flux  measurement  could in  principle  be used  to constrain  an
uncertain input parameter.

For example, \citet{PGS2008} define the dependence of $\phi(^8$B) on the
S-factors under discussion here,
\begin{eqnarray}
\phi(^8 \mathrm{B}) \propto  (1 + \delta \mathrm{S}_{11} )^{-2.73}
                                                   (1 + \delta \mathrm{S}_{33})^{-0.43}
                                                   (1 + \delta \mathrm{S}_{34})^{0.85} &&\nonumber \\
                                                 \times   (1 + \delta \mathrm{S}_{17})^{1.0}
                                                   (1 + \delta \mathrm{S}_{e7} )^{-1.0}
                                                   (1 + \delta \mathrm{S}_{1~14} )^{-0.02},~~~&&
\end{eqnarray}
where S$_{11}$ denotes the S-factor for p+p reaction, etc.,
and $\delta S_{11} \equiv \Delta \mathrm{S}_{11}/\mathrm{S}_{11}(0)$ denotes its fractional uncertainty.
This review gives the best current values
for the needed $\delta S$s.                                   

\subsection{Screening of stellar and laboratory reactions}
\label{sec:screen}
One must take into account differences in the atomic environments
to correctly relate screened laboratory  and solar cross sections, 
\textit{$\sigma_\mathrm{s}^\mathrm{lab}$}$(E)$ and
\textit{$\sigma_\mathrm{s}^\mathrm{solar}$}$(E)$, to each other or to the
underlying bare cross section $\sigma _\mathrm{b}(E)$.  Screening enhances solar cross sections by
reducing the Coulomb barrier that reacting ions must overcome. 
As light nuclei in the solar core are almost
completely ionized, the solar electron screening correction $f_0$,
\begin{equation}
 f_0(E) \equiv {\sigma _\mathrm{s}^\mathrm{solar}(E) \over \sigma _\mathrm{b}(E)},
\end{equation}
can be treated in a weak--screening approximation \cite{Sal54}.
The impact of the modified potential,
\begin{equation}
V(r) = {\alpha Z_1 Z_2 \over r} \exp\left(-{r \over R_D} \right),
\end{equation}
on reactions depends on the ratio of the Coulomb potential
at the Debye radius $R_D$ to the temperature,
\begin{equation}
    f_0 \sim \;\mathrm{exp} \left({Z_1 \,Z_2 \;\alpha \over R_D k T}\right)
    =\;\mathrm{exp} \left( {0.188\,Z_1 \,Z_2 \;\zeta \;\rho_0
    ^{1/2}\;T_6^{-3/2} } \right),
        \label{eq:eq7}
\end{equation}
where $\zeta R_D = \left[ k T / (4 \pi \alpha \rho) \right]^{1/2}$,
$\rho$ is the number density of nucleons, $\rho_0$ is a dimensionless density
measured in g/cm$^3$,
$\zeta = \left[ \sum\limits_i~X_i ~
 \left(Z_i^2/A_i \right) +  \left( f_0^\prime / f_0 \right) \sum\limits_i~X_i \left( Z_i / A_i \right) \right]^{1/2}$,
$X_i$ is the mass fraction of nuclei of type $i$,
and $T_6$ is the dimensionless temperature in units of 10$^6$ K. The factor
$f_0^\prime/f_0 \sim 0.92$ corrects for the effects of electron degeneracy
in the solar core \cite{Sal54}. 

The weak-screening approximation requires the average interaction energy
between particles to be  smaller than the average particle kinetic energy 
\cite{Kob95,Bai95}.  This places a constraint on the argument
of Eq.~(\ref{eq:eq7}), $Z_1 \,Z_2 \alpha / \left( R_D k T \right) \ll$ 1, that is satisfied 
in the solar core  if $Z_1 Z_2 \lsim 10$ \cite{Gruzinov98}, a condition met by the
low-Z reactions of the pp chain and CNO bi-cycle.   However corrections 
to the Salpeter formula are expected at some level.  
Nonadiabatic effects have been suggested as one source, e.g.,  
when a high Gamow energy guarantees
reacting nuclei having velocities significantly higher than the typical
ion velocity, so that the response of slower plasma ions might be suppressed.  At the
time of Solar Fusion I such dynamic corrections
were a source of controversy.  Dynamic corrections were first discussed 
by \citet{Mit77} and later studied by \citet{CSK88}.
Subsequent work showed that Salpeter's formula would be valid
independent of the Gamow energy due to the nearly precise thermodynamic
equilibrium of the solar plasma \cite{BS97,Gruzinov98,GB98}.  The arguments,
summarized in Solar Fusion I, were significantly
extended in 2002 by \citet{BBGS02}, who pointed
out a number of contradictions in investigations claiming larger corrections,
and showed that a field theoretic approach led to the expectation
of only small ($\sim$ 4\%) corrections to the standard formula, for solar conditions. 
However controversies have not entirely died out \cite{Mao09}.

The Salpeter correction relates the solar and bare cross sections, 
\textit{$\sigma_\mathrm{s}^\mathrm{solar}$}$(E)$ and \textit{$\sigma_\mathrm{b}$}$(E)$.
As the reactions
studied in the laboratory generally involve target nuclei bound in
neutral atoms and molecules, not bare ions, a second step is needed to
extract \textit{$\sigma_\mathrm{b}$}$(E)$ from laboratory data.  As in the Sun, electrons
in the laboratory target tend to reduce the barrier, so that the screened
cross section $\sigma_\mathrm{s}^\mathrm{lab}(E)$ will exceed that for bare ions $\sigma_\mathrm{b}(E)$.
The enhancement is given by \cite{Ass87}
\begin{equation}
 f_\mathrm{lab}(E) \equiv {\sigma _\mathrm{s}^\mathrm{lab}(E) \over \sigma _\mathrm{b}(E)} \sim
\mathrm{exp} \left[ {\pi \eta(E) U_{e} \over E} \right] \ge 1~~\mathrm{for}~~U_{e} \ll E,
    \label{eq13}
\end{equation}
where $U_e$ is an electron--screening potential energy. This
energy can be estimated from the difference in
atomic binding energies between the compound atom and the
projectile plus target atoms of the entrance channel.  Because the
correction depends on the ratio $U_e/E$, one expects screening
corrections to be most important for very low projectile energy.

In contrast with the case of solar screening, a great deal
can be done experimentally 
\cite{Ass87,Rol95,Rol01,Eng88,Eng92,Ang93,Prati94,Gre95} 
to test our understanding of electron
screening in terrestrial targets.  Studies of reactions involving light nuclei
\cite{Eng88,Strieder2001} revealed an upturn in cross section at low energies,
as predicted by Eq.~(\ref{eq13}).  For
example, results for $^3$He(d,p)$^4$He \cite{Aliotta2001} could
be represented by Eq.~(\ref{eq13}) for a screening potential
 $U_e$ = 219$\pm $15 eV.  While this potential is 
 significantly larger than the one
obtained from the adiabatic approximation, $U_{ad}$ = 119 eV, the
analysis requires one to assume an energy dependence
of the bare cross section $\sigma _\mathrm{b}(E)$.   This adds a difficult-to-quantify theoretical 
uncertainty to the extracted potential. 
It may be possible
to remove much of this uncertainty through an indirect measurement
of $\sigma_\mathrm{b}(E)$ by the Trojan Horse Method \cite{lattuada01,spitaleri01,Strieder2001,Tumino03}.

There exist various surrogate environments that have been exploited by
experimentalists to test our understanding of plasma screening effects.
Screening in d(d,p)t has been studied for gaseous targets and for
deuterated metals, insulators, and semiconductors \cite{Raiola2004}.   For a summary of the
results see \citet{hpr}:  it is believed that the quasi-free valence
electrons in metals create a screening environment quite similar
to that found in stellar plasmas.  Experiments in metals have confirmed
important predictions of the Debye model, such as the temperature
dependence $U_e(T) \propto T^{-1/2}$.

The tendency of experimentally determined values of $U_e$ to exceed
theoretical estimates by a factor $\sim$ 2 has been noted by 
\citet{Ass87,Rol95,Rol01}.  Various possible explanations
have been considered \cite{Sho93,BBH97,FZ99,HB02,Fio03}. 
A possible solution of the laboratory screening problem was proposed
in \citet{LSBR96} and in \citet{BFMH96}, that
the stopping of ions in matter differs at low energy
from that obtained by extrapolating  from stopping power 
tables at higher energies \cite{AZ77}. Smaller stopping powers
were indeed verified experimentally \cite{GS91,Rol01} and explained theoretically 
\cite{BD00,bert04}. 

Screening corrections for laboratory reactions are important in extracting
S-factors in cases where data extend to very low energies.   In this review two cases of
interest are $^3$He+$^3$He $\rightarrow$ p+p+$^4$He,
where the lowest data point is at $E$ = 16 keV, and $^{14}$N(p,$\gamma$)$^{15}$O,
where measurements extend down to 70 keV.

\subsection{Fitting and extrapolating S-factors}
S$(0)$ (and its derivatives S$^\prime$(0) and S$^{\prime \prime}$(0))
needed in Eq.~(\ref{eq:taylor})
could be taken from a polynomial fit to data.
A quadratic form often provides an excellent representation of
the data up to a few hundred keV.   However, as the procedure
is purely empirical, it provides no theoretical justification
for extrapolating beyond the last known data point.
For example, a quadratic fit to the laboratory data for
${}^{7}$Be(p,$\gamma$)${}^{8}$B would miss the upturn
in the S-factor at low energy expected from theory, as this
increase occurs beyond the range of existing data.  For
this reason, we restrict our use of empirical fitting
functions to cases where the data sets encompass the
full range of energies relevant to astrophysics.

\subsubsection{Theory constraints: model-based methods}
\label{sec:theory1}
One class of important theoretical constraints makes use of the peripheral
nature of non-resonant radiative capture reactions close
to the threshold.  If the reaction occurs at separations much larger than
the sum of the nuclear radii, one can derive the coefficients for the 
Taylor series for S$(E)$ independent of models, as
only the asymptotic forms of the 
bound and scattering initial- and final-state wave functions
are relevant. This idea has been exploited in several ways.

\citet{wil1981} used Bessel function expansions of
Coulomb wave functions and a hard-sphere approximation to derive
an expansion
of the low-energy logarithmic derivative, 
\begin{equation}
{1 \over \mathrm{S}(E)} {d\mathrm{S}(E) \over dE} = a + b E.
\end{equation}
This approach was further developed by \citet{muk2002},
who considered variables such as the remnant Coulomb barrier, the
initial and final centrifugal barriers, and the binding energy (but not the
interactions of the colliding nuclei in the entrance channel).  They found
that the near-threshold behavior of S$(E)$ could be sensitive to
such parameters.
Baye and collaborators, employing zero-energy 
solutions of the Schr\"{o}dinger equation
and their energy derivatives, showed that model-independent values for the
coefficients in the Taylor expansion for S$(E)$ around $E=0$
could be extracted from the
asymptotic normalization coefficient (ANC) of the bound state wave
function and the scattering lengths of the scattering states,
thus including effects from interactions in the continuum
\cite{bay2000a,bay2004,bay2005,bb2000}. 

Despite the successful application of the Taylor series expansion for S$(E)$,
it was noticed that the series has a restricted domain of convergence,
determined by the binding energy $E_{B}$ of the final state.  This is 
a consequence of a pole in the relevant radial integral at $E=-E_{B}$
\cite{jen1998a,jen1998b,bay2000a}.
This limitation becomes particularly severe for weakly bound nuclei:
for ${}^{7}$Be(p,$\gamma$)${}^{8}$B,
$|E_{B}| \sim 138$~keV barely reaches the domain of experimental data.
Thus the alternatives of a Laurent expansion of the S-factor
in the photon energy $E_{\gamma} = E+E_{B}$,
an expansion of $(E+E_{B})S(E)$, and the
explicit treatment of the pole have been explored as alternatives
in the analysis of experimental data \cite{cyb2004,cyburt08}.
See also \citet{typ2005} for explicit expressions of the cross sections
without the convergence limitation.

Model-based calculations of fusion cross sections also provide
a template for fitting and extrapolating experimental
data. Models can be constrained
by the known properties of the system under study and can be applied over
a wide range of energies.  While they often predict the energy dependence of S($E$)
accurately, in many cases an overall renormalization is needed
to give the correct magnitude of the S-factor. The 
need for this scaling is qualitatively understood, as model
calculations of interior wave functions are generally done in restricted
spaces, and thus lack high-momentum (and certain low-momentum)
components of the true wave function, with consequences for the
normalization.  (The goal of predicting both the shape and normalization
of S-factors is motivating the development of quasi-exact {\it ab initio} methods,
as discussed below.)

Modeling approaches involve various levels of complexity.
The simplest microscopic reaction theories are the potential models, in 
which the internal structure of the colliding
nuclei is ignored.
The dynamics of the process is reduced to a single 
coordinate, the distance vector between the two nuclei.  The potential-model Hamiltonian
is typically a phenomenological one, e.g., a Woods-Saxon potential, with
parameters that can be determined by fitting data, such as the
elastic cross section.

More realism is provided by cluster models
like the resonating group method (RGM) or the generator-coordinate
method (GCM), which take into account the many-body substructure
of the reacting nuclei. These models employ fully antisymmetrized many-body
wave functions of the compound system, though constructed in a restricted model
space. The full wave function is described as a superposition
of many-body cluster wave functions of fixed internal structure
moving against each other. The interaction is described by
phenomenological nucleon-nucleon potentials with parameters
that are adjusted for each
reaction under consideration.

Another description of fusion reaction cross sections comes from the R-matrix. 
Space is divided into two regions, the interior
where nuclear forces are important, and the exterior where the
interaction between the nuclei is assumed to be only
Coulombic. The full scattering wave function connecting
different channels $i$ is expanded
in partial waves with total angular momentum $J$.
The Schr\"{o}dinger equation for the interior
Hamiltonian is solved, with
boundary conditions at the channel radii $a_{i}$ encoding
the correct asymptotic behavior.
The solutions of the Schr\"{o}dinger
equation determine the level energies $E_{\lambda}$
and reduced widths $\gamma_{\lambda i}$ that appear
in the expression for the R-matrix 
\begin{equation}
R_{ij}(E)= \sum_{\lambda =1}^N 
\frac{\gamma_{\lambda i} \gamma_{\lambda j}}{E-E_\lambda},
\end{equation}
for each $J$,
in the standard approach of \citet{lan1958}.
Simple expressions relate the reaction cross sections at energy $E$ 
to the R-matrix.
The cross section should be insensitive
to the choice of the channel radii.
In most applications the R-matrix is viewed as a parameterization
of measured reaction cross sections in terms of fitted
level energies and reduced widths. A connection
to an underlying reaction model is not required.   The R-matrix 
allows one to properly account for penetrability effects, and to
adjust the complexity of the fitting in response to various practical
considerations, such as the energy range of interest.

R-matrix resonance parameters (level energies
and reduced widths) are not directly comparable
to the experimental quantities due to level shifts
associated with the chosen boundary conditions. 
Generalizing earlier ideas of \citet{bar1971} and 
\citet{ang2000}, an alternative
parametrization of R-matrix theory has been developed by 
\citet{bru2002}
where all level shifts vanish and the partial widths and level
energies are identical to the observed parameters. 
This approach simplifies the incorporation of known nuclear
properties in the fitting procedure and the comparison with
experimental resonance properties.

\subsubsection{Theory constraints: {\it ab initio} methods}
\label{theoryabinitio}
{\it Ab initio} methods -- defined here as methods that provide a quasi-exact
solution to the many-body Schr\"{o}dinger equation, such as the hyperspherical harmonic expansion (HH)
and Green's function Monte Carlo (GFMC) methods, 
or that express observables in terms of a controlled expansion, such as effective field theory --
play two critical roles.
Two reactions discussed in this review, p+p $\rightarrow$ d+e$^+$+$\nu_e$
and $^3$He+p $\rightarrow$ $^4$He+e$^+$+$\nu_e$, are presently beyond the reach of
experiment.  Thus we are entirely dependent on theory for the corresponding
S-factors.  The convincing demonstration that  the rate for p+p $\rightarrow$ d+e$^+$+$\nu_e$
can be calculated to a precision of  $\lsim$ 1\% is one of the important achievements
of {\it ab initio} nuclear theory, as described in Sec. \ref{sec:s11}.

Furthermore, {\it ab initio} methods potentially could be applied to all other reactions
in the pp chain (and, farther in the future, to the CNO bi-cycle) to provide a more reliable
basis for extrapolating data.  One of the impressive examples of progress to
date, the agreement between and data for d(p,$\gamma)^3$He
and theory (calculations employing variational HH wave functions 
in combination with an electromagnetic current operator with both one- and
two-body components),
is discussed in Sec. \ref{sec:dp} and illustrated in Fig. \ref{fig:f_pd}.

{\it Ab initio} methods break into two broad categories, potential-based calculations
and effective field theory expansions.  The former are distinguished from model-based methods
discussed in Sec. \ref{sec:theory1} in two regards.  First, they use a realistic interaction that fits two-body
scattering data in detail, as well as certain bound-state properties of 
the lightest nuclei.   Thus the
interaction has both a rich operator structure and an explicit treatment of the short-distance
repulsive core.  Second, they combine this potential,
\begin{equation}
H_{A}=\sum_{i=1}^{A}t_{i}\,\,+\,\sum_{i<j}^{A}v_{ij}^{\mathrm{phen}
}+\sum_{i<j<k}^{A}v_{ijk}^{\mathrm{phen}}\,,  \label{HA}
\end{equation}
with numerical techniques that can accurately treat an interaction of such complexity and with
such disparate spatial scales,
producing a quasi-exact solution of the many-body
Schr\"{o}dinger equation.  The form of the three-body potential in Eq.~(\ref{HA}), which contributes for
A $\geq$ 3 but plays a less important role than the dominant two-body
potential, is typically taken from theory.
Once the wave functions are obtained, they can be
combined with electroweak transition operators to produce estimates of
observables.  The transition operators include both one-body terms 
determined from the coupling of
the single nucleon to the electroweak current, and two-body corrections, 
typically derived from one-boson-exchange
diagrams.  Examples of the potential approach, including
discussions of the associated issue of 
transition operators, are found in Secs. \ref{sec:s11}, \ref{sec:dp}, and \ref{sec:hep}.

The second approach is based on effective field theory (EFT).  EFTs
exploit the gap between the long-wavelength properties of nuclei that govern
nuclear reactions near threshold, and the short-range interactions in the NN 
potential that make an exact solution of the Schr\"{o}dinger equation technically
difficult.  The calculations are restricted to a limited basis describing the
long-wavelength behavior, and the omitted degrees of freedom are absorbed into effective
operators that can be organized in powers of $Q/\Lambda_\mathrm{cut}$, where
$Q$ is the momentum characterizing the physics of interest and $\Lambda_\mathrm{cut}$
is the momentum characterizing the omitted physics.  If carried out completely,
no simplification is achieved, because the low-momentum EFT Lagrangian has
an infinite number of such operators.  EFT becomes useful when there is a
significant gap between $Q$ and $\Lambda_\mathrm{cut}$, so that only a small number
of the effective operators corresponding to the leading powers in $Q/\Lambda_\mathrm{cut}$
must be retained,  to reproduce long-wavelength
observables to a specified accuracy.  The coefficients of the leading operators can
then be determined by fitting data: if enough constraints exist to fix all of the needed
low-energy constants, then accurate predictions can be made about new processes.
The application of this method to p+p $\rightarrow$ d+e$^+$+$\nu_e$
and $^3$He + p $\rightarrow$ $^4$He + e$^+$ + $\nu_e$ 
is described in some detail in Secs. \ref{sec:s11} and \ref{sec:hep}, respectively.  This approach can also be
applied to d(p,$\gamma)^3$He. 

One of the potential-based methods now being developed for reactions should be highlighted
because of its established success in predicting bound-state properties throughout
most of the $1p$ shell.  The quantum Monte Carlo (QMC) approach combines the
variational Monte Carlo (VMC) and GFMC
methods \cite{P08}.  The VMC calculation produces an approximate wave function by minimizing the energy
of a variational wave function including elaborate two- and three-body correlations.
The GFMC method is then employed to make the needed small
improvements to the VMC result required for a true
solution to the Schr\"{o}dinger equation.

The GFMC method requires a local potential, so its use has been
restricted to the Argonne $v_{18}$ NN potential \cite{wir95},
denoted AV18.  There is also an important three-nucleon interaction,
determined by fitting 17 bound- and narrow-state
energies for $A\leq 8$ \cite{PPWC01}.  The high quality of the 
QMC predictions for  energies of bound states
and sharp resonances in nuclei with
$A\leq 12$, and for charge radii, electromagnetic moments, and other
observables, has been thoroughly established \cite{PPWC01,PVW02,PWC04}.  

Recent VMC-based calculations of capture cross sections
using realistic potentials \cite{NWS01,N01,mnsw06}
represent a first step in extending the QMC program to reactions.  These calculations used VMC
wave functions for bound states in $^3$H, $^3$He, $^4$He, $^6$Li,
$^7$Li, and $^7$Be, as well as an exact deuteron.  Initial states in
the reactions $\mathrm{d}(\alpha,\gamma)^6\mathrm{Li}$,
$^3\mathrm{H}(\alpha,\gamma)^7\mathrm{Li}$, and
$^3\mathrm{He}(\alpha,\gamma)^7\mathrm{Be}$ were computed as products
of the reactant VMC wave functions and a correlation, matched to experimental
phase shifts, to describe the
relative motion of the interacting nuclei.  Work has focused, in particular, on building in the proper
long-range clustering of the final states, as this is important in
reproducing the proper energy dependence of S-factors.  Results for
$^3\mathrm{H}(\alpha,\gamma)^7\mathrm{Li}$ closely match the
measured absolute S-factor.  However, the prediction for
$^3\mathrm{He}(\alpha,\gamma)^7\mathrm{Be}$ lies below the data by
about a factor of 1.3 to 1.45.

Better QMC calculations of those and other cross sections are
possible.  VMC wave functions were used partly because of the
technical difficulty of computing quantities off-diagonal in the
energy eigenstates using GFMC; this problem has now been solved, and
electroweak matrix elements between discrete levels have been computed
\cite{PPW07,MPPSW08}.  Scattering wave functions are also now being
computed directly from the NN+NNN potential, with successful
calculations of low-energy neutron-$^4$He scattering wave functions
reported by \textcite{NPWCH07} using particle-in-a-box formulations of
the QMC methods.

While we have used the QMC approach to illustrate the progress in
quasi-exact approaches, there are other important efforts underway to
compute cross sections beyond A=4 from realistic NN potentials.
Examples include the {\it ab initio} no-core shell model both alone
\cite{navratil06a,navratil06b} and in combination with the resonating
group method \cite{quaglioni09}; the Lorentz integral transform method \cite{efros07};
and the unitary correlation operator
method \cite{neff08}.  The hypersherical harmonics method, which
will be discussed in connection with the d(p,$\gamma)^3$He and hep reactions, is also being extended
to heavier systems.

We anticipate that quasi-exact methods will soon be practical for many
scattering and capture processes in light nuclei.
Calculations based on exact solutions of accurate interactions will
predict not only the energy dependences of solar fusion reactions but also absolute cross sections.
Theory may thus provide a firm basis for
validating and extrapolating data and for
resolving systematic differences between measured data sets.

\subsubsection{Adopted procedures}
These are the procedures we adopt for fitting and extrapolating data:
\begin{itemize}
\item In two cases, p+p $\rightarrow$ d+e$^+$+$\nu_e$ and
$^3$He+p $\rightarrow$ $^4$He+e$^+$+$\nu_e$, S-factor estimates depend
entirely on theory.  The goal in such cases should be the application of both
potential and EFT or EFT-inspired methods, yielding consistent results with
quantified uncertainties.  As detailed in Sec. \ref{sec:s11}, one is close to 
achieving this for S$_{11}$, with two methods providing consistent answers
and uncertainties of $\lsim$ 1\%, and with a third method
(EFT) potentially reaching similar precision, if ancillary measurements
can better determine the needed low-energy constant.   In the case of S$_\mathrm{hep}$,
a less critical cross section, the further developments of methods like Green's
function Monte Carlo will provide an important check on the current state-of-the-art,
a variational calculation in which a correlated hyperspherical
harmonics expansion was used.
\item In cases where data exist through the energy range of astrophysical 
interest, much can be done independent of theory.  
A polynomial representation of S($E$), e.g., values for
S(0), S$^\prime$(0), and S$^{\prime \prime}$(0), could be obtained by directly fitting the data.

However, as S($E$) represents the bare cross section,
theory may still be needed to remove the effects of screening in the terrestrial target.
As detailed above, there is some confidence that theory determines the functional
form of the screening (Eq.~(\ref{eq13})), so that such effects can be subtracted given
sufficient low-energy data to fix the numerical value of the screening potential (which theory
appears to predict less reliably).  This issue arises in S$_{33}$.
\item In cases where data exist but are not adequate to fully characterize the cross section
in the region of astrophysical interest, we advocate the use of fitting functions
motivated by theory to extrapolate data, with data determining the normalization.
To the extent that well-justified models differ in their predictions, 
additional uncertainties must be assigned to S(0) and its derivatives.   
Judgment is required in assessing the models and determining how they 
should be applied, e.g., the range in $E$ over which a given model is likely
to be valid.   Each
working group was asked to consider such issues, and to present and justify
the procedures it followed to assess associated fitting uncertainties.
\end{itemize} 

\subsection{Treatment of uncertainties}
The treatment of uncertainties -- the statistical and systematic errors in data and the
impact of imperfect theory in fitting and extrapolating data -- is 
discussed in some detail in the Appendix.  There are cases where several
high-quality data sets exist, each with errors that presumably reflect both
the statistical and evaluated systematic uncertainties of the experiment, that disagree by
more than the error bars would indicate.  In treating such cases, an error-bar
``inflation factor" is commonly introduced, to account for the apparent
underestimation of systematic errors.  We have done so
following Particle Data Group (PDG) conventions  \cite{Amsler:2008zzb}, with one minor
modification described in the Appendix.  Uncertainties quoted in this review 
correspond to one standard deviation (68\% confidence level).

As discussed in the Appendix, there are alternative prescriptions for apportioning the
unidentified systematics -- and thus the inflations -- 
among the experiments that disagree.
However our group concluded that the PDG procedure was the
best choice both for technical reasons and because the procedure is
widely used in the physics community.

\section{THE \lowercase{pp} REACTION} 
\label{sec:s11}
The rate for the initial reaction in the pp chain,  p+p$\rightarrow \mathrm{d}+e^{+}+\nu _{e}$,
is too small to be measured in the laboratory. Instead, this cross
section must be calculated from standard weak
interaction theory.

As in Solar Fusion I, the adopted value and range for the logarithmic derivative
is taken from
\textcite{Bahcall:1968wz},
\begin{equation}
\mathrm{S}^{\prime }_{11}(0)= \mathrm{S}_{11}(0)~(11.2\pm 0.1)\,\mathrm{MeV}^{-1}.
\label{logderivative}
\end{equation}
This result is in excellent agreement with those obtained from linear fits to the modern
potential-model calculations of \citet{Sch98}, which yield values of 11.14 MeV$^{-1}$ and
11.16 MeV$^{-1}$ for the full and impulse-approximation calculations.
As the Gamow peak
energy is $\sim$ 6 keV for temperatures characteristic of the Sun's center, 
the linear term generates 
a $\lsim$ 8\%
correction to the $E=0$ value. The 1\%
uncertainty in Eq.~(\ref{logderivative}) corresponds to a 
$\lsim$ 0.1\% uncertainty in the total reaction rate. This is negligible
compared to other uncertainties described below. Therefore, in the
following, we focus on S$_{11}(0)$.

At zero relative energy S$_{11}(0)$ can be
written \cite{Bahcall:1968xb,Bahcall:1968wz}, 
\begin{equation}
\mathrm{S}_{11}(0)=6\pi ^{2}m_{p}\alpha \ln 2\,{\frac{\overline{\Lambda }^{2}}{\gamma ^{3}
}}\left( {\frac{G_{A}}{G_{V}}}\right) ^{2}{\frac{f_{pp}^{R}}{
(ft)_{0^{+}\rightarrow 0^{+}}}},
\end{equation}
where $\alpha $ is the fine-structure constant; $m_{p}$ is the proton mass; $
G_{V}$ and $G_{A}$ are the usual Fermi and axial-vector weak coupling
constants; $\gamma =(2\mu B_{d})^{1/2}=0.23161$~fm$^{-1}$ is the deuteron
binding wave number; $\mu $ is the proton-neutron reduced mass;  $B_{d}$
is the deuteron binding energy; $f_{pp}^{R}$ is the phase-space factor for
the pp reaction with radiative corrections; $(ft)_{0^{+}\rightarrow 0^{+}}$
is the $ft$ value for superallowed $0^{+}\rightarrow 0^{+}$ transitions; and  $
\overline{\Lambda }$ is proportional to the transition matrix element
connecting the pp and deuteron states.

Inserting the current best values, we find
\begin{eqnarray}
\mathrm{S}_{11}(0)&=&4.01\times 10^{-25}\,\mathrm{MeV~b}\,\left( {\frac{(ft)_{0^{+}
\rightarrow 0^{+}}}{3071\,\mathrm{s}}}\right) ^{-1} \nonumber \\
&\times& \left( {\frac{
G_{A}/G_{V}}{1.2695}}\right) ^{2} 
\left( {\frac{f_{pp}^{R}}{0.144}}\right)
\left( {\frac{\overline{\Lambda }^{2}}{7.035}}\right) .
\label{firstestimate}
\end{eqnarray}
We now discuss the best estimates and the uncertainties for each of the
factors appearing in Eq.~(\ref{firstestimate}).

We take $(ft)_{0^{+}\rightarrow 0^{+}}=(3071.4\pm 0.8)$ s, the
value for superallowed ${0^{+}\rightarrow 0^{+}}$ transitions that has been
determined from a comprehensive analysis of experimental rates 
corrected for radiative and Coulomb
effects \cite{Hardy:2008gy}. This value determines
the weak mixing matrix element $\left\vert
V_{ud}\right\vert =0.97418(27)$, the value adopted by the 
PDG \cite{Amsler:2008zzb}. This $ft$ value is also consistent with $%
(3073.1\pm 3.1)$ s used in Solar Fusion I.

For $G_{A}/G_{V}$, we use the PDG value $G_{A}/G_{V}=1.2695\pm 0.0029$ which
is consistent with $1.2654\pm 0.0042$ used in Solar Fusion I.

For the phase-space factor $f_{pp}^{R}$, we have taken the value without
radiative corrections, $f_{pp}=0.142$ \cite{Bahcall:1968wz} and increased it
by 1.62\% to take into account radiative corrections to the cross section 
\cite{Kurylov:2002vj}. The main source of error is from neglected diagrams
in which the lepton exchanges a weak boson and a photon with different
nucleons. These diagrams are estimated to modify $f_{pp}^R$ by $\sim 0.1\%$, based
on scaling the similar nucleus-dependent correction in superallowed $\beta $
decay \cite{Kurylov:2002vj}.  It would be useful to check this estimate
through direct computations. We adopt $f_{pp}^{R}=0.144(1\pm 0.001)$,
which is consistent with $0.144(1\pm 0.005)$ used in Solar Fusion I.

The dominant uncertainty in S$_{11}(0)$ comes from the normalized Gamow-Teller
(GT) matrix element $\overline{\Lambda }$.   A great deal of theoretical
work since Solar Fusion I has focused on reducing this uncertainty. 
In Solar Fusion I $\overline{\Lambda }$ was
decomposed into $\overline{\Lambda }=\Lambda \left( 1+\delta \right) $,
where $\Lambda$ represents the contribution of the one-body transition
operator and $\Lambda \delta$ that from two-body corrections. %
%The separation is based on whether the
%transition involves only one nucleon or two nucleons.
%
% KK's Remark:  I don't understand the meaning of the
% the following phrase:
% "within a resolution scale set by the pion mass"
%
$\Lambda $ thus involves an evaluation of the Gamow-Teller operator
between the initial-state pp wave
function and the final-state deuteron wave function. $\Lambda
^{2}=6.92 (1\pm 0.002_{-0.009}^{+0.014})$ 
was adopted, where the first and second uncertainties reflect, respectively,
variations in empirical values of the deuteron and low-energy pp scattering parameters, and 
the model dependence of the nuclear
potential \cite{Kamionkowski:1993fr}.
The value and uncertainty of the exchange current
contribution,  $\delta =0.01_{-0.01}^{+0.02}$, was determined from the range of values
of published calculations, following the
conservative recommendation of \citet{Bahcall:1992hn}.

Two major steps have contributed to reducing the uncertainty on $\overline{
\Lambda }$ since Solar Fusion I. The first is a much deeper understanding
of the correlation between the uncertainties in $\Lambda $ and $
\delta \Lambda $:  the overall uncertainty in $\overline{\Lambda }$
can be described by a universal parameter that can be fixed by a
single measurement.  The study of \citet{Sch98} demonstrated this
phenomenologically in the context of potential-model approaches,
while later analysis via EFT provided a more formal
justification \cite{Butler:2000zp,Par03}. The second step is the
use of the precisely known tritium $\beta $\ decay rate $\Gamma _{\beta
}^{T}$, as first proposed by \citet{Carlson:1991ju}, to fix this universal
parameter. This has been done in both potential models \cite%
{Sch98} and in the hybrid EFT approach \cite{Par03}. We
briefly describe these developments.

\subsection{Progress in potential models}
The most elaborate calculation for the pp fusion process in the
potential-model approach (see Sec. \ref{theoryabinitio})
was carried out by \citet{Sch98}. 
A comparison of the results for five representative modern potentials --
potentials designed to accurately reproduce nucleon-nucleon scattering data --
yielded $\Lambda ^{2}=6.975\pm 0.010$.
This study demonstrated the importance of using the
tritium $\beta$ decay rate to constrain the
two-body GT transition operator.  (Both the Fermi and GT operators contribute to
tritium $\beta $ decay, but the former can be reliably calculated
because of the conserved vector current and the smallness of isospin
breaking effects, $\sim$ 0.06\%.)  If one adjusts the uncertain strength of
the exchange current so that the tritium $\beta$ decay rate is reproduced, 
the variation in S$_{11}(0)$ that otherwise would come from
the choice of the phenomenological potential is largely removed.
Predictions for five representative high-precision phenomenological
potentials fall in a narrow interval 7.03 $\lsim \overline{\Lambda }^{2} \lsim$7.04
 \cite{Sch98}.

We note two other sources of model dependence that
contribute to the overall uncertainty in $\overline{\Lambda }$. First, as
three-body potentials and currents  contribute to the tritium $\beta$ decay rate,
uncertainties in modeling such effects will influence the extracted constraint on 
the two-body currents needed for S$_{11}(0)$.
The best estimate of the consequences of this uncertainty
for S$_{11}(0)$, $\sim 0.8\%$, comes from the chiral (or pionful) EFT* approach
described below. Second, the experimental uncertainties
in the effective range parameters for nucleon-nucleon scattering will
propagate to $\overline{\Lambda }$.  We have assigned a 0.5\% uncertainty
in $\overline{\Lambda}^2$ to this source, pending future work in EFT
to better quantify this uncertainty.  By adding in quadrature these uncertainties
of 0.8\% and 0.5\% and the smaller uncertainty associated with the above potential
range, $\overline{\Lambda }^2=7.035 \pm 0.005$, we obtain the potential model
estimate
\begin{equation}
\overline{\Lambda }^2=7.035(1  \pm 0.009).
\end{equation}

\subsection{Progress in effective field theory (EFT)}
The application of EFT, described in Sec. \ref{theoryabinitio}, to the calculation of the
pp  fusion rate (and several other electroweak processes in light
nuclei) is one of the notable developments since Solar Fusion I. 
There have been two lines of EFT calculations of 
pp fusion, described below.

\subsubsection{Hybrid EFT (EFT*)}
Electroweak nuclear transitions in EFT
\begin{equation}
\mathcal{M}^{\mathrm{EFT}}=\,<\!\Psi _{f}^{\mathrm{EFT}}|\sum_{i}^{A}%
\mathcal{O}_{i}^{\mathrm{EFT}}+\sum_{i<j}^{A}\mathcal{O}_{ij}^{\mathrm{EFT}%
}|\Psi _{i}^{\mathit{EFT}}\!\!>\,, 
\end{equation}
require initial and final nuclear wave functions and
the transition operators to be derived from EFT. However, this has not yet been
achieved in EFT with dynamical pions for pp fusion. Instead, a hybrid
approach \cite{Par03} called EFT* (or MEEFT) has been developed
in which
transition operators are taken from chiral perturbation theory ($\chi$PT), but
sandwiched between phenomenological wave functions, $\Psi _{i}^{\mathrm{%
phen}}$ and $\Psi _{f}^{\mathrm{phen}}$, generated by a potential model.
As discussed below, this approach is a substantial improvement over the earlier calculation
of \citet{Park:1998wq}.

For the low-energy GT transition that governs
pp fusion, the one-body transition operators $\mathcal{O}_{i}^{\mathrm{EFT}%
}$ are well known, while the two-body operators $\mathcal{O}_{ij}^{%
\mathrm{EFT}}$ contain only one unknown low-energy
constant (LEC).  This LEC, denoted by $\hat{d}^{R}$, parameterizes
the strength of contact-type four-nucleon coupling to the axial current. %
\citet{Par03} chose to determine $\hat{d}^{R}$ from the tritium 
$\beta $-decay rate $\Gamma _{\beta }^{T}$. The fact that $\Psi ^{\mathrm{%
phen}}$ is not exactly an eigenstate of the EFT Hamiltonian can in principle
be a source of concern, but it is plausible that the mismatch affects primarily
the short-distance behavior of the wave function, so that the procedure of
fixing the relevant LEC(s) to data can remove
most of the inconsistency: While $\mathcal{L}_{\chi \mathrm{PT}}$ by
construction is valid only well below $\Lambda _{\mathrm{QCD}}$, the use of
the phenomenological Hamiltonian, Eq.~({\ref{HA}}), introduces high momentum
components above $\Lambda _{\mathrm{QCD}}$. To test this procedure, one can
introduce a cutoff $\Lambda _{\mathrm{NN}}$ to eliminate
high-momentum components in the two-nucleon relative wave function,
fitting the LEC as a function of this parameter.  One
expects, if the fitting of the LEC reasonably accounts for missing or
inconsistent short-distance physics, little 
$\Lambda _{\mathrm{NN}}$ dependence would be found in the calculated pp
fusion rate. The residual dependence on
$\Lambda _{\mathrm{NN}}$, when this cutoff is varied over a physically reasonable range, provides
a measure of the model independence of an EFT* calculation.

The \citet{Par03} calculation included up to next-to-next-to-next-to-leading
order (N$^3$LO) terms in chiral expansion, and after fitting $\hat{d}^{R}$ 
%and as long as the two-body current operator $\widehat{d}%
%^{R}$ is used to absorbed the model dependence of short distance wave
%function, robust predictions can still be obtained.
to $\Gamma _{\beta }^{T}$, yielded $\overline{\Lambda }%
^{2}=7.03(1\pm 0.008)$.  The uncertainty was estimated from the 
changes in $\overline{\Lambda}^2$ when $\Lambda_\mathrm{NN}$ is varied
over an energy range typical of vector meson masses,  500 to 800 MeV.
A rough estimate based on higher order chiral
contributions was also made.  Specifically,
the contributions of the first four chiral orders 
to $\overline{\Lambda}$ follow the pattern
(1+0.0\%+0.1\%+0.9\%), while the fifth-order term
is estimated to be $\sim$ 0.4\%.
We assume that the
second- and third-order terms
are accidentally small, while the fourth- and fifth-order terms
reflect the convergence of the expansion in
$m_\pi/\Lambda_{QCD} \sim 1/7$.
Three-body currents contribute 
in sixth order.
We therefore use the size of the fifth-order
term, 0.4\%, as a measure of the uncertainty due to neglected 
higher order contributions (including three-body
currents).

Full EFT calculations that use $\Psi ^{\mathrm{EFT}}$
instead of $\Psi ^{\mathrm{phen}}$, thus eliminating operator-wave function
inconsistencies, are an important goal.   Progress
toward this goal includes recent constructions of
EFT-based nuclear interactions; see, \textit{e.g.}, \citet{Epelbaum:2005pn}
and \citet{Gazit:2008ma}.

\subsubsection{Pionless EFT}
This  approach can be applied to processes where the 
characteristic momentum $p$ is much smaller than the pion mass $m_{\pi }$ \cite%
{Kaplan:1996xu,Bedaque:1998kg,Chen:1999tn}, which is the case for solar pp
fusion. Pions can then be integrated out, so that all nucleon-nucleon
interactions and two-body currents are described by point-like contact
interactions with a systematic expansion in powers of $p/m_{\pi }$. The one- 
and two-body contributions individually depend on the
momentum cut-off but the sum does not. Thus, $\Lambda $ and $\Lambda \delta $
in pp fusion are correlated. In pionless EFT only one two-body
current (with coupling $L_{1,A}$) is needed in the description of
deuteron weak breakup
processes, through next-to-next-to-leading order (NNLO) in the $p/m_{\pi }$ expansion \cite{Butler:2000zp}. This
two-body current is a GT operator. Other two-body currents are either
missing due to conservation of the vector current,  or involve matrix elements
suppressed due to the pseudo-orthogonality of the initial- and final-state
wave functions. This means the universal number $L_{1,A}$ encodes the
two-body contributions for all low-energy weak deuteron breakup processes,
so that a single measurement will fix the rates of all such processes. The other approaches 
discussed above share this feature.

The computation of $\overline{\Lambda }$ in pionless EFT was carried out to
the second order by  \citet{Kong:2000px} and \citet{Ando:2008va} and then to the fifth order 
by \citet{Butler:2001jj}. Constraints on $L_{1,A}$ from two nucleon systems \cite%
{Butler:2002cw,Chen:2002pv} yield $\overline{\Lambda }^{2}=6.99\pm 0.21$.
The MuSun experiment \cite{MuSun} is taking data on $\mu$ capture on deuterium. The
experimental goal is to
constrain $\overline{\Lambda }^{2}$ to $\lsim$ 1.5\% for
pionless EFT \cite{Chen:2005ak} and chiral EFT* \cite{Ando:2001es}.

\subsubsection{Comment on Mosconi {\it et al.}}
%Mosconi et al. [PRC 75, 044610 (2007)]
\citet{Mosconi:2007tz} have compared $\nu $-d reaction cross
sections for various models that differ in their treatments
of two-body transition operators, concluding from this comparison that the results
obtained in potential models, EFT*, and pionless EFT have uncertainties as
large as 2-3\%. Although they address only $\nu$-d cross
sections, a comment is in order here because this process is closely 
related to that for pp fusion.  \citet{Mosconi:2007tz} reach their conclusions
by examining the scatter of unconstrained calculations of the $\nu$-d
cross section.   However, all  state-of-the-art calculations use
$\Gamma _{\beta }^{T}$ to reduce two-body current and other uncertainties,
as we have detailed here.  Once this
requirement is imposed, the scatter in the calculated value of $\nu $-d
cross sections is significantly reduced.

\subsection{Summary}
We have seen that the various approaches discussed above
yield accurate and very consistent
values for $\overline{\Lambda}^2$.
The remaining factors in Eq~(18) also have uncertainties, but
these are common to all the calculations.
Adding all the uncertainties in quadrature,
we find that the current best estimates for S$_{11}(0)$  are 
\begin{eqnarray}
4.01(1\pm 0.009) \times 10^{-25} &\mathrm{MeV~ b}&~~~ \mathrm{potential~models}\nonumber\\
4.01(1\pm0.009) \times 10^{-25}  &\mathrm{MeV~ b}&~~~\mathrm{EFT^*}\nonumber\\
3.99(1\pm 0.030) \times 10^{-25} &\mathrm{MeV ~b}&~~~\mathrm{pionless~EFT}.
\end{eqnarray}
The larger uncertainty in the pionless EFT result is 
due to the relatively weak constraints on $L_{1,A}$
that can be imposed within two-nucleon systems
but, as mentioned, this situation will soon be improved.
The agreement of the central values
obtained in the potential model and EFT$^*$ 
indicates the robustness of the results
as long as the two-body current is constrained by
tritium $\beta$ decay. 
Meanwhile, the agreement of the error estimates in the two approaches 
is primarily due to the fact that, as explained above, 
the dominant part of the uncertainty has been estimated
using the same argument.
Based on the result obtained in the potential model and EFT$^*$,
we adopt as the recommended value
\begin{equation}
\mathrm{S}_{11}(0)=4.01(1\pm0.009)\times 10^{-25}\,{\rm MeV~b}.
\end{equation}
We adopt the \citet{Bahcall:1968wz} value for S$^\prime_{11}(0)$
\begin{equation}
\mathrm{S}^\prime_{11}(0)=\mathrm{S}_{11}(0) (11.2\pm0.1)\,{\rm MeV}^{-1}\,,
\end{equation}
\citet{Bahcall:1968wz} also estimated dimensionally that
$S^{\prime \prime}_{11}(0)$ would enter at the  level of $\sim$ 1\%, for
temperatures characteristic of the solar center.  As this is now
comparable to the overall error in S$_{11}$, we recommend
that a modern calculation of S$^{\prime \prime}_{11}(0)$ be undertaken.

\section{THE \lowercase{d(p},$\gamma$)$^3$H\lowercase{e} RADIATIVE CAPTURE REACTION}
\label{sec:dp}
The radiative capture of protons on deuterium is the second reaction
occurring in the pp chain.  Because this reaction is so much faster than 
the pp weak rate discussed in the previous section, it effectively instantaneously
converts deuterium to $^3$He, with no observable signature.  Thus 
uncertainties in its rate have no consequences for solar energy generation.
By comparing the pp and d(p,$\gamma$)$^3$He rates, one finds that the
lifetime of a deuterium nucleus in the solar core is $\sim$ 1 s, and that the
equilibrium abundance of deuterium relative to H is maintained at
 $\sim$ 3 $\times$ 10$^{-18}$.

However, the d(p,$\gamma)^3$He reaction plays a more
prominent role in the evolution of protostars.  As a cloud of
interstellar gas collapses on itself, the gas temperature rises
to the point of d(p,$\gamma)^3$He ignition,
$\sim$ $10^6$ K.
The main effect of the onset of  deuterium burning is to slow down
the contraction and, in turn, the heating. As a consequence, the
lifetime of the proto-star increases and its observational properties
(surface luminosity and temperature) are frozen until the original
deuterium is fully consumed \cite{Stahler88}. Due to the slow
evolutionary timescale, a large fraction of observed proto-stars
are in the d-burning phase, while only a few
are found in the earlier, cooler, rapidly evolving phase. A
reliable knowledge of the rate of d(p,$\gamma)^3$He down to a few
keV (the Gamow peak in a proto-star) is of fundamental importance for
modeling proto-stellar evolution.  

The pd reaction also plays an important role in Big Bang nucleosynthesis,
which begins when the early universe has cooled to a temperature of $\sim$ 100 keV.
The uncertainty in the pd reaction in the relevant energy window
(25-120 keV) propagates into uncertainties in the deuterium,
$^3$He and $^7$Li abundances, scaling roughly as
\begin{equation}
{\mathrm{d} \over \mathrm{H}}\propto R_{\mathrm{pd}}^{-0.32}~~~~~~~{^3\mathrm{He} \over \mathrm{H}} \propto R_{\mathrm{pd}}^{0.38}~~~~~~~
{^7\mathrm{Li} \over \mathrm{H}}\propto R_{\mathrm{pd}}^{0.59},
\end{equation}
where $R_{\mathrm{pd}}$ is the value of  S$_{12}$
relative to the fiducial value in~\citet{cyburt04}.  Thus a 10\% error in
the pd capture rate propagates into roughly 3.2\%, 3.8\% and 5.9\%
uncertainties in the light element primordial abundances, d, $^3$He and $^7$Li,
respectively.

\begin{figure}
\begin{center}
\includegraphics[width=8.7cm]{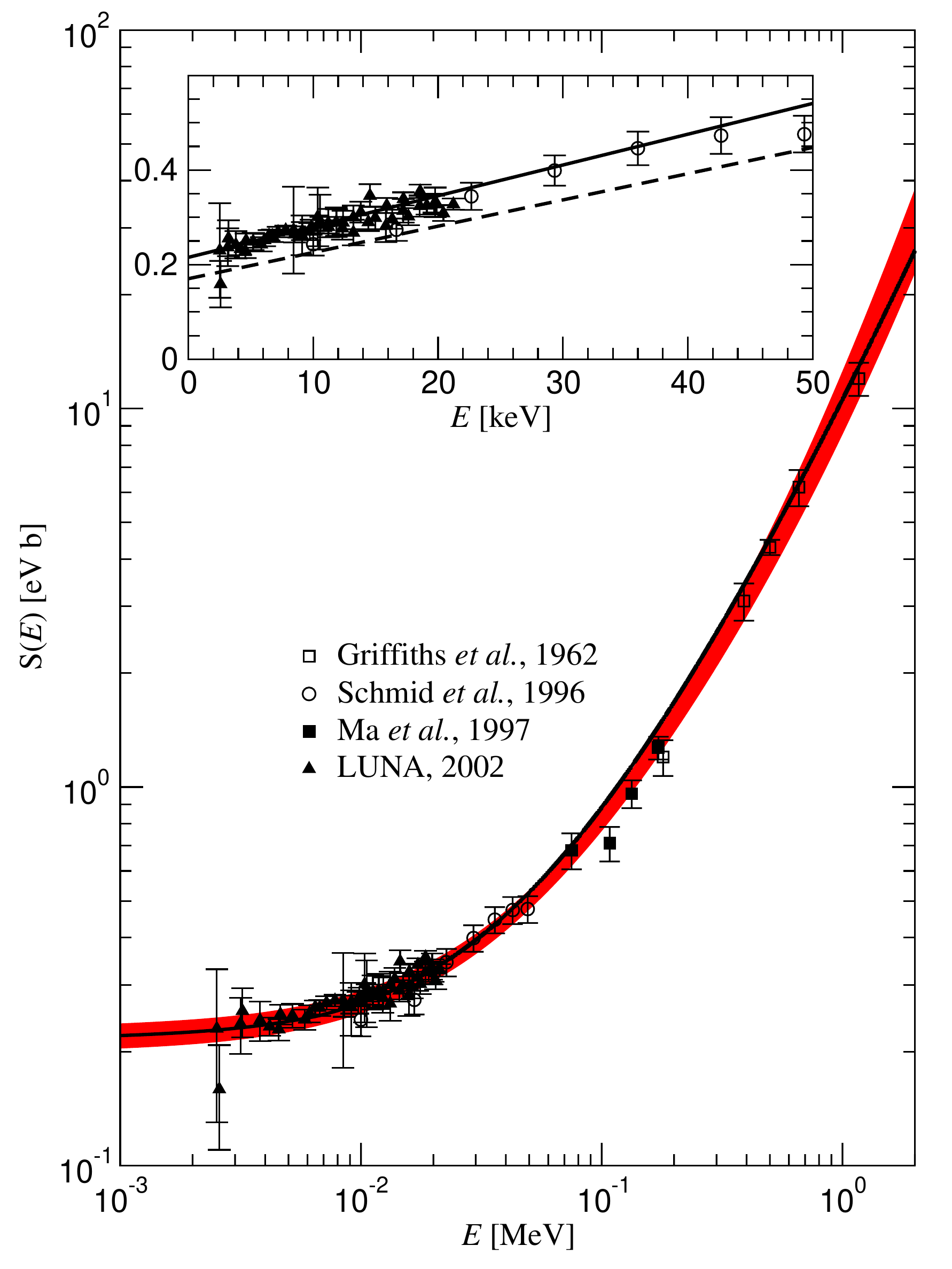}
\caption{(Color online) The astrophysical S$_{12}$-factor datasets~\cite{Gri62,Sch96,ma97,luna}
are plotted together with theoretical predictions of \citet{Marcucci05}.
The solid line represents the ``full'' theoretical calculation, while the 
red band represents the 68\% lower and upper bounds of the adopted 
quadratic best fit to the four experimental datasets (see text and Eq. (\ref{eq:s12quad}) for more explanation).
In the insert, the S$_{12}$-factor of the $^2$H(p,$\gamma$)$^3$He reaction
in the energy range 0-50 keV,
obtained with the Argonne $v_{18}$ two-nucleon and Urbana IX three-nucleon
Hamiltonian model in the impulse approximation (dashed line) and with inclusion of interaction
currents (solid line), is compared with the experimental
results.}
\label{fig:f_pd}
\end{center}
\end{figure}

\subsection{Data sets}
The extensive experimental data sets for pd radiative capture
include total cross sections and spin polarization observables at
center-of-mass energies $E$ ranging from several
tens of MeV to a few keV, covering all the relevant astrophysical energies.  
In the regime $E \lsim 2$ MeV (below the
deuteron breakup threshold), the relevant experimental data
include \citet{Gri62,Gri63,bai70,Sch95,Sch96,ma97,luna}.
The \citet{Gri63} and \citet{bai70} low energy data may be $\sim~15$\% too high because of the use of incorrect
stopping powers \cite{Sch95,Sch96,ma97}.
Also, the \citet{Sch95,Sch96} data sets may have not propagated their energy-dependent systematic
uncertainties. In Fig. \ref{fig:f_pd}, the data for S$_{12}$ used for the best fit in Sec. \ref{sec:dpC} 
are plotted together with theoretical predictions of \citet{Marcucci05}.
The observed linear dependence of S$_{12}$ on $E$ at low energies,
as well as the angular distributions of the cross
section and polarization observables, indicate that the
d(p,$\gamma)^3$He reaction proceeds predominantly through s- and
p-wave capture, induced, respectively, by magnetic ($M1$) and electric
($E1$) dipole transitions.  The $M1$ transitions (proceeding through
$^2$S$_{1/2}$ and $^4$S$_{3/2}$ pd channels) are especially
interesting, as the one-body $M1$ operator cannot connect the main
s-state components of the pd and $^3$He wave functions at low
energies.  Because of this ``pseudo-orthogonality'' only the small
components of the wave functions contribute in the impulse approximation
(IA).  In contrast, as exchange-current operators are not similarly hindered,
their matrix
elements are exceptionally large relative to those obtained with the
one-body $M1$ operator.  The suppression of matrix elements calculated
in the IA and their consequent enhancement by exchange-current
contributions are a feature common to other $M1$-induced processes in
A=3 and 4 systems, such as the nd and n$^3$He
radiative captures at thermal neutron energies.

\subsection{Theoretical studies}
The most extensive and recent theoretical studies of the
d(p,$\gamma)^3$He reaction at low energies have been
carried out by \citet{Marcucci05}.  The calculated S$_{12}$, shown
in Fig.~\ref{fig:f_pd}, is in excellent agreement with data.
To describe the pd continuum and $^3$He bound states, these authors used
variational wave functions built in a correlated-hyperspherical-harmonics
(CHH) basis for a Hamiltonian consisting of the 
Argonne $v_{18}$ two-nucleon~\cite{wir95} and the Urbana IX~\cite{Pud95} 
three-nucleon potentials.  This Hamiltonian is known to reproduce
a variety of three-nucleon bound- and
scattering-state properties, including binding energies,
charge radii, elastic and inelastic cross sections, and low-energy polarization
observables, while the accuracy of  the CHH variational method is
comparable to that of other quasi-exact methods \cite{Nogga03}.
 
The nuclear electromagnetic current consists of one-body terms (the IA currents),
originating from the convection and spin-magnetization currents
of individual protons and neutrons, and two- and three-body
exchange currents, constructed from the corresponding
potentials by a procedure that preserves current conservation (CC).
The method by which this is achieved has been improved over the
years \cite{Riska84,Sch98}, and its latest implementation is discussed
at length by \citet{Marcucci05}.  The currents are still model dependent, of 
course, as CC
places no constraints on their transverse components.  

The calculated value for S$_{12}$(0)
including exchange-current contributions
is 0.219 eV b, in excellent agreement with the value extrapolated
from the LUNA measurements (0.216$\pm$ 0.010 eV b), and evaluations by
\citet{cyburt04} (0.227 $\pm$ 0.014 eV b), \citet{descouvemont04} (0.223$\pm$ 0.007 eV b) and
\citet{serpico04} (0.214$\pm$ 0.007 eV b).   In \citet{descouvemont04} 
systematic
and statistical errors are combined before following a standard fitting procedure. However,
as this artificially reduces the impact of systematic errors,
their cited uncertainties have been underestimated.  \citet{serpico04} properly separates
systematic and statistical errors in their treatment, but do not cite
68\% confidence limits, also yielding an error that is too small.
The evaluation by \citet{cyburt04} separates systematic and statistical uncertainties
and cites errors consistent with 68\% confidence limits,
yielding realistic uncertainties.

\subsection{Summary}
\label{sec:dpC}
In this report, we evaluate the \citet{luna}, \citet{Gri62}, \citet{Sch96} 
and \citet{ma97} data, determining S$_{12}(E)$ as a function 
of the center-of-mass energy by fitting the four data sets
by a quadratic polynomial in $E$.  
We adopt this fitting procedure, despite our earlier arguments favoring fitting
formulas that are motivated by theory, because the energy window of interest
is fully covered by the experiments.   This yields
\begin{equation}
\mathrm{S}_{12}(0) = 0.214^{+0.017}_{-0.016}~ \mathrm{eV~ b},
\end{equation}
 in agreement with previous
evaluations.  The error is larger here, because of the exclusion of 
the \citet{bai70} data. 

We also determined the 68\% upper and lower bounds for the quadratic parameterizations,
valid for $E \lsim$ 1 MeV, the range spanned by the data we considered.
The results are (see also Fig.~\ref{fig:f_pd})
\begin{eqnarray}
\mathrm{S}_{12}^\mathrm{lower}(E)  &=& 0.1983 + 5.3636~\left({E \over \mathrm{MeV}}\right)  \nonumber \\
&+& 2.9647\left({E \over \mathrm{MeV}}\right)^2~\mathrm{eV~b}  \nonumber \\
\mathrm{S}_{12}^\mathrm{best}(E) &=& 0.2145 + 5.5612\left({E \over \mathrm{MeV}}\right) \nonumber \\
 &+& 4.6581\left({E \over \mathrm{MeV}}\right)^2~\mathrm{eV~b} \nonumber \\
\mathrm{S}_{12}^\mathrm{upper}(E) &=& 0.2316 + 5.7381\left({E \over \mathrm{MeV}}\right) \nonumber \\
&+& 6.5846\left({E \over \mathrm{MeV}}\right)^2~\mathrm{eV~b}.
\label{eq:s12quad}
\end{eqnarray}
The results determine the S-factor and its uncertainty in the vicinity of the
solar Gamow peak.  In particular, for a temperature characteristic of the Sun's center,
1.55 $\times$ 10$^7$ K,
\begin{equation}
\mathrm{S}_{12}(E_0=6.64~\mathrm{keV}) = 0.252 \pm 0.018~ \mathrm{eV~ b},
\end{equation}
so that the estimate uncertainty is $\sim$ 7.1\%.

\section{THE $^3$H\lowercase{e}($^3$H\lowercase{e,2p})$^4$H\lowercase{e} REACTION}
\label{sec:s33}
 The $^3$He($^3$He,2p)$^4$He reaction is the termination of the ppI
 cycle and thus, as Solar Fusion I describes in more detail,
 uncertainties in this cross section played a prominent role in early
 speculations about a nuclear astrophysics solution to the solar
 neutrino problem.  As an increase in S$_{33}(E)$ would reduce the
 branchings to the ppII and ppIII cycles -- thus also reducing the
 neutrino fluxes measured by Davis -- the possibility of an
 undiscovered narrow resonance at energies beyond the reach of early
 experiments was raised by \citet{Fet72} and \citet{Fow72}.  This
 motivated efforts to measure S$_{33}(E)$ at lower energies, and
 particularly stimulated the efforts of the LUNA collaboration in the
 1990s to map the cross section in the solar Gamow peak
 \cite{Greife94, Arpesella96, Junker98, Bonetti99}.  The principal
 result since Solar Fusion I is the completion of this program by
 \citet{Bonetti99}, extending measurements to the lower edge of the
 Gamow peak at 16 keV, making S$_{33}(E)$ the most directly constrained
 S-factor within the pp chain.
 
 S$_{33}(E)$ remains of significant importance, as it controls the
 ppI/ppII+ppIII branching ratio and thus the
 ratio of the pp/pep to $^7$Be/$^8$B neutrino fluxes.  This ratio is
 important to future strategies to better constrain neutrino
 oscillation parameters and matter effects, through comparison of
 high-energy (matter influenced) and low-energy (vacuum) fluxes.
 The ratio of S$_{33}$ to S$_{34}$ enters in computing the neutrino energy losses of the
 Sun, and thus influences the connection between the Sun's photon
 luminosity and its total energy production.  
 
\subsection{Data sets and fitting}
We consider data available at the time of Solar Fusion I
\citep{Greife94,Arpesella96,Junker98,Bacher65,Dwar71,Krauss87} as well
two new data sets:
the extreme low energy data of LUNA \citep{Bonetti99} and
results from the OCEAN
experiment \citep{Kudomi04} at energies slightly above the solar Gamow
region.  In order to follow the recommended fitting prescription
discussed in the Appendix, one needs a detailed discussion
of systematic uncertainties, particularly common mode systematics.
This requirement reduces the datasets considered to just four
experiments.  The earliest of these originates from the
Muenster group \citep{Krauss87}, followed by the two LUNA publications
\citet{Junker98} (which supersedes \citet{Arpesella96}) and 
\citet{Bonetti99}; and the OCEAN effort \citet{Kudomi04}. \citet{Krauss87} 
and \citet{Kudomi04} identified a common systematic error for their
respective data sets while the LUNA group provided statistical and
systematical errors at each experimental energy measured. In order to
use a uniform treatment we calculated an average systematic error for
the latter data sets. Larger systematic errors were noted only at the
lowest energies (due to uncertainties in stopping power) where the
total error is dominated by statistics.

Past efforts have fit data to an S-factor including screening
corrections, with the bare S-factor a polynomial up to quadratic
order, 
\begin{eqnarray}
\label{eqn:s33fit}
\mathrm{S}_{33}(E)\! &=&\! \mathrm{S}_{33}^\mathrm{bare}(E) \exp{\!\left(\frac{\pi\eta(E) U_e}{E}\right)} \\
\nonumber \mathrm{S}_{33}^\mathrm{bare}(E)\! &=& \! \mathrm{S}_{33}(0) + \mathrm{S}_{33}^{\prime}(0)E + \frac{1}{2}
\mathrm{S}_{33}^{\prime\prime}(0)E^2.
\end{eqnarray}
Although model calculations of S$_{33}^\mathrm{bare}(E)$ are available (see, e.g., \citet{Typel91}),
a phenomenological representation for the bare S-factor is appropriate because
the data extend to the Gamow peak.  There is no need for a theoretical
model to guide an extrapolation, apart from the functional form of the
screening potential.  

The selected data for this review cover the range from the solar
Gamow peak to 350 keV, providing a limited range with which to perform
a four parameter fit to the S-factor including electron
screening (S$_{33}(0)$, S$_{33}^\prime(0)$, S$_{33}^{\prime\prime}(0)$, $U_e$). We
test the robustness of the fit parameters, by varying the order of the
polynomial for the bare S-factor.  Our results are in
Table~\ref{tab:s33}.  

\begin{table}[h]
\caption{Table of fit parameters and their total errors for
constant, linear, and quadratic representations of the bare S-factor. }
\label{tab:s33}
\begin{center}
\begin{tabular}{|l||c|c|c|}
\hline\hline
parameter & constant & linear & quadratic \\ \hline\hline S$_{33}$(0) (MeV
b) & $4.84\pm0.13$ & $4.95\pm0.15$ & $5.32\pm0.23$ \\ \hline
S$_{33}^{\prime}(0)$ (b) & N.A. & $-1.06\pm0.51$ & $-6.44\pm1.29$ \\ \hline
S$_{33}^{\prime\prime}(0)$ (MeV$^{-1}$ b) & N.A. & N.A. & $30.7\pm12.2$ \\
\hline $U_e$ (eV) & $395\pm50$ & $360\pm55$ & $280\pm70$ \\ \hline
\hline $\chi^2_{tot}$ & 35.4 & 34.1 & 31.8 \\ \hline
$\chi^2_{tot}/dof$ & 0.40 & 0.39 & 0.37 \\ \hline\hline
\end{tabular}
\end{center}
\end{table}

\begin{figure}
\begin{center}
\includegraphics[width=8.7cm]{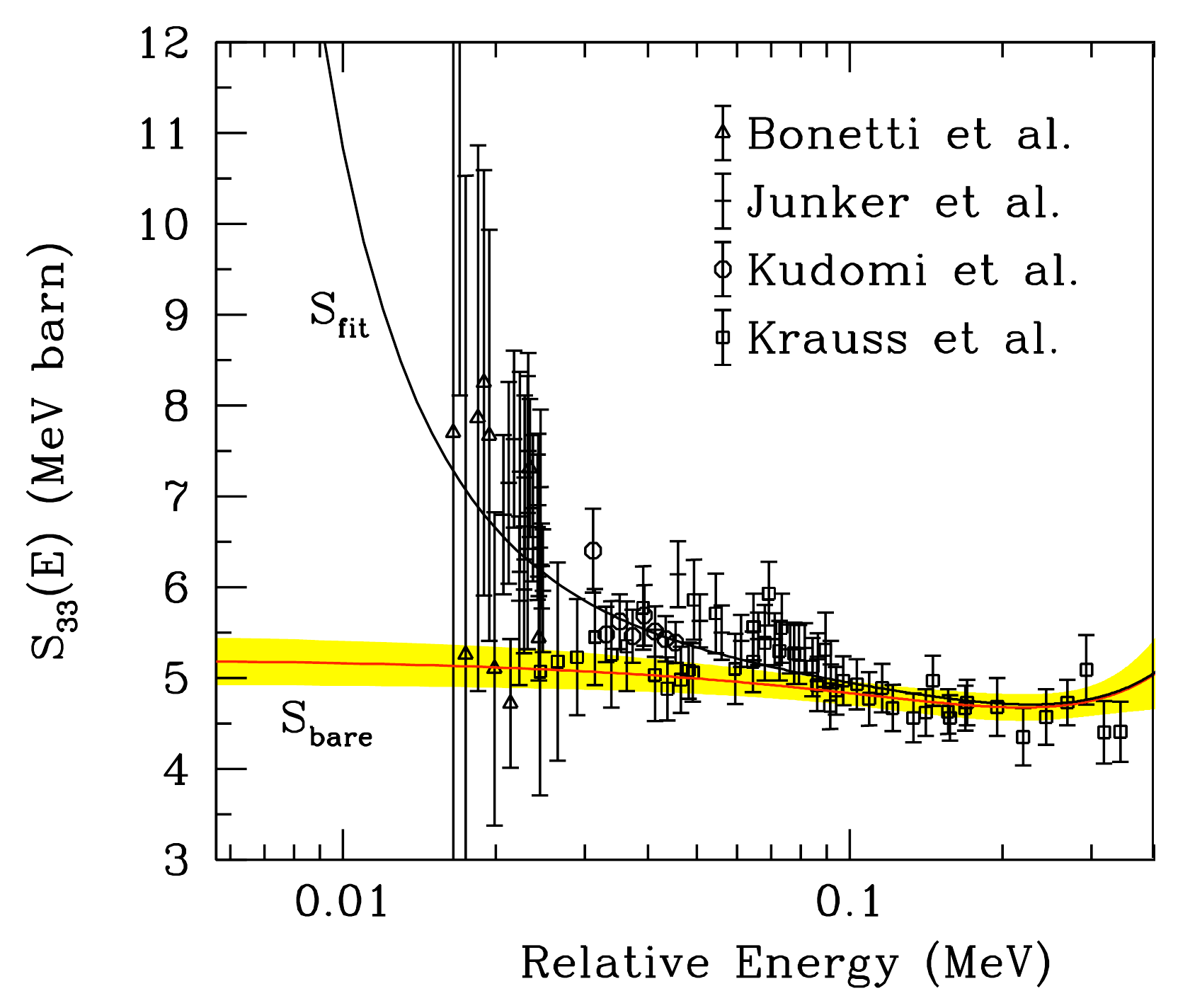}
\caption{(Color online) The data, the best quadratic+screening result for S$_{33}(E)$,
and the deduced best quadratic fit (red line) and allowed range (yellow band) for S$^\mathrm{bare}_{33}$.  
See text for references.}
\label{fig:s33}
\end{center}
\end{figure}

Our quadratic fit agrees quite well with the fit derived by
\citet{Krauss87}, adopted in the reaction rate compilation
of \citet{CF88}.  However, there is a significant spread in fit
parameter values for the different order polynomial fits, with slight
decreases in the total $\chi^2$.  One can also see this spread in fit
results from other groups~\citep{Junker98,Bonetti99,Kudomi04}.  This
suggests that the data do not have the resolving power to accurately
determine all fit parameters: there are strong correlations
for the choices of data and fitting functions made here.  Adopting any
{\em single} fit will underestimate the uncertainties due to the
degeneracy between parameter values.  From Bayes's theorem,
assuming that the S-factor in this region ($E<350$ keV) can be described
without cubic terms, we can derive constraints on the parameters by
weighting each fit in Table~\ref{tab:s33} by its total $\chi^2$ value.
This method takes into account the spread from fit-to-fit.  We find

\begin{eqnarray}
\mathrm{S}_{33}(0) &=& 5.21\pm0.27 {\rm \ MeV \ b} \\
\nonumber \mathrm{S}_{33}^{\prime}(0) &=& -4.90\pm3.18 {\rm \ b} \\
\nonumber \mathrm{S}_{33}^{\prime\prime}(0) &=& 22.4\pm17.1 {\rm \ MeV^{-1} \ b} \\
\nonumber U_e &=& 305\pm90 {\rm \ eV}.
\end{eqnarray}

The results reveal that existing data cannot strongly constrain all
of the fitting parameters separately, and in particular do not sharply constrain $U_e$.  
To improve constraints on the screening potential one
will need more precise data from near the Gamow peak, as well as new measurements up to the
MeV range (with well documented systematics) to better determine the higher-order
terms in the quadratic fit.  New theory efforts in
determining the shape of this S-factor would also be beneficial,
as new low energy $^3$He-$^3$He elastic scattering data could be
used as an additional constraint. 

However, our principal concern is the precision with which S$^\mathrm{bare}_{33}(E)$ can be 
determined in the vicinity of the Gamow peak, not the separate parameters.
From the fit's correlation matrix we find
\begin{eqnarray}
\mathrm{S}_{33}^\mathrm{best}(E) = 5.21 - 4.90 \left( {E \over \mathrm{MeV}} \right)
+ 11.21 \left( {E \over \mathrm{MeV}} \right)^2\mathrm{MeV~b} \nonumber \\
\delta \mathrm{S}_{33}(E) = \left[0.075 - 1.516 \left( {E \over \mathrm{MeV}} \right)
+ 14.037 \left( {E \over \mathrm{MeV}} \right)^2 \right.~~~ \nonumber \\
\left. -15.504 \left({E \over \mathrm{MeV}} \right)^3 +
 71.640 \left({E \over \mathrm{MeV}} \right)^4 \right]^{1/2}\mathrm{MeV~b}~~~ \nonumber
\end{eqnarray}
where
\begin{equation}
\mathrm{S}_{33}^\mathrm{bare}(E) \equiv \mathrm{S}_{33}^\mathrm{best}(E) \pm \delta \mathrm{S}_{33}(E).
\label{eq:s33f}
\end{equation}
Because these results were obtained with a phenomenological fitting function, their reliability
has been establish only for the energy range covered by the data employed in the fit.
Thus Eq.~(\ref{eq:s33f}) should be used for energies $E \lsim$ 350 keV.
For a temperature 1.55 $\times$ 10$^7$ K corresponding
to the Sun's center, we find at the Gamow peak
\begin{equation}
S_{33}^\mathrm{bare}(E_0 = 21.94~\mathrm{keV}) = 5.11 \pm 0.22 \mathrm{~MeV~b},
\end{equation}
so that the estimated uncertainty is 4.3\%.

\section{THE $^3$H\lowercase{e}($\alpha$,$\gamma$)$^7$B\lowercase{e} REACTION}
\label{sec:s34}
When Solar Fusion I appeared, the most recent $^3$He($^4$He,$\gamma)^7$Be measurement was 10 years old.  The four new measurements that have been published since that time, in response to a
challenge by John Bahcall, are the focus of this section.  

For energies of interest, $E~\lsim~1$ MeV, $^3$He($^4$He,$\gamma)^7$Be is a nonresonant
reaction, predominantly external direct capture \cite{christy61} by electric dipole emission from s- and 
d-wave initial states to the two bound states of $^7$Be.  Reaction measurements have been 
made by detecting the prompt $\gamma$-rays, the $^7$Be activity, and the $^7$Be recoils.   
Below we discuss the measurements, the theory needed to extrapolate the measurements to 
astrophysical energies, and our determination of S$_{34}(0)$.   

\subsection{Experimental measurements}  
Groups at the Weizmann Institute~\cite{weiz} and at the University of Washington-Seattle~\cite{sea} carried out cross section measurements in the center-of-mass energy range $E$ = 0.42 to 0.95 MeV and 0.33 to 1.23 MeV, respectively, using gas cells with Ni entrance windows.  The LUNA collaboration~\cite{luna1,luna2,luna3} (see also \citet{luna4}) carried out low-background measurements from $E$ = 0.093 to 0.170 MeV at the LUNA facility in the Gran Sasso underground laboratory,  and a European collaboration~\cite{erna} (here called ERNA)  made measurements from $E$ = 0.65 to 2.51 MeV, both with windowless gas cells.  

An important concern in Solar Fusion I was whether $^3$He($^4$He,$\gamma)^7$Be measurements made by detecting the $^7$Be activity might be affected by background $^7$Be produced by contaminant reactions. Possibilities include $^6$Li(d,n)$^7$Be or $^{10}$B(p,$\alpha)^7$Be, which could occur given proton or deuteron contamination in the $^4$He beam in combination
with $^6$Li or $^{10}$B contamination in the gas cell, for example, in the foil or beam stop.  Only 
one of the older experiments - that of Osborne - involved measurements of both 
prompt $\gamma$s and $^7$Be activity (see Solar Fusion I for older references).  While the 
Osborne experiment found agreement between the $^3$He($^4$He,$\gamma)^7$Be cross 
sections determined
by the two methods, in general the cross section determined from activity-based experiments
was somewhat larger
than that determined from prompt-$\gamma$ experiments.

In the new experiments, all but the Weizmann group measured both prompt $\gamma$s and $^7$Be activity, while ERNA also measured $^7$Be recoils.  In each of these experiments, the cross sections deduced by the different methods were consistent, leading to upper limits on nonradiative capture of 2-5\% from $E$ = 0.09 to 2.5 MeV.  This is consistent with theoretical calculations that indicate much smaller rates expected for $E0$ capture and other electromagnetic processes that could produce $^7$Be without accompanying energetic prompt $\gamma$s \cite{e0}.  All new experiments except that of the Weizmann group employed $^4$He beams and $^3$He targets, thus minimizing potential problems with background $^7$Be production.  In the new experiments sensitive checks
ruled out contaminant $^7$Be production at lower levels.  Thus we see no reason to 
doubt the new activity measurements.  

$^7$Be activity measurements provide a direct determination of the total cross section.  In contrast, as
prompt $\gamma$-ray yields are anisotropic, one must take into account detector geometry and the
anisotropy to determine a total cross section.
% could you check above?  original seemed a bit obscure to 
[The $\sim$ 30\% capture branch to the 429-keV first excited state of $^7$Be has usually been 
determined from the isotropic 429 keV $\rightarrow$ ground state yield.]  Unfortunately, no 
angular distribution measurements exist at the needed level of precision.  The theoretical 
angular distributions of \citet{tombrello63a} (see also \citet{kim81}) were used to correct the 
% is this kim81?
prompt LUNA data, while the UW-Seattle data agree better with an assumed isotropic $\gamma_0$ angular distribution than with theory.  As the prompt anisotropy corrections can be 
comparable to the overall quoted cross section uncertainty, we decided to 
exclude the prompt data from our analysis.  We do this in part because little additional
precision would be gained by combining the highly correlated prompt and activation data.  
Hence we base our analysis on activation data, plus the ERNA recoil data.

The ERNA data and the older data of \citet{parker63} extend well above 1 MeV, where measurements may provide information useful for constraining theoretical models of S$_{34}(E)$.  Of these two data sets, only ERNA shows evidence for a significant rise in S$_{34}(E)$ above 1.5 MeV (see Fig. 1 of \citet{erna}).  

\subsection{Theory}
Relative (but not
absolute) S-factors at energies below 1 MeV vary by only a few
percent among credible models, with small differences arising from
non-external contributions and initial-state phase shifts.  The two 
bound states of $^7$Be populated by 
$^3\mathrm{He}(\alpha,\gamma)^7\mathrm{Be}$ direct capture have large 
overlaps with
$^3\mathrm{He}+\,^4\mathrm{He}$ cluster configurations.   The
Pauli principle requires radial nodes in these overlaps, guaranteeing 
a small (but nonzero) short-range
contribution because of cancellation in the matrix-element integral.

Considerable accuracy below 1 MeV can be achieved by
a pure external-capture model, with hard-sphere scattering at a radius chosen
to reproduce measured phase shifts.  In such a model $^3\mathrm{He}$ and 
$^4\mathrm{He}$ are treated as point particles, and
final states are modeled only by their long-range asymptotic parts.    
This is the approach of the \citet{tombrello63a}
model, used to fit S$_{34}$ in Solar Fusion I.  A more realistic treatment
of contributions from 2.8 to 7.0 fm is provided by potential models
\cite{dubovichenko95,kim81,buck88,buck85,mohr93,mohr09}, which generate
wave functions from a Woods-Saxon or similar potential, constrained by
measured phase shifts.  
% The bib file contains identical citations labeled kim and kim81
% I used one of those, and relabeled all \cite{kim} to \cite{kim81}

Microscopic models take explicit account of nucleon short-range correlations.
In the resonating-group method
(RGM) a simplified nucleon-nucleon interaction
is tuned to observables in the system being investigated (e.g., energies 
of the $^7\mathrm{Be}$ bound states), and the phase shifts are computed, 
not fitted.  The RGM wave functions are sums of states consisting of 
simple cluster substructure; in most $^7\mathrm{Be}$ calculations, they are
antisymmetrized products of Gaussians for
$^4\mathrm{He}$ and $^3\mathrm{He}$, multiplied by a function of the
coordinate describing cluster separation.

The RGM calculations of \citet{kajino86} and the potential-model of
\citet{langanke86} (which employed antisymmetrized many-body wave functions)
predicted the energy dependence of
the $^3\mathrm{H}(\alpha,\gamma)^7\mathrm{Li}$ reaction quite accurately, prior to the
precise measurement of \citet{brune94}.  On the other hand, there is
some variation of the computed
$^3\mathrm{He}(\alpha,\gamma)^7\mathrm{Be}$ S-factors among RGM
models using different interaction types and different Gaussian
widths within the clusters.  This variation has been shown to
correlate with measures of the diffuseness
of the $^7\mathrm{Be}$ ground state \cite{kajino86,csoto00}.
Substantial changes in the S-factor and phase shifts also occur when
$^6\mathrm{Li}$+p configurations are added to the RGM wave functions
\cite{mertelmeier86,csoto00}.

Calculations using highly accurate nucleon-nucleon potentials are now
possible.  In \citet{N01}, both
bound states were computed using the variational Monte Carlo method,
while the relative motion of the initial-state nuclei was modeled by
one-body wave functions from the earlier potential-model studies.
This approach should provide additional realism to the nuclear
wave function at short range, and it features initial states that fit
the measured phase shifts.  It produced very nearly the same S$_{34}(E)$ energy
dependence as \citet{kajino86}, and an absolute S$_{34}(0)$
that is lower by about 25\%.  

Through a numerical coincidence, the branching ratio for captures to
the two final states is very nearly constant at low energy
\cite{kajino86}.  This circumstance and the external-capture nature of the reaction suggest that laboratory data can be extrapolated to low energy by
fitting a single rescaling parameter that multiplies a model S$_{34}(E)$ to match
the data.  
Such a rescaling does not have a strong physical justification for
microscopic models, as they do not have undetermined spectroscopic
factors.
However, rescaled microscopic models should be at least
as accurate as potential models and more accurate than the hard-sphere model. 

A different approach was followed in \citet{cyburt08}, where a parameterized function fit was made to 
three of the four modern data sets over a wider
energy interval than we used to determine our recommended S$_{34}(0)$ (see below), with 
the result S$_{34}(0)$ = 0.580 $\pm$ 0.043 keV b.  Their fitting function is motivated by recent work
emphasizing external capture and subthreshold poles in low-energy
S-factors \cite{jen1998a,jen1998b,muk2002}, and it matches
expressions for zero phase shift derived in \citet{muk2002}.  For
S$_{34}$, the d-waves have small phase shifts, and the function
describes d-wave capture quite well.  In the more-important s-wave
capture, the function does not match detailed models of S$_{34}(E)$,
irrespective of fitted parameters; its closeness to the expressions of
\citet{muk2002} suggests that some other functional form is needed to
account for nonzero phase shifts.  

\subsubsection{Model selection for S$_{34}(0)$ determination}
To determine S$_{34}(0)$ from experimental capture data, we use the microscopic models  of \citet{kajino86} and \citet{N01} (Kim A potential), rescaled to
fit the data below $E$ = 1 MeV (see below).  We selected these two models based on several factors.  
\begin{enumerate}
\renewcommand{\labelenumi}{\roman{enumi})}
\item
They both accurately reproduce the s-wave phase shifts (as given by the phase-shift analysis of
\citet{tombrello63b}) and the long-range asymptotics of the $^7\mathrm{Be}$ bound states.  The Kajino model reproduces the phase shifts without having been fitted to them. 
\item They contain more short-range physics than hard-sphere or potential models,
which may extend the energy range over which they describe the reaction
correctly.
\item They agree well with each other even though they were generated by very different
computational approaches.  
\item They reproduce the measured energy dependence of S$_{34}(E)$ well, up to at least $E$ = 1.5 MeV (see Fig.~\ref{s34fig}, also Fig. 3 of \citet{erna}).  
\item They calculate other 
electromagnetic observables in  $^7\mathrm{Li}$ and $^7\mathrm{Be}$, that are in reasonable agreement with experiment. 
\end{enumerate} 

\subsubsection{Region of S$_{34}(E)$ fitting}
We restricted the energy range for fitting to $E \leq 1 \ \mathrm{MeV}$.  The scatter among models (which
differ mainly at short range) becomes much larger at energies above 1
MeV, suggesting that the calculations are most reliable at lower energies, where poorly-constrained short-range contributions to S$_{34}(E)$ are minimized.  In
\citet{N01}, the contribution of $^3\mathrm{He}$-$^4\mathrm{He}$
separations less than 4 fm was about 4\% of S$_{34}(0)$ and about 8\% of
S$_{34}(1\ \mathrm{MeV})$.  
Since a uniform 4\% at all energies could be absorbed into the rescaling,
the difference between short-range
contributions at 0 and 1 MeV suggests 4\% as a conservative estimate of the
rescaling error.  

\subsubsection{Theoretical uncertainty in the S$_{34}(0)$ determination}
\label{theoryerror}
We estimate a theoretical uncertainty in the S$_{34}(0)$
determination by rescaling several models to the capture data in the same manner used to determine the recommended value of S$_{34}$(0), and examining the resulting spread in S$_{34}(0)$ values.  We restrict our
consideration to microscopic models that reproduce the s-wave phase
shifts, choosing those of \citet{walliser84}, \citet{csoto00} (only
those with $^3\mathrm{He}+{}^4\mathrm{He}$ clusterization),
\citet{N01}, and new variants of
the \citet{N01} calculation possessing phase shifts perturbed from
the empirical values.  

The full spread among the chosen set of
models is $\pm 0.030$ keV b, relative to the \citet{kajino86} and \citet{N01} (Kim A potential) fits.    
We somewhat arbitrarily recommend two-thirds of this value; i.e.,
$\pm 0.02$ keV b, as an approximate 1-$\sigma$ theoretical error.
The scatter among these models is not independent of the rescaling
uncertainty estimated above; hence, we have not included an explicit rescaling contribution in this estimate.

\subsubsection{S-factor derivatives}
The data do not provide a useful constraint on low-energy
derivatives of S$_{34}(E)$.  Microscopic models that
reproduce the phase shifts and simpler models that focus on wave-function 
asymptotics produce values of S$_{34}^\prime(0)$/S$_{34}(0)$
in the range $-0.55$ to $-0.79\ \mathrm{MeV^{-1}}$.  These values depend on both the model and the method of estimation.  Only \citet{wil1981},
\citet{walliser83}, and \citet{walliser84} published enough information to allow one to extract
an estimate for
S$_{34}^{\prime\prime}$, yielding S$_{34}^{\prime\prime}(0)/\mathrm{S}_{34}(0)$ = 0.26 to 0.43 MeV$^{-2}$.
We base our recommendations on the \citet{N01} (Kim A) model, which yields effectively
S$_{34}^\prime(0)/\mathrm{S}_{34}(0)=-0.64\ \mathrm{MeV^{-1}}$ and S$_{34}^{\prime\prime}(0)/\mathrm{S}_{34}(0)$ = 0.27 MeV$^{-2}$ from a quadratic fit below 0.5 MeV.

\subsubsection{Comment on phase shifts}
As the bound-state $^7$Be wave functions have known
asymptotic forms, differences of the low-energy S$(E)$
among models arise from differing s-wave phase shifts and from
short-range contributions.  The short-range contributions, which are
difficult to compute convincingly, are probed by capture experiments above 1 MeV.  With the
exception
of \citet{mohr93} and \citet{mohr09},
phase-shift fitting
for studies of the
$^3\mathrm{He}(\alpha,\gamma)^7\mathrm{Be}$ reaction has been based
almost entirely on the phase-shift analysis of \citet{tombrello63b}.
While this phase-shift analysis provides a useful constraint, it
depends mainly on a single experiment from the early 1960s, and it
does not include an error estimation.  The modern \citet{mohr93} experiment 
extended to lower energies, but it has no published error estimate or
phase-shift analysis.  
%New scattering and capture experiments above 1 MeV could be useful for future model building.

\begin{figure}
\includegraphics[width=8.5cm]{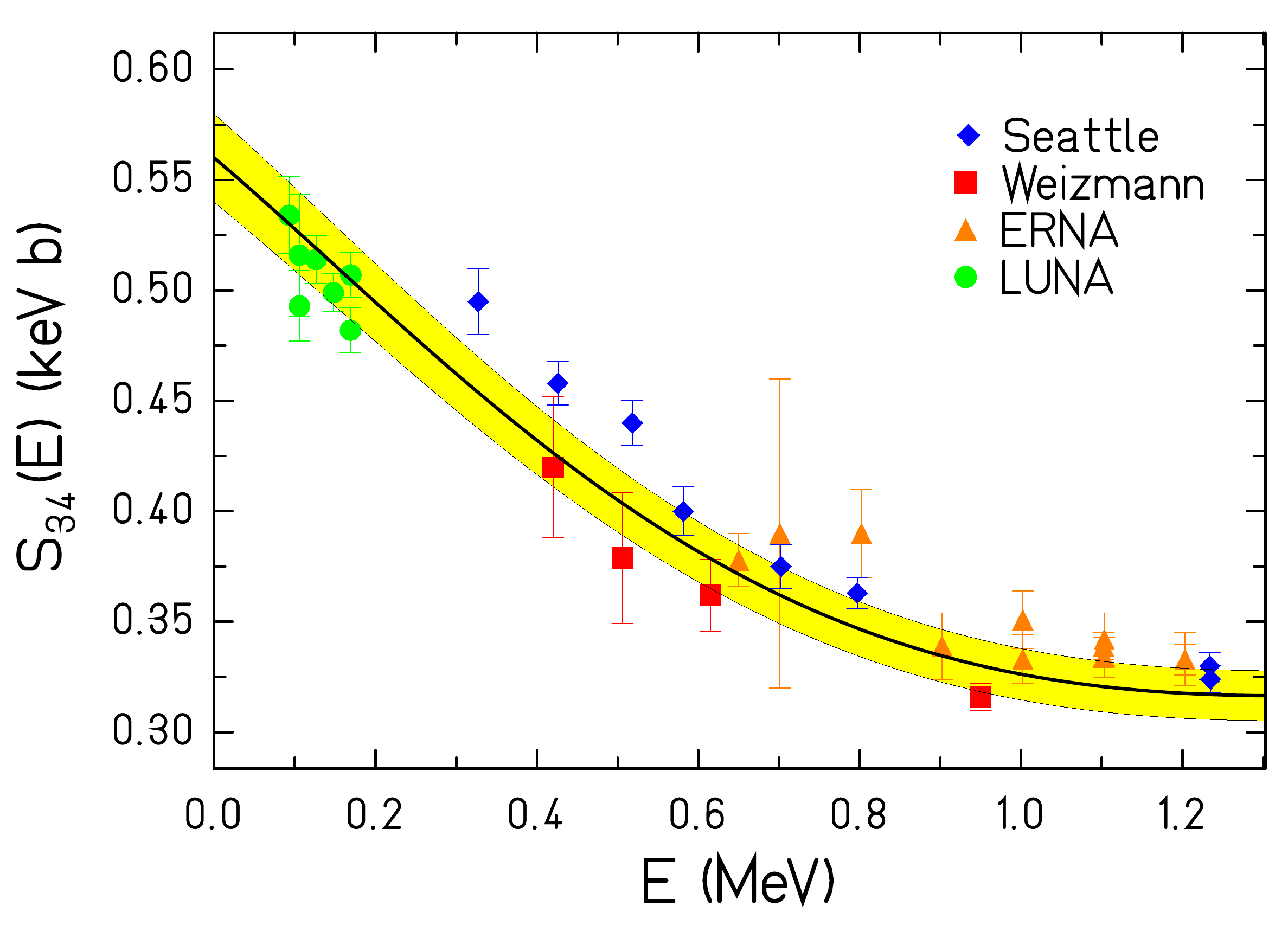}
\caption{(Color online) S$_{34}(E)$ vs. $E$.  Data points: LUNA - green circles; Weizmann - red squares; UW-Seattle - blue diamonds; ERNA - brown triangles.  Solid curve - best fit scaled Nollett theory to the data with $E$ $\leq$ 1.002 MeV.  The yellow band indicates the $\pm$1-$\sigma$ error band.  Data are shown with statistical-plus-varying-systematic errors only; overall systematic errors are not included.}
\label{s34fig}
\end{figure}

\subsection{S$_{34}(0)$ determination}
Figure~\ref{s34fig} shows the low energy data with $E$ $\leq$ 1.23 MeV, and the fit obtained by scaling the Nollett (Kim A potential) theory to best match the data with $E$ $\leq$ 1.002 MeV.  We used the analytic function
\begin{eqnarray}
\mathrm{S}_{34}(E) &=& \mathrm{S}_{34}(0)~e^{-0.580E} \nonumber \\
&\times& (1  -0.4054E^2 +   0.577E^3 -0.1353E^4),~~~~~
\end{eqnarray}
where $E$ is in units of MeV.  Below one MeV this expression is valid to better than 0.3\%, on average.

The best-fit curve in Fig.~\ref{s34fig} was obtained by fitting each data set separately with the scaled theory, and then fitting the set of four S$_{34}(0)$ values to determine the mean S$_{34}(0)$ value and its error.  
\begin{table}
\caption{Experimental S$_{34}(0)$ values and 1-$\sigma$ uncertainties determined from fits of the scaled Nollett (Kim A potential) theory to published data with $E \leq 1.002$ MeV. Total errors are quoted, including inflation factors, and systematic errors of  LUNA: $\pm$ 2.9\%; Weizmann: $\pm$ 2.2\%; UW-Seattle: $\pm$ 3.0\%; ERNA: $\pm$ 5.0\%. }  
\label{s34table}
\begin{ruledtabular}
\begin{tabular}{lccc}

Experiment   & S$_{34}(0)$ & Error & Inflation  \\
 & (keV b) & (keV b) & Factor \\
\hline
LUNA & 0.550  & 0.017   & 1.06 \\
   Weizmann & 0.538  & 0.015   & 1.00 \\
   UW-Seattle & 0.598  & 0.019   & 1.15 \\
   ERNA  & 0.582  & 0.029   & 1.03 \\
\hline
   Combined result & 0.560  & 0.016   & 1.72 \\
\end{tabular}
\end{ruledtabular}
\end{table}

As can be seen from Table~\ref{s34table}, the fits to the individual data sets are good, indicating consistency with the theoretical energy dependence, within the limited energy ranges of each set.  The fit to the combined set of four S$(0)$ values is of marginal quality, indicating a lack of good agreement in the absolute normalizations of the different experiments.  The combined fit has $\chi^2/dof$ = 2.3 ($dof$ = 3), corresponding to P($\chi^2,dof)$ = 0.07.  All of the errors given in Table~\ref{s34table} include the inflation factors determined from the goodness of fit (see the Appendix, Sec.~\ref{errorsappendix}).  Fits to these data using the scaled theory of Kajino yield slightly smaller $\chi^2$ values, and reproduce the low-energy UW-Seattle data somewhat better; however, the 
mean S$_{34}(0)$, 0.561 keV b, is essentially identical to the result obtained with Nollett's theory.  

We have focused here on measurements published since Solar Fusion I.  We do so because in general they are better documented than the older ones, and address issues such as contaminant $^7$Be production in a quantitative manner that lends greater confidence to the results.  One may judge from the Kajino-fit analysis presented in~\citet{sea}, that including older measurements would lower the mean S$(0)$ by at most 0.01 keV b or so.  Thus including the older measurements would not change our result significantly.    

Given the marginal quality of the mean experimental S$_{34}(0)$ fit, we round off the values given above, 
and quote a ``best" result,
\begin{equation}
\mathrm{S}_{34}(0) = 0.56  \pm 0.02\mbox{(expt)} \pm 0.02\mbox{(theor)}  \hspace{0.1cm}\mbox{keV b},
\label{s0}
\end{equation}
based on activation data and the ERNA recoil data, and taking the theoretical error from Sec.~\ref{theoryerror}.

Our best S$_{34}(0)$ estimate may be compared to the value S$_{34}(0)$ = 0.53 $\pm$ 0.05 keV b given in Solar Fusion I.

New capture experiments below 1 MeV would be most valuable for reducing the experimental uncertainty in 
S$_{34}(E)$,
particularly ones that maximize overlap with the existing modern data sets.
New scattering and capture experiments above 1 MeV, as well as precise angular distribution measurements, could be useful for constraining future theoretical calculations.  \footnote{Note added in proof: Recent 
fermionic molecular dynamics (FDM)  calculations \cite{Neff10} of S$_{34}(E)$ 
are in excellent agreement, in both absolute magnitude and energy dependence, with the experimental
data shown in Fig.~\ref{s34fig} and with the high-energy ERNA data up to 2.5 MeV.  The FDM is a
nearly {\it ab initio} microscopic method employing realistic effective interactions.}

\section{THE $^3$H\lowercase{e(p,e}$^+$$\nu_e$)$^4$H\lowercase{e} REACTION}
\label{sec:hep}
The hep reaction
\begin{equation}
\rm{p}+{}^3{\rm He}\rightarrow {}^4{\rm He}+e^++\nu_e
\label{eq:hep}
\end{equation}
is the source of the pp chain's most energetic
neutrinos, with an endpoint energy of 18.8 MeV.  The
Super-Kamiokande and SNO collaborations have
placed interesting limits on the hep neutrino flux by searching 
for these neutrinos in the energy window above the $^8$B neutrino endpoint,
even though the expected flux is very low (see Fig. \ref{fig:CNOEC}).  The
hep rate is beyond the reach of current experiments:  this process is induced by the weak
interaction and further suppressed by a Coulomb barrier and by other aspects
of the nuclear physics, as explained below. 
Thus theory provides our only estimate of S$_\mathrm{hep}$. 

The calculation of S$_\mathrm{hep}$ is a difficult challenge.
The leading one-body (1B) Gamow-Teller (GT)
transition operator cannot connect the main s-state components
of the p+$^3$He and $^4$He initial- and final-state wave functions.\footnote{
While the radial wave functions of the four nucleons in $^4$He can all be 1s,
with the various single-particle states distinguished by spin and isospin, this is not the case
for the three protons in p+$^3$He: the Pauli principle
requires that one must be radially excited.  The GT
transition operator does not alter radial quantum numbers, only spin and isospin.
Thus the GT matrix element between p+$^3$He and $^4$He is suppressed due
to the s-wave orthogonality.}
Hence, at the 1B level the
reaction proceeds through the small components of the $^3$He and $^4$He
wave functions, such as  d-state components.  Consequently, the relative
importance of other transition operators, such as
axial meson-exchange currents (MEC), is enhanced,
as is the contribution  from p-wave p+$^3$He capture, normally
kinematically suppressed at solar temperatures.
The situation is further complicated by the fact that the axial
1B and MEC ``corrections'' have opposite signs, making
s-wave hep capture even more suppressed.

\subsection{hep calculations}
Some of the features mentioned above are shared by
the hen process (n+$^3$He $\rightarrow ^4$He$+\gamma$),
in particular the strong suppression of 1B contributions.
The possibility of  deducing S$_\mathrm{hep}$ from the known hen cross section
was explored in early studies:  while these reactions are not
isospin mirrors, there is a close relationship between the isovector spin
contribution to hen and the GT contribution to hep.  However the
hep S-factors determined in these studies differed,
in some cases, by orders of magnitude.

In an attempt to understand the origin of this large uncertainty, fully microscopic
calculations of both the hep and hen reactions were
performed by \citet{Carlson:1991ju} and \citet{Sch92},
using a realistic Hamiltonian with two- and three-nucleon
interactions. Among the approximations made in the 
\citet{Sch92} calculation were the description of the p+$^3$He
initial state as s-wave and the omission of the dependence of the
weak operators on the lepton pair momentum.
Corrections to the 1B GT operator were evaluated, with
the largest two-body (2B) contributions coming from
the excitation  of intermediate $\Delta$-isobars.
The $\Delta$-isobar degrees of freedom were explicitly included
in the nuclear wave functions, using a scaled-down approach to the
full $N+\Delta$ coupled-channel problem known as
the transition-correlation operator method.
\textcite{Carlson:1991ju} and \textcite{Sch92} found that effects
such as the different initial-state interactions
 for n+$^3$He and p+$^3$He were so substantial that
 the known hen cross section was not a useful constraint on hep.
 Two estimates were given for the hep S-factor at zero energy \cite{Sch92},
 \begin{equation}
 \mathrm{S}_\mathrm{hep}(0)=\left\{ \begin{array}{c} 1.4 \\ 3.1 \end{array} \right\} \times 10^{-20}~\mathrm{keV~b},
 \label{hep_early}
 \end{equation}
 depending on the method used to fix the  weak $N-\Delta$
 coupling constant, $g_{\beta N \Delta}$: the larger of the results
 corresponds to the na\"ive quark model prediction for $g_{\beta N \Delta}$,
 while in the smaller,  $g_{\beta N \Delta}$ was
 determined empirically from tritium $\beta$ decay.
 The Solar Fusion I best value for S$_\mathrm{hep}$ is the average of the values in 
 Eq.~(\ref{hep_early}).

This problem was revisited nearly a decade later, following improvements
in the description of bound and continuum four-body
wave functions.  The wave functions of
\citet{Mar00a}
were obtained with the
correlated-hyperspherical-harmonics (CHH)
variational method~\cite{Viv95,Viv98},
using the Argonne $v_{18}$ (AV18) two-nucleon~\cite{wir95}
and Urbana IX (UIX) three-nucleon interactions~\cite{Pud95}.
The method produced binding energies of $^3$He
and $^4$He and the singlet and triplet p+$^3$He scattering lengths
in excellent agreement with experiment.

The \textcite{Mar00a} calculation included all s- and p-wave capture channels
in the p+$^3$He initial state and all multipole contributions
in the expansion of the weak vector and axial-vector transition operators.
The weak operators corresponding to the space component of the 1B weak vector 
current and the time
component of the 1B axial current, both of order $v/c$,
have significant
exchange-current corrections of the same order from pion-exchange.  These two-body
operators were constructed to satisfy (approximately) the constraints of
current conservation and PCAC (partial conservation of the axial-vector current).   Corrections to the allowed GT
operator include both  $(v/c)^2$ 1B 
and exchange-current contributions.
The treatment of the latter followed \textcite{Carlson:1991ju} and \textcite{Sch92} in
using the transition-correlation operator scheme and in fixing
$g_{\beta N\Delta}$ to the experimental GT strength in tritium $\beta$ 
decay.

Table \ref{tab:hep1} gives the resulting S$_\mathrm{hep}$ at three
center-of-mass energies.  The energy dependence is rather weak.
The  p waves have a significant effect, accounting for
about one-third of the total cross section at $E$=0.
Despite the delicacy of the calculation,
\textcite{Mar00a} concluded that the degree of model dependence was moderate:
the calculations were repeated for the older
Argonne $v_{14}$~\cite{Wir84} two-nucleon
and Urbana VIII~\cite{Wir91} three-nucleon interactions, but the predictions for
S$_\mathrm{hep}$ differed only by 6\%. 
The best estimate of \textcite{Mar00a}, 
S$_\mathrm{hep}=(10.1\pm 0.6)\times 10^{-20}~ \mathrm{keV~b}$,
is about four times the value given in Solar Fusion I.

\begin{table}
\caption{S$_\mathrm{hep}$ in units of $10^{-20}$ keV~b, calculated
with CHH wave functions generated from the AV18/UIX Hamiltonian
\protect\cite{Mar00a} for three p+$^3$He
center-of-mass energies $E$.
The ``One-body'' and ``Full'' labels denote calculations with the
one-body  and full (one- and two-body)
nuclear weak transition operators. Contributions from the
$^3$S$_1$ channel and from all s- and p-wave channels are listed
separately.}
\label{tab:hep1}
\begin{tabular}{lcccccc}
\hline\hline
 & \multicolumn{2}{c}{$E=0$ keV} & \multicolumn{2}{c}{$E=5$ keV} &
\multicolumn{2}{c}{$E=10$ keV} \\
 & ~~$^3$S$_1$~~ & ~~s+p~~ & ~~$^3$S$_1$~~ & ~~s+p~~ & ~~$^3$S$_1$~~ & ~~s+p~~ \\
\hline
One-body~~ & 26.4 & 29.0 & 25.9 & 28.7 & 26.2 & 29.2 \\
Full & 6.38 & 9.64 & 6.20 & 9.70 & 6.36 & 10.1 \\
\hline\hline
\end{tabular}
\end{table}

A further development came with the use of heavy-baryon chiral
perturbation theory (HBChPT) to derive the needed electroweak current operators systematically,
with \textcite{Par03} carrying out the expansion to next-to-next-to-next-to leading order
(N$^3$LO), thereby generating all possible operators to this order.  These
operators represent the short-range physics that resides above the scale of the
EFT, which \textcite{Par03} defined via a Gaussian regulator with a cutoff $\Lambda$,
a parameter that was varied in the calculations between 500 and 800 MeV (see Sec. \ref{sec:s11}).
S$_{\mathrm{hep}}$ was
obtained by calculating the matrix elements of these EFT current operators
with phenomenological wave functions,
obtained using the AV18/UIX Hamiltonian and the CHH method.  (See Sec. \ref{sec:s11}
for a more extended discussion of such hybrid EFT$^*$ approaches.)

%\begin{table}
%\caption{Contributions of the S- and P-wave capture channels
%to S$_\mathrm{hep}$ (in $10^{-20}$ keV~b)
%at $E$=0 \protect\cite{Par03}. The last column gives the results
%obtained by~\protect\textcite{Mar00a}.}
%\label{tab:hep2}
%\begin{center}
%\begin{tabular}{lcccc}
%\hline\hline
%$\Lambda$ (MeV) & ~~500~~ & ~~600~~ & ~~800~~ & ~~\protect\textcite{Mar00a}~~ \\
%\hline
%$d_R$ & 1.00(7) & 1.78(8) & 3.90(10) & \\
%\hline
%$^1$S$_0$ & 0.02 & 0.02 & 0.02 & 0.02 \\
%$^3$S$_1$ & 7.00 & 6.37 & 4.30 & 6.38 \\
%$^3$P$_0$ & 0.67 & 0.66 & 0.66 & 0.82 \\
%$^1$P$_1$ & 0.85 & 0.88 & 0.91 & 1.00 \\
%$^3$P$_1$ & 0.34 & 0.34 & 0.34 & 0.30 \\
%$^3$P$_2$ & 1.06 & 1.06 & 1.06 & 0.97 \\
%\hline
%Total & 9.95 & 9.37 & 7.32 & 9.64 \\
%\hline\hline
%\end{tabular}
%\end{center}
%\end{table}
%
To this order, the resulting currents are 1B and 2B: three-body operators 
arise at order N$^4$LO.  The expansion reproduces the one-pion exchange-current
corrections to the space component of the vector current
and charge component of the
axial current, as
dictated by chiral symmetry,
while the time component of the vector current has no MEC corrections.
The MEC contributions to the axial GT operator
include both a
one-pion-exchange term and a (non-derivative)
two-nucleon contact-term.  The low-energy constant determining the strength
of the contact term must be determined from an observable.
Following the treatment of $g_{\beta N\Delta}$ by \textcite{Mar00a},
this was done by fitting the GT transition strength extracted from tritium
$\beta$ decay.

%It is the basic premise of EFT that physics of ranges shorter than the relevant
%scale is to be lodged in local operators,
%and that the short-range contributions
%are embodied in the CT coefficient $\dR$.
%Stated differently,
%once a proper renormalization procedure for $\dR$ is implemented,
%different values of the cutoff and different short-range behavior of the
%wave functions are expected to only shift the value of $\dR$,
%leaving the final results unchanged, and therefore model-independent.
%It also means that the formal mismatch between phenomenological wave functions
%and EFT current operators 
%can be essentially removed,
%provided that the wave functions can reproduce all the long-ranged
%effective range parameters such as binding energies and scattering lengths.

%The above was omitted because the same issue is discussed in some 
%detail in the EFT* section of S_{11}  - WH

Table \ref{TabL1A} gives
the values determined by \textcite{Par03} for  S$_\mathrm{hep}$(0) and 
for the GT matrix element between the $^3$S$_1$ p$+^3$He initial and
the $^4$He final states, as a function of $\Lambda$.
By fixing the strength of the
contact term to an observable, one hopes in such hybrid EFT$^*$ approaches
to remove most of the calculation's
cutoff dependence.  Heuristically, the contact term
compensates for high-momentum components in the
phenomenological wave functions that would not be there had both operators
and wave functions been derived rigorously from EFT, with a common
cutoff.  However, the table shows that significant cutoff-dependence remains in the 
total amplitude
because of cancellation between the
1B and 2B contributions: the variation in S$_\mathrm{hep}$ is $\sim$ 15\%.
This is taken as the uncertainty in the \textcite{Par03} estimate for S$_\mathrm{hep}$,
S$_\mathrm{hep}(0) = (8.6\pm 1.3)\times 10^{-20}$ keV b.  The result is consistent 
with that of~\textcite{Mar00a}. 

\begin{table}
\caption{\label{TabL1A}\protect The hep GT matrix element
$\overline{L}_1(q;A)$ (in fm$^{3/2}$) for the transition from the initial $^3$S$_1$ p$+^3$He
state to the final $^4$He state, as a function of the cutoff $\Lambda$ \cite{Par03},
at $E$=0.  $\overline{L}_1(q;A)$ is evaluated at $q=19.2$ MeV,
the momentum carried out by the lepton pair.
S$_\mathrm{hep}$ (in $10^{-20}$ keV~b) is also given.}
\begin{tabular}{lrrr} \hline \hline
$\Lambda$ (MeV) & ~~~~500~~ & ~~~~600~~ & ~~~~800~~ \\ \hline
$\overline{L}_1(q;A)$: 1B                    & $-0.081$ & $-0.081$ & $-0.081$ \\
$\overline{L}_1(q;A)$: 2B (no contact term)    & $0.093$  & $0.122$  & $0.166$  \\
$\overline{L}_1(q;A)$: 2B (with contact term)    & $-0.044$ & $-0.070$ & $-0.107$ \\
$\overline{L}_1(q;A)$: 2B-total              & $0.049$  & $0.052$  & $0.059$  \\
\hline
S$_\mathrm{hep}$  & 9.95 & 9.37 & 7.32 
\\ \hline \hline
\end{tabular}
\end{table}

The prediction of~\textcite{Par03}
was used by \textcite{Bah06} and by \textcite{PGS2008} in
% Bah06 reference changed to Ap J Suppl from Ap J
their latest determinations of the hep neutrino flux,
$\phi_\nu(\mathrm{hep})=(8.22\pm 1.23)\times 10^3$ cm$^{-2}$ s$^{-1}$,
where the error reflects again the 15\% uncertainty
quoted above. The value for $\phi_\nu(\mathrm{hep})$
is in agreement with the Super-Kamiokande~\cite{SK} and
SNO~\cite{Aha06} upper limits at
90\% confidence level, $40\times 10^3$ and
$23\times 10^3$ cm$^{-2}$ s$^{-1}$, respectively.

\subsection{Summary}
Given the two consistent calculations presented above, 
with the internal checks on the sensitivities to input
wave functions and to cutoffs, and given the compatibility with the limits established by
 Super-Kamiokande and SNO,
we recommend
\begin{equation}
\mathrm{S}_\mathrm{hep}(0)=(8.6\pm 2.6)\times 10^{-20} \,{\rm keV~b},
\label{eq:hepS}
\end{equation}
where the uncertainty is obtained by doubling
the cutoff-dependence found in the~\textcite{Par03} calculation.  One anticipates
that the cutoff dependence would be reduced if the operator expansion
were carried out beyond N$^3$LO.   Thus such a program could increase
confidence in Eq.~(\ref{eq:hepS}) and narrow the uncertainty,
even without a fully consistent treatment
of both operators and wave functions.

Other ancillary calculations that could strengthen confidence in this S-factor estimate include
\begin{itemize}
\item  new studies of the hep reaction in which a broad spectrum of Hamiltonian
models are explored, as was done by \textcite{Sch98} for the pp reaction;
\item  study of related electroweak reactions where rates are known, such as muon 
capture, as was done 
by \textcite{Mar02} and \textcite{Gaz08} for 
$\mu^-+\,^3{\rm He}\rightarrow\,^3{\rm H}+\nu_\mu$; and
\item further work to understand the relationship between the
suppressed processes hep and hen.
\end{itemize}

\section{ELECTRON CAPTURE BY $^7$B\lowercase{e}, \lowercase{pp}, and CNO NUCLEI}
\label{sec:ec}
Electron capture is the source of line features in the solar neutrino spectrum, and represents an important pathway for energy production in the pp chain.  Solar electron-capture lifetimes differ substantially from laboratory values because light nuclei are highly ionized and because the continuum electron density is large.  

The relative rates of $^7$Be  electron capture and $^7$Be(p,$\gamma$)$^8$B determine the ppII/ppIII
branching ratio and thus the ratio of the $^7$Be and $^8$B neutrino fluxes.  The electron capture 
proceeds by the mirror transition to the ground state of $^7$Li (3/2$^-$) and by an
allowed transition to the first excited state (1/2$^-$, 478 keV).
By normalizing the solar rate to the known terrestrial decay rate, the nuclear physics dependence of
the solar rate can be eliminated.  The ratio of rates depends on the relative electron probability
densities averaged
over the nucleus.  This requires a calculation of the atomic probability densities
governing the K and L terrestrial electron capture rates,
the continuum electron probability densities at the nucleus for the solar rate, and corrections to the
solar rate resulting from incomplete ionization.  The solar continuum calculation was done by \citet{Bahcall62}, 
and estimates of the bound-electron contributions have been made by \citet{Iben67}, \citet{Moeller}, and
\citet{Bahcall:1994cf}.  The solar continuum calculations have typically been done
by employing the Debye-H\"uckel
approximation for plasma screening.  Electrons within the local Debye
sphere screen the nuclear potential, thus lowering the electron density at the nucleus and
the electron capture rate, while protons penetrating that radius would enhance the rate.

Our recommended rate is based on the calculation of \citet{Moeller}, 
with updates including the currently adopted $^7$Be half-life of 53.22 $\pm$ 0.06 days, a
total-to-continuum capture ratio of 1.217 $\pm$ 0.002 \cite{Bahcall:1994cf}, and a terrestrial L/K capture ratio
of 0.040 $\pm$ 0.006 \cite{LtoK}.  We use the original estimate of \citet{Bahcall62} for
the terrestrial K-electron probability at the nucleus.  The result,
\begin{eqnarray}
R(^7\mathrm{Be}+e^-) &=& 5.60 (1 \pm 0.02) \times 10^{-9} (\rho/\mu_e) ~~~~~~~~ \nonumber \\
&\times&~T_6^{-1/2}[1+0.004(T_6-16)] ~\mathrm{s}^{-1},
\label{eq:bec}
\end{eqnarray}
valid for 10 $<T_6<$ 16, is identical to Eq.~(26) of Solar Fusion I.  
Here $\rho$ is the density in units of g/cm$^3$, $T_6$ is
the temperature in units of 10$^6$K, and $\mu_e$ is the mean molecular weight per electron.
The assigned uncertainty of 2\% is dominated by possible corrections to the Debye-H\"uckel 
approximation for charge fluctuations (reflecting the small number of electrons within
the Debye sphere), and by breakdowns in the adiabatic
approximation, as evaluated by \citet{Johnson} in self-consistent
thermal Hartree calculations.
The small rate enhancement they found, 1.3\%, is incorporated
into and dominates the error in Eq.~(\ref{eq:bec}).

Despite the lack of changes since Solar Fusion I, there have been developments in two 
areas, each concerned with screening corrections.  First, a series of
precise measurements of the terrestrial electron capture rate have been carried out to 
assess the dependence of screening on target chemistry,  which could alter
the L/K ratio (because of L-capture sensitivity to changes in the
valence electrons).  Over the past decade such changes, first
suggested by \citet{Segre}, have been explored in a series of half-life measurements in
which $^7$Be was implanted in metals and insulators, or encapsulated in fullerene
\cite{Ray99,Norman01,Ray02,Ohtsuki04,Das05,Ray06,Limata06,Wang06,Nir-El07}.
The pattern of results is somewhat confused, with claims of variations up to 1.1\%, but
with other studies limiting effects
to levels $\lsim$ (0.2-0.4)\% \cite{Limata06,Nir-El07}, despite use of host materials
with substantially different electron affinities.  Our tentative conclusion is that the
uncertainty assigned in Eq.~(\ref{eq:bec}) is sufficient to allow for likely variations in
terrestrial screening corrections.

Second, questions about the adequacy of solar plasma screening corrections, detailed in Solar
Fusion I, have not died out.
\citet{Quarati:2007ie,Quarati:2008hd} reconsidered the plasma fluctuation contributions to the electron-capture rate of $^7$Be, concluding that corrections of 7 - 10\% are required.  The {\it ansatz} of \citet{Quarati:2007ie} was previously considered and rejected by \citet{BBGS02}, however.   The influence of protons on the rate of $^7$Be electron capture in the Sun was claimed to be more significant by \citet{Belyaev:2006qt} than was previously thought.  \citet{Davids:2007jf}, however, reject their argument, pointing out that only the previously 
investigated electromagnetic contributions of protons play a role, and that the approximations under which a putative
three-body electromagnetic contribution was calculated are invalid.

The electron captures on p+p and on CNO nuclei compete with the corresponding $\beta$ decays,
and thus these rates have been conventionally normalized to solar $\beta$ decay rates.  As electron
capture and $\beta$ decay depend on the same allowed nuclear matrix element, the ratio is
independent of the nuclear physics.
The result from Solar Fusion I,
from \citet{Bahcall:1968wz}, is
\begin{eqnarray}
R^\mathrm{Tree}(\mathrm{p}e\mathrm{p}) &=& 1.102 (1 \pm 0.01) \times 10^{-4} (\rho/\mu_e)~~~~~~~~~~~~~~ \nonumber \\
&\times&~T_6^{-1/2} [1+0.02(T_6-16)] R^\mathrm{Tree}(\mathrm{pp}),
\label{eq:pep}
\end{eqnarray}
where the superscript ``Tree" indicates that the relationship omits
radiative corrections, which are discussed below.   The range of validity is 10 $<T_6<$ 16.

\begin{figure}
\includegraphics[width=9cm]{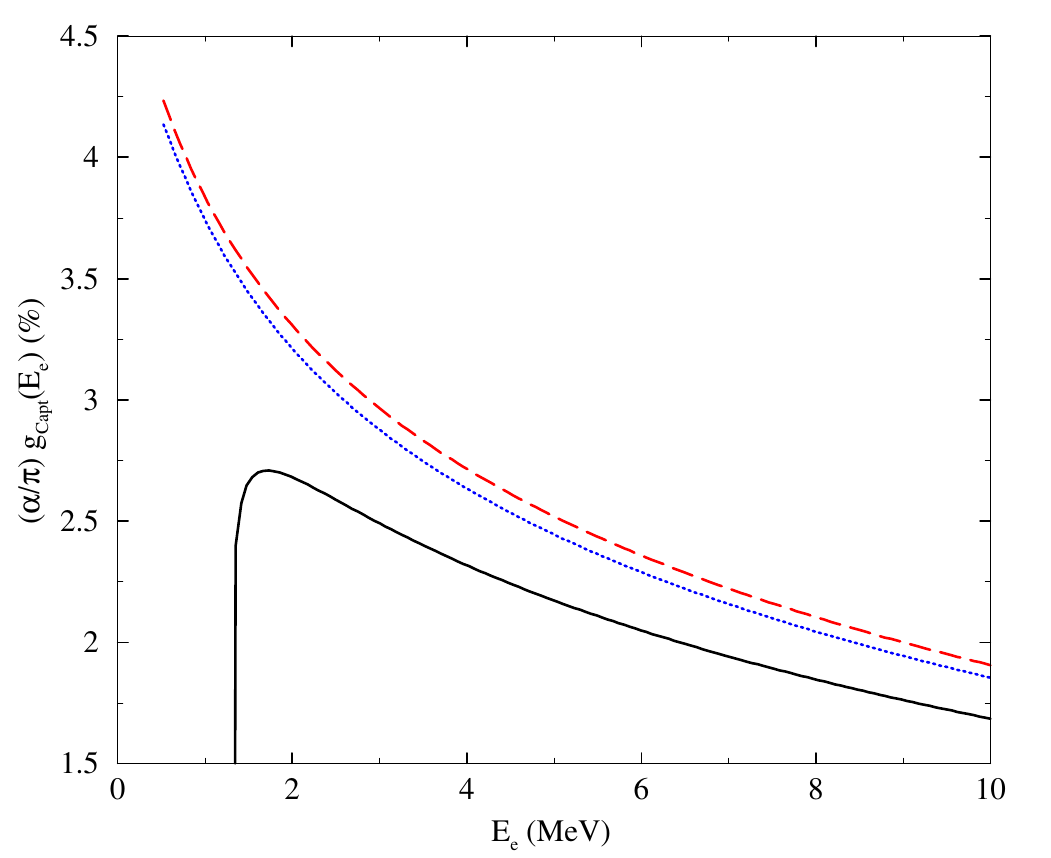}
\caption{(Color online) Calculated radiative 
corrections for p+p+$e^-$  $\rightarrow$  d +$\nu_e$ (dashed line) and 
$^7{\rm Be}+e^-  \rightarrow  {}^7{\rm Li}+\nu_e$ (dotted line).  The solid line 
is for p+$e^-  \rightarrow  n + \nu_e $. Figure from  \citet{Kurylov:2002vj}.}
\label{fig:KRV}
\end{figure}

Radiative corrections were evaluated by
\citet{Kurylov:2002vj} for the two pp-chain reactions under discussion,
\begin{eqnarray}
\mathrm{p+p}+e^-&\rightarrow& \mathrm{d}+\nu_e \\
^7{\rm Be}+e^- & \rightarrow & ^7{\rm Li} +\nu_e.
\end{eqnarray}
The radiative corrections were given as
\begin{eqnarray}
{\Gamma_{\rm Capt} \over \Gamma_{\rm Capt}^{\rm Tree}} = \left[1 + \frac{\alpha}{\pi}g_{\rm Capt}(E_e,Q)\right]
\equiv C^\mathrm{rad}(E_e,Q),
\end{eqnarray}
where $\Gamma_{\rm Capt} $ is the total decay width, $\Gamma_{\rm Capt}^{\rm Tree}$ is the tree-level width without radiative corrections, and $g_{\rm Capt}(E_e,Q)$ is a calculated factor that depends on both 
the total energy $E_e$ of the captured electron and the $Q$-value of the transition.   Figure \ref{fig:KRV} shows the resulting correction factors.

Because Eq.~(\ref{eq:bec}) corresponds to a ratio of stellar and terrestrial electron capture rates,
the radiative corrections should almost exactly cancel: although the initial atomic state in
the solar plasma differs somewhat from that in a terrestrial experiment, the short-range
effects that dominate the radiative corrections should be similar for the two cases.
[Indeed, this is the reason the pp and $^7$Be electron
corrections shown in Fig. \ref{fig:KRV} are nearly identical.]  However the same argument 
cannot be made for the ratio of pep electron capture to pp $\beta$ decay, as the 
electron kinematics for these processes differ.
With corrections Eq.~(\ref{eq:pep}) becomes
\begin{eqnarray}
R(\mathrm{p}e\mathrm{p}) &=& {\langle C^\mathrm{rad}(\mathrm{p}e\mathrm{p}) \rangle \over \langle C^\mathrm{rad}(\mathrm{pp})\rangle}1.102 (1 \pm 0.01) \times 10^{-4} (\rho/\mu_e) \nonumber \\
&\times&~T_6^{-1/2} [1+0.02(T_6-16)] R(\mathrm{pp}),
\label{eq:pep2}
\end{eqnarray}
where the radiative corrections have been averaged over reaction kinematics. 
\citet{Kurylov:2002vj} found a 1.62\% radiative
correction for the $\beta$ decay rate, 
$\langle C^\mathrm{rad}(\mathrm{pp}) \rangle \sim$ 1.016 (see
discussion in Sec. \ref{sec:s11}), while $\langle C^\mathrm{rad}(\mathrm{p}e\mathrm{p}) \rangle
\sim $1.042.   Thus  $\langle C^\mathrm{rad}(\mathrm{p}e\mathrm{p}) \rangle /\langle C^\mathrm{rad}(\mathrm{pp})\rangle \sim$ 1.026, so that our final result becomes
\begin{eqnarray}
R(\mathrm{p}e\mathrm{p}) &=&1.130 (1 \pm 0.01) \times 10^{-4} (\rho/\mu_e) \nonumber \\
&\times&~T_6^{-1/2} [1+0.02(T_6-16)] R(\mathrm{pp}).
\label{eq:pep3}
\end{eqnarray}

While certain improvements could be envisioned in the \citet{Kurylov:2002vj}
calculation -- for example, in the matching
onto nuclear degrees of freedom at some characteristic scale $\sim$ GeV --
rather large changes would be needed to impact the overall rate at the relevant 1\% level.
For this reason, and because we have no obvious basis for estimating the theory
uncertainty, we have not included an additional theory uncertainty in Eq.~(\ref{eq:pep3}).  
However, scrutiny of the presently unknown hadronic and nuclear effects in $g_{\rm Capt}(E_e,Q)$
would be worthwhile.  As one of the possible strategies 
for more tightly constraining the neutrino mixing angle $\theta_{12}$ is a measurement 
of the pep flux, one would like to reduce theory uncertainties as much as possible.
%wording above suggested by M. Ramsey-Musolf

\begin{figure}
\includegraphics[width=8.8cm]{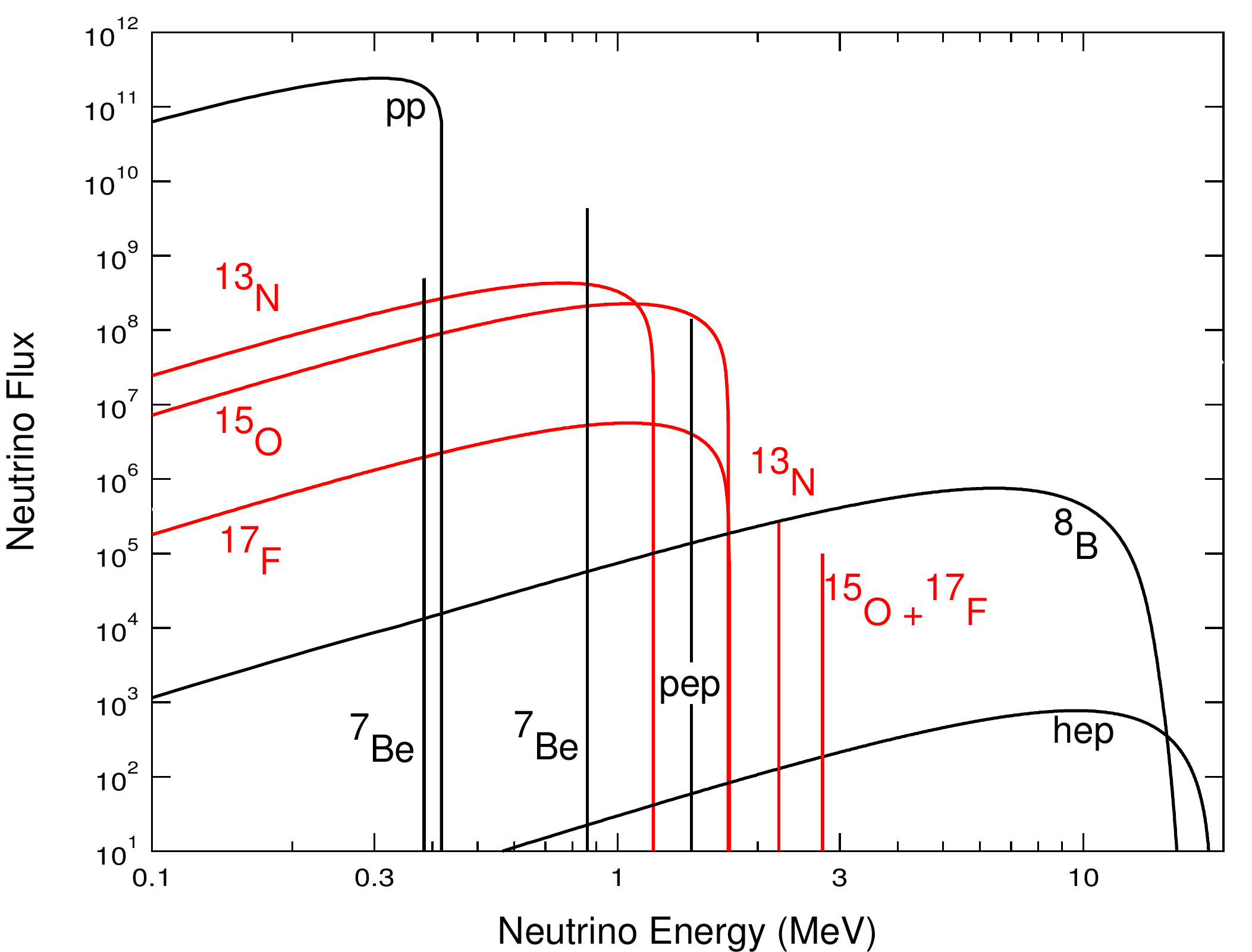}
\caption{(Color online) Solar neutrino fluxes based on the ``OP'' calculations of \citet{Bahcall:2004yr}, 
with the addition of the new line features from CNO reactions.  Line fluxes are in cm$^{-2}$ s$^{-1}$ 
and spectral
fluxes are in cm$^{-2}$ s$^{-1}$ MeV$^{-1}$.  Figure adapted from  \citet{Stonehill:2003zf}.}
\label{fig:CNOEC}
\end{figure}

The  electron capture decay branches for the CNO isotopes $^{13}$N, $^{15}$O, and $^{17}$F were 
first estimated by \citet{Bahcall:1990tb}.   In his calculation, only capture from the continuum was 
considered.  More recently, \citet{Stonehill:2003zf}  have re-evaluated these line spectra by including 
capture from bound states.  Between 66\% and 82\% of the electron density at the nucleus is from 
bound states.  Nevertheless, the electron-capture component is more than three orders of magnitude 
smaller than the $\beta^+$ component for these CNO isotopes, and it has no effect on energy 
production.  However, the capture lines are in a region of the neutrino spectrum otherwise unoccupied 
except for $^8$B neutrinos, and they have an intensity  that is comparable to the $^8$B neutrino intensity 
per MeV (Fig. \ref{fig:CNOEC}), which may provide a spectroscopically cleaner approach to measuring the 
CNO fluxes than the continuum neutrinos do.

The recommended values for the ratio of line neutrino flux to total neutrino flux are listed in Table \ref{tab:EC}.  

\begin{table}
\caption{\label{tab:EC}%
The ratios of neutrino line intensity to the total intensity, after integration over the solar
model.}
\begin{ruledtabular}
\begin{tabular}{lcr}
Source & $R_{\rm line}/R_{\rm total}$   & Ref.  \\
%\textrm{Left\footnote{Note a.}}&
%\textrm{Centered\footnote{Note b.}}&
%\multicolumn{1}{c}{\textrm{Decimal}}&
%\textrm{Right}\\
\colrule
p+p & $2.35\times10^{-3}$~\footnote{includes a 2.6\% radiative correction from 
\citet{Kurylov:2002vj}} & \citet{Bahcall:1990tb}  \\
$^3$He+p & $4\times10^{-8}$~\footnote{to $^4$He ground state}  &\citet{Bahcall:1990tb} \\
 & $\leq 7\times10^{-7}$~\footnote{to $^4$He excited state}  &\citet{Bahcall:1990tb} \\
$^7$Be & 0.8951~\footnote{to $^7$Li ground state} & see text \\
& 0.1049~\footnote{to $^7$Li excited state} &  \\
$^8$B & $2\times10^{-7}$ & \citet{Bahcall:1990tb}  \\
$^{13}$N &  $7.9\times10^{-4}$ & \citet{Stonehill:2003zf}  \\
$^{15}$O & $4.0\times10^{-4}$  & \citet{Stonehill:2003zf}  \\
$^{17}$F & $5.9\times10^{-4}$  & \citet{Stonehill:2003zf}  \\
\end{tabular}
\end{ruledtabular}
\end{table}
%Hamish, it appears that your version used an older table: you had previously
%updated it to include normalization to the total intensity.  That results in some very small changes,
%and alters footnotes.  I have made the new changes for 7Be.  But please check this over
%for me, please -- I think the above is probably right.

The ratio depends weakly on temperature and density, and thus on radius in the Sun.  The values 
given are for the SSM and do not depend significantly on the details of the model.  
The branching ratio for $^7$Be decay to the first excited state in the laboratory is a weighted average 
of the results from \citet{Balamuth:1983zz}, \citet{Donoghue:1983zz}, \citet{Mathews:1983zza}, 
\citet{Davids:1983zz}, \citet{Norman:1983zz,Norman:1983erratum}, 
and an average of earlier results, $10.37\pm0.12$\% (see \citet{Balamuth:1983zz}).
The adopted average, $10.45\pm0.09$\% decay to the first excited state, is corrected by a 
factor 1.003 for the average electron energy in the solar plasma, 1.2 keV \cite{Bahcall:1994cf},
to yield a recommended branching ratio of  $10.49\pm0.09$\%.

\section{THE $^7$B\lowercase{e(p},$\gamma$)$^8$B  REACTION}
\label{sec:s17}
The $^7$Be(p,$\gamma$)$^8$B reaction at low energies is predominantly 
nonresonant $E1$, s- and d-wave capture into the weakly-bound ground state of 
$^8$B \cite{robertson73}.  At solar
energies the reaction proceeds by external direct capture, with
matrix-element contributions dominated by
$^7$Be-p separations on the order of tens of fermis.  The
energy dependence near the Gamow peak cannot be determined from simple
extrapolations of higher energy data, but must be taken from models.  The 
narrow 1$^+$ resonance at $E_\mathrm{p}$ = 720 keV as well as resonances at higher energies are 
usually treated separately, and have little influence on solar rates.   

In Solar Fusion I only one direct $^7$Be(p,$\gamma$)$^8$B measurement was 
found to be sufficiently well documented to allow an independent assessment of the systematic
errors.  Consequently the recommended S$_{17}(0)$ was based on a single experiment,
that of \citet{filippone}.  Since Solar Fusion I
new direct $^7$Be(p,$\gamma$)$^8$B measurements have been carried out at Bordeaux/Orsay
\cite{hamm1,hamm2}, the Weizmann Institute \cite{weiz2,weiz3} (see also \citet{weiz1}), 
Bochum \cite{strieder} and the 
University of Washington-Seattle/TRIUMF \cite{junghans1,junghans2,junghans3}.  
These modern measurements form the basis 
for our Solar Fusion II S$_{17}$(0) recommendation. 

Other new measurements include two performed with $^7$Be 
beams \cite{gialanella00, bardayan09}.  Although inverse measurements of this sort are 
much more difficult, they offer the attraction of different systematic errors.   However, these 
experiments did not reach a precision useful for our purposes and  thus play no role
in our current assessment.

In addition to direct measurements, S$_{17}(0)$ has been determined
indirectly from Coulomb dissociation, as summarized below in Sec. \ref{coulbreakup},
and from peripheral heavy-ion transfer and breakup reactions.  General aspects
of such techniques are discussed in 
Sec. \ref{sec:indirect}.

\subsection{The direct $^7$Be(p,$\gamma$)$^8$B reaction}
All modern $^7$Be(p,$\gamma$)$^8$B experiments have employed the same basic method 
of counting $\beta$-delayed $\alpha$s from the decay 
of $^8$B to determine the reaction yield. 
However, different experimental techniques were used, and different levels of precision were 
achieved in the procedures for converting measured yields into 
cross sections and S-factors. Below we discuss the most important issues.

\subsubsection{Beam-target overlap}
In a conventional experiment with a beam area smaller than the target area, it can be difficult 
to determine accurately the overlap of the beam with the target, due to
non-uniformities in the areal density of typical targets.
This is frequently the case for radioactive target experiments, as
target designs are often quite compact, with cross sections comparable to 
the beam area, in order to minimize unused target material.   This potential
problem has been avoided in the most recent $^7$Be(p,$\gamma$)$^8$B experiments 
by using small-area targets irradiated by uniform beam fluxes.   The reaction yield is then
proportional to the product of the beam flux and the total number of $^7$Be atoms.  The latter
quantity can be 
determined accurately from the $^7$Be decay radioactivity.  As the target density may 
have tails extending to large radii, and as the beam density may not be perfectly uniform, it is 
necessary to 
carry out ancillary measurements to demonstrate the accuracy of this technique.  Measurements can
include separate determinations of the radial dependence of the beam density and the target density, 
and/or the radial dependence of the product of the beam and target densities.  While the Bochum,
Weizmann, and UW-Seattle/TRIUMF
experiments all used the small-area 
%might check that I got these references right (need RMP style)
target/uniform-beam-flux method, only the latter two experiments provided sufficient information to permit
an independent assessment of procedures.

\subsubsection{$^8$B backscattering}
A systematic error in $^7$Be(p,$\gamma$)$^8$B measurements that was identified after Solar Fusion I is 
the loss of $^8$B reaction products due to backscattering out of the target
\cite{weissman98,strieder98}.  This loss is particularly significant for high-Z target backings and low 
proton energy.  The \citet{filippone} and Bordeaux/Orsay experiments used Pt backings, for which 
the backscattering corrections are significant. In the Bordeaux/Orsay experiment, calculated 
backscattering corrections were applied to the data, while the \citet{filippone} experiment was performed 
prior to the identification of $^8$B backscattering as a serious concern.  \citet{junghans2} 
estimated that the backscattering correction for the \citet{filippone} data would be between -2\% and -4\%
(a factor of two smaller than the estimate given in \citet{weissman98}).  Here we ignore this correction 
because it is well within the overall precision claimed in the \citet{filippone} experiment and because
it is incomplete, as effects due to  target thickness nonuniformity (unknown)
and surface composition have not been included.

For the other modern experiments, $^8$B backscattering losses are not an issue: the Bochum
experiment used a low-Z backing, while the UW-Seattle/TRIUMF experiments used 
an intermediate-Z backing 
and demonstrated by direct measurement that backscattering losses were very small.  The 
Weizmann experiment used implanted targets with an intermediate-Z substrate.

\subsubsection{Proton energy loss corrections}
Low-energy data must be corrected by energy-averaging to account for 
proton energy loss in the target.  This 
requires knowledge of the energy loss profile of the target and the target composition, as well as the
monitoring of possible carbon buildup during bombardment.  The most detailed determination of 
these quantities was made in the UW-Seattle/TRIUMF experiments, 
where the target profile was 
determined from the narrow ($\Gamma <<$ 1 keV) $^7$Be($\alpha, \gamma$)$^{11}$C resonance 
at $E_{\alpha}$ = 1377 keV.  In \citet{junghans3} a more detailed resonance profile analysis of the 
previously published data was presented, allowing for possible depth-dependent target composition.
The varying systematic errors on the low energy ``BE3" thick-target data were increased over the 
original results in \citet{junghans2} due primarily to larger assumed dE/dx uncertainties. 

In the \citet{filippone} experiment, the energy loss profile of the target was deduced from the 
measured shape of the 12-keV wide $^7$Li(p,$\gamma$) resonance at $E_\mathrm{p}$ = 441 keV, 
assuming the $^7$Li and $^7$Be distributions in the target were the same.  In the Bordeaux/Orsay
experiment, Rutherford backscattering and (d,p) measurements were used to determine the 
target composition and proton energy loss.  In the Bochum and Weizmann
experiments, the $\Gamma$ = 36 keV $^7$Be(p,$\gamma$) resonance at $E_\mathrm{p}$ = 720 keV was used to 
determine the proton energy loss.  The Weizmann experiment used implanted 
targets with known composition, verified by direct secondary ion mass spectrometry measurements.  
In the \citet{filippone} and Bochum measurements, limits on the composition were inferred from the fabrication process.

Other important factors include determination and monitoring of the $^7$Be target activity, corrections 
for sputtering losses, and determination of the  efficiency for $\alpha$ detection.  For the implanted target
of the Weizmann experiment,
target sputtering losses were shown to be negligible.  The UW-Seattle/TRIUMF
experiments have the most extensive error analysis of the modern experiments.  Measurements 
were made
with two targets of different thicknesses (labeled BE1 and BE3) and with two different methods for
determining the detection efficiency for $\alpha$s.  The resulting statistical 
and systematic errors are the smallest yet achieved.

\subsection{Theory}
\label{s17theory}
Among the many theoretical models that have been published, the simplest are those
in which the interaction between the $^7$Be nucleus and proton are described
by a Woods-Saxon or similar potential
 \cite{tombrello65,aurdal70,robertson73,barker80,kim87,krauss93,riisager93,
bertulani96,typel97,nunes97a,nunes97b,nunes98,davids03,esbensen04}. 
The main constraints on such models are the ground-state energy, the energies of
low-lying resonances, and s-wave scattering lengths \cite{angulo03}.
Charge symmetry has been used to obtain potentials from
$^7\mathrm{Li}+\mathrm{n}$ scattering lengths and the
$^7\mathrm{Li}(\mathrm{n},\gamma)^8\mathrm{Li}$ cross section, but persistent
difficulties in simultaneously reproducing the absolute cross sections for
$^7\mathrm{Be}(\mathrm{p},\gamma)^8\mathrm{B}$ and
$^7\mathrm{Li}(\mathrm{n},\gamma)^8\mathrm{Li}$ may reflect the
greater sensitivity of neutron capture to the inner part of the wave
function \cite{barker80,esbensen04}.  Among potential models, only
those of \citet{nunes97a,nunes97b,nunes98} include coupling to
inelastic channels, open above the 430 keV threshold for
excitation of $^7$Be.  No significant effect was found, consistent with
results of microscopic models.

Potential models yield a reasonably accurate description of the
external part of the direct capture.
% being also useful for calculations of $^8$B Coulomb dissociation.  
The wave function at $r <
5$ fm is not tightly constrained in potential
models but contributes to
the capture at all energies, particularly
above 500 keV
\cite{csoto97,jen1998b}.  However, one requirement is the existence of a node
in s-wave scattering states, as the scattered wave function must be orthogonal
to those of the closed He core
assumed in the description of $^7$Be \cite{aurdal70}.
Model spectroscopic factors have been
taken from shell-model studies, fixed to
match transfer-reaction results (including the asymptotic normalization
coefficients discussed in Sec. \ref{sec:indirect}), or
determined by rescaling computed S-factors to match capture data.

R-matrix models of direct capture
\cite{barker95,barker00} resemble potential models in their
lack of explicit $^7$Be substructure, their need for
fitting constraints, their apparent fidelity at large $^7$Be-p
separation, and their relative lack of short-range details.
Similar data are fitted and similar results produced.  The R-matrix
as applied to direct capture differs from the discussion in Sec. \ref{sec:theory} only in its 
need for radiative widths and  attention
to the long-range tails of bound states \cite{barker95}.  

``Microscopic'' models explicitly containing eight nucleons can
include substructure within $^7$Be and configurations not reducible to
$^7\mathrm{Be}$+p, calculated from the (effective) nucleon-nucleon 
interaction.   The antisymmetry between the last or
scattering proton and those within $^7$Be is maintained.  Fully microscopic
calculations to date generally apply versions of
the resonating
group method (RGM) to significantly simplify
the many-body problem
\cite{descouvemont88,Johnson,descouvemont94,csoto95,csoto97,d04}.
For S$_{17}$ the interaction is usually tuned to
reproduce the proton separation energy of $^8$B, but may also be
adjusted to reproduce the scattering length of $^7\mathrm{Be}+\mathrm{p}$ in
the $S=2$, $L=0$ channel that dominates capture at zero energy
\cite{d04}.  RGM models do roughly as well as potential models in the
external ($> 5$ fm) region while providing a more realistic description of structure in
the internal region.  Nonetheless, RGM results depend on the choice of
nucleon-nucleon interaction and on the data used to fix parameters.  RGM
predictions of absolute cross sections tend to be high relative to measured values.
Thus RGM results are frequently rescaled, so that theory is used only to predict
the energy dependence of S-factors, in extrapolating higher energy data to
the region of the Gamow peak.

Other microscopic approaches have used
effective interactions in combination with the shell model, adapted
to treat weakly-bound and unbound states of $p$-shell nuclei 
\cite{bennaceur99,halderson06}.  These studies focused on
spectroscopic properties of $A=8$ nuclei rather than the radiative 
capture.  While this approach is not as well developed as the RGM 
method, it has produced low-energy S-factors similar to those of the RGM and
other models.  The absolute S-factor of  \citet{bennaceur99} is in
good agreement with the data, while that of \citet{halderson06} is $\sim$ 40\%
larger than experiment.

Ideally microscopic calculations would be carried out with realistic nucleon-nucleon interactions,
but this is challenging due the complexity of the interaction and the need for very large
spaces.  The only published example is that of
\citet{navratil06a,navratil06b}, in which the overlap integrals between $^8$B and
$^7\mathrm{Be}$+p were computed from seven- and eight-body wave functions
obtained with the {\it ab initio} no-core shell model (NCSM).
Due to the finite range of the harmonic oscillator
basis, the long tails of the $^7\mathrm{Be}$+p overlaps were corrected
by matching their logarithmic derivatives 
to Whittaker functions at intermediate
distances.  These overlaps were then used as final states, with
initial scattering states drawn from previous potential-model studies.
The resulting S$_{17}(0)$, 22.1 eV b, is close to the experimental value.
The calculated S$_{17}(E)$ is relatively insensitive to
the choice of initial state for $E < 100$ keV, but more so at higher energies
(e.g., with variations of 20\% at 1.6 MeV).

The envelope of predicted energy dependences of theoretical models 
has about a 30\% spread over the energy range fitted below.
While efforts have been made to fit
S$_{17}(E)$ with as little theoretical input as possible, some degree of model
input appears necessary \cite{cyb2004}.

We adopt the RGM calculation of \citet{d04} as the standard to
extrapolate the experimental data to energies of astrophysical interest.
Among available RGM calculations, this one is the most complete
numerically.  Of the two $NN$ interactions used in \citet{d04}, the
Minnesota interaction was judged to describe light nuclei more
accurately. The predicted S$_{17}(0)=24.69\ \mathrm{eV~b}$ is 19\% larger
than our recommended value, while the calculated shape of S$_{17}(E)$ provides
a marginally better fit to the data, compared to other models
we considered. Other $^8$B and $^8$Li properties computed in this
model also match experiment reasonably well.  Nevertheless, the substantial
theoretical error bar assigned to our end result of Sec. \ref{s17results} -- 
to remove much of the dependence on choice of model -- dominates 
the overall uncertainty in our value for S$_{17}(0)$.

Low-order polynomial representations of S$_{17}(E)$ that span both the solar Gamow peak
and energies
where data are available have poor convergence due to a pole in
the S-factor at $-138$ keV
\cite{jen1998b,jen1998a,wil1981}.   Thus instead we
fit the models over a more limited energy range important to stellar fusion,
0 to 50 keV.   A quadratic expansion then provides a good representation.
This procedure yields S$_{17}^\prime(0)/$S$_{17}(0)$ between
$-1.4$/MeV and $-1.83$/MeV for the models used in our fitting.  We recommend as
a best value and probable range 
\begin{equation}
{\mathrm{S}_{17}^\prime(0) \over \mathrm{S}_{17}(0)}=(-1.5\pm 0.1)/\mathrm{MeV}.
\end{equation}
The
corresponding values for  S$_{17}^{\prime\prime}(0)/\mathrm{S}_{17}(0)$ vary from 7.2/MeV$^2$ to
20.4/MeV$^2$; we recommend 
\begin{equation}
{\mathrm{S}_{17}^{\prime\prime}(0) \over \mathrm{S}_{17}(0)} =
(11\pm 4)/\mathrm{MeV}^{2}. 
\end{equation}
The ranges are consistent with other published values
where derivatives were defined by similar procedures
\cite{barker83,descouvemont88,kolbe88,bennaceur99}. Published
values outside our recommended ranges  \cite{wil1981,baye85,Johnson,jen1998b,baye98,Adel98,bay2000a,bb2000}  are either mathematical
derivatives at $E=0$ or fits over a wider energy interval.  For
the adopted \citet{d04} model with MN potential, the corresponding
numbers are S$_{17}^\prime(0)/$S$_{17}(0)=-1.51$/MeV and
S$_{17}^{\prime\prime}(0)/$S$_{17}(0) = 13.5$/MeV$^2$.

\subsection{$^8$B Coulomb dissociation measurements}
\label{coulbreakup}
Estimates of direct (p,$\gamma$) capture cross sections can be derived from Coulomb 
Dissociation (CD) measurements (see Sec. \ref{sec:indirect}).  Because of the
complexity of the associated analysis and the absence of convincing benchmarks
for the CD method, 
the Solar Fusion I authors concluded that it would be premature to use
information from the CD of $^8$B in deriving
a recommended value for S$_{17}(0)$. However, the CD of $^8$B was identified
as a prime test case for this method, because this reaction can be
studied both directly and indirectly, is characterized by a low proton
binding energy, and is dominated by $E1$ transitions. Three groups have performed 
CD experiments with radioactive $^8$B
beams of incident energies between 44 and 254 A MeV. A
comparison of their results to those from radiative proton capture
allows one to assess the
precision that might be possible with the CD method.

Exclusive CD measurements were performed at 47 A MeV
\cite{motobayashi94, iwa96} and 52 A MeV \cite{kik97, kikuchi98} at RIKEN,
at 83 A MeV at MSU \cite{dav01a,davids01a}, and at 254 A MeV at
GSI \cite{iwasa99,sch03,schuemann06}. For the RIKEN and GSI experiments,
the most recent publications supersede the previously published
ones. The RIKEN experiment measured the CD of $^8$B in complete
kinematics including $\gamma$-rays, but had to cope with a large
background induced by reactions in the He bag between the target
and the fragment detectors. The MSU experiment suffered from a low
detection efficiency, particularly at high p-$^7$Be relative
energies. The GSI experiment eliminated background by
reconstruction of the break-up vertex and utilized a focusing
spectrometer with large momentum acceptance that provided high
geometric detection efficiency. These considerations suggest that
the GSI measurement of \citet{schuemann06} represents the most complete
experimental study of $^8$B CD to date.  

The extraction of S$_{17}(E)$ from the differential CD cross
section d$\sigma/\mathrm{d}E$, which varies rapidly with energy, is not
trivial. The poor energy resolution in CD experiments, together
with the influence of experimental cuts, require careful
simulations of this distribution using a theoretical model. In
addition to the dominant single $E1$ photon exchange, other
potentially important factors are $E2$ transitions, nuclear
break-up, and higher-order corrections. All of these effects are
expected to be smaller at the higher energy of the GSI experiment than at
the lower energies of the RIKEN and MSU experiments.
However, a proper analysis of the GSI experiment requires
relativistic modeling, a step so far taken only in
perturbation theory \cite{bertulani05,ogata09}.

For the RIKEN case, \citet{kik97} presented differential cross
sections $\mathrm{d}\sigma/\mathrm{d}\theta_8$, where $\theta_8$ is the scattering
angle of the excited $^8$B$^*$ system reconstructed from the
$^7$Be and p momentum vectors, relative to that of the incoming
$^8$B. The measured distribution was compared to first-order
perturbative calculations that included $E1$ and both nuclear and
Coulomb $\ell=2$ transition amplitudes. At low relative energies,
the authors found good agreement of their measured distributions
with those from a model that assumes only a dipole
contribution. Later investigations of
the same data employed more sophisticated reaction models, stressing
the importance of all the effects mentioned above
 \cite{alt03,alt05,esbensen05,summers05,ogata06,goldstein07}.  For example, the
value of S$_{17}$(0) obtained from the
continuum-discretized coupled-channels (CDCC) analysis of
\citet{ogata06} is 13\% larger
than that determined in the first-order calculation of \citet{kikuchi98}.

At MSU, inclusive measurements were performed to test the
prediction that interference between $E1$ and $E2$ transitions in
the CD of $^8$B would produce asymmetries in the longitudinal
momentum distributions of the emitted fragments \cite{esbensen96}.
Longitudinal momentum distributions of the $^7$Be fragments from
the break-up of $^8$B on Pb and Ag targets at beam energies of 44
and 81 A MeV were measured \cite{davids98,davids01a}. Asymmetries in
these distributions were incontrovertibly observed and were
interpreted with both first-order perturbative and CDCC
calculations. The $E2$ strengths deduced from first order
perturbation theory were found to be somewhat smaller than or
consistent with all published models of $^8$B structure. Later,
the longitudinal momentum distributions of the emitted protons
were studied in the exclusive MSU measurement at 83 A MeV
\cite{dav01a,davids01a} and found to be consistent with the $^7$Be
distributions observed in the inclusive measurement. The
S$_{17}(E)$ distribution was extracted from $\mathrm{d}\sigma/\mathrm{d}E$
\cite{davids03} with a requirement that $\theta_{8}<1.8^\circ$,
corresponding classically to an impact parameter of 30 fm; a small
$E2$ contribution derived from the inclusive measurements was
taken into account.

\begin{figure}[bt]
\includegraphics[width=9.0cm]{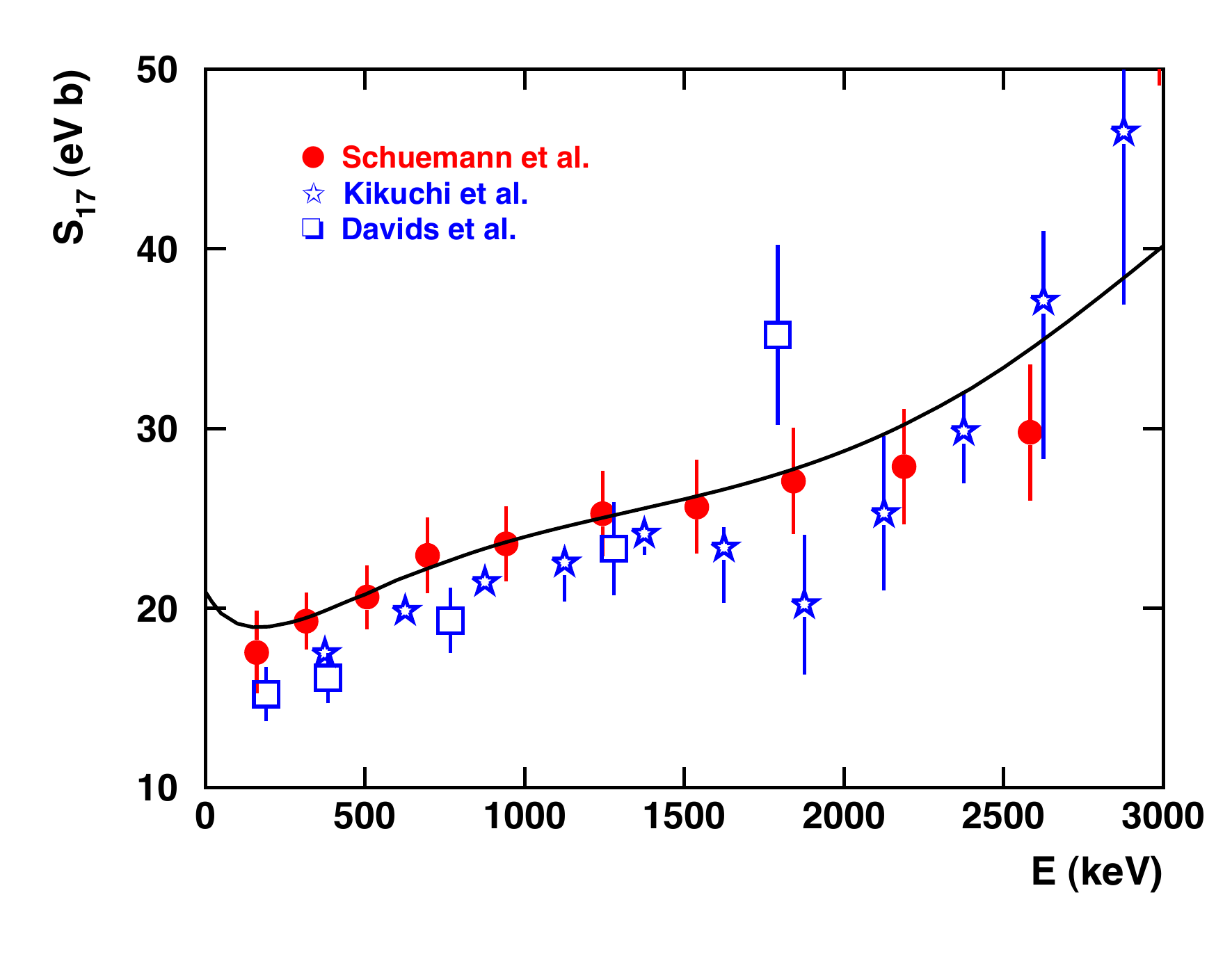}
\caption{(Color online) S$_{17}$ values from CD experiments. Full
red circles:  latest analysis of the GSI CD experiment
\protect\cite{schuemann06}; open blue stars:  \protect\citet{kikuchi98}
analyzed in first-order perturbation theory; open blue squares:
\protect\citet{davids03}. The error bars include statistical and estimated
systematic errors.  The curve is taken from the
cluster-model theory of \citet{descouvemont04}, normalized to
S$_{17}(0)=20.8$ eV b.} \label{s17_cd}
\end{figure}

\citet{schuemann06} published the most extensive set of differential
cross sections for the GSI experiment. All distributions were
gated by $\theta_8 < 1^\circ$, corresponding to an impact
parameter of 18.5 fm. The measured distributions were compared to
theoretical ones filtered by the experimental efficiency and
resolution using a GEANT-3 simulation. The event generator employed
a simple first-order perturbation-theory description of
Coulomb break-up with only $E1$ transitions included. The authors
chose this simple model for its ease in numerical calculations and
for its fidelity in reproducing,  e.g., the inclusive $\theta_8$
distribution (Fig.11 of \citet{schuemann06}) and the surprisingly
symmetric $\theta^\mathrm{p}_{cm}$ distributions of the protons in the
$^8$B$^*$ reference system (Fig. 13 in \citet{schuemann06}).
Consequently, S$_{17}(E)$ was deduced from this model under the
assumption that, contrary to theoretical expectations, $E2$
transitions could be ignored. The data points resulting from all
three CD experiments are shown in Fig.~\ref{s17_cd}. (Note that the
RIKEN data points were taken from the first-order
perturbation-theory analysis by \citet{kikuchi98}.)

%****************************  Fig.1 ***************************************
%*************************************************************************

The different assumptions made in analyzing the experiments as
well as the number and precision of the CD S$_{17}(E)$ data points
prevent a precise determination of the shape, which therefore
has to be taken from the radiative-capture measurements. In
Fig.~\ref{s17_cd} we display the best-fit curve for the direct
$(\mathrm{p},\gamma)$ data,  including the dominant $E1$ multipole but not the $M1$
contribution (see Sec.~\ref{s17results}).

It is difficult to quantitatively assess the impact 
of the different theories and energy ranges used
in analyzing the three CD experiments on the derived S$_{17}(0)$ values.
The resulting values are ${21.4}
\pm {2.0}$ eV b for the RIKEN experiment, as reanalyzed by
\citet{ogata06};  ${20.6} \pm {1.4}$ eV b for the GSI experiment; 
and ${17.8}^{+1.4}_{-1.2}$ eV b for the MSU experiment.
Empirically these values are consistent with the range Solar
Fusion I defined for direct measurements, 
S$_{17}(0) = 19 ^{+4}_{-2}$ eV b.
Moreover, the good agreement between the shapes of
the GSI CD and the radiative capture data eliminates the concern
about systematically different slopes of S$_{17}(E)$ derived from
the respective methods. However, we believe it would be premature
to include the CD results in our determination of a
recommended value for S$_{17}(0)$, as a better
understanding of the role of $E2$ transitions and higher order
effects in $^8$B breakup at various energies is needed.
Further discussions can be found in Sec. \ref{sec:indirect}.

\subsection{Direct $^7$Be(p,$\gamma$)$^8$B analysis and S$_{17}$(0) determination}
\label{s17results}
\begin{figure}
\includegraphics[width=8.5cm]{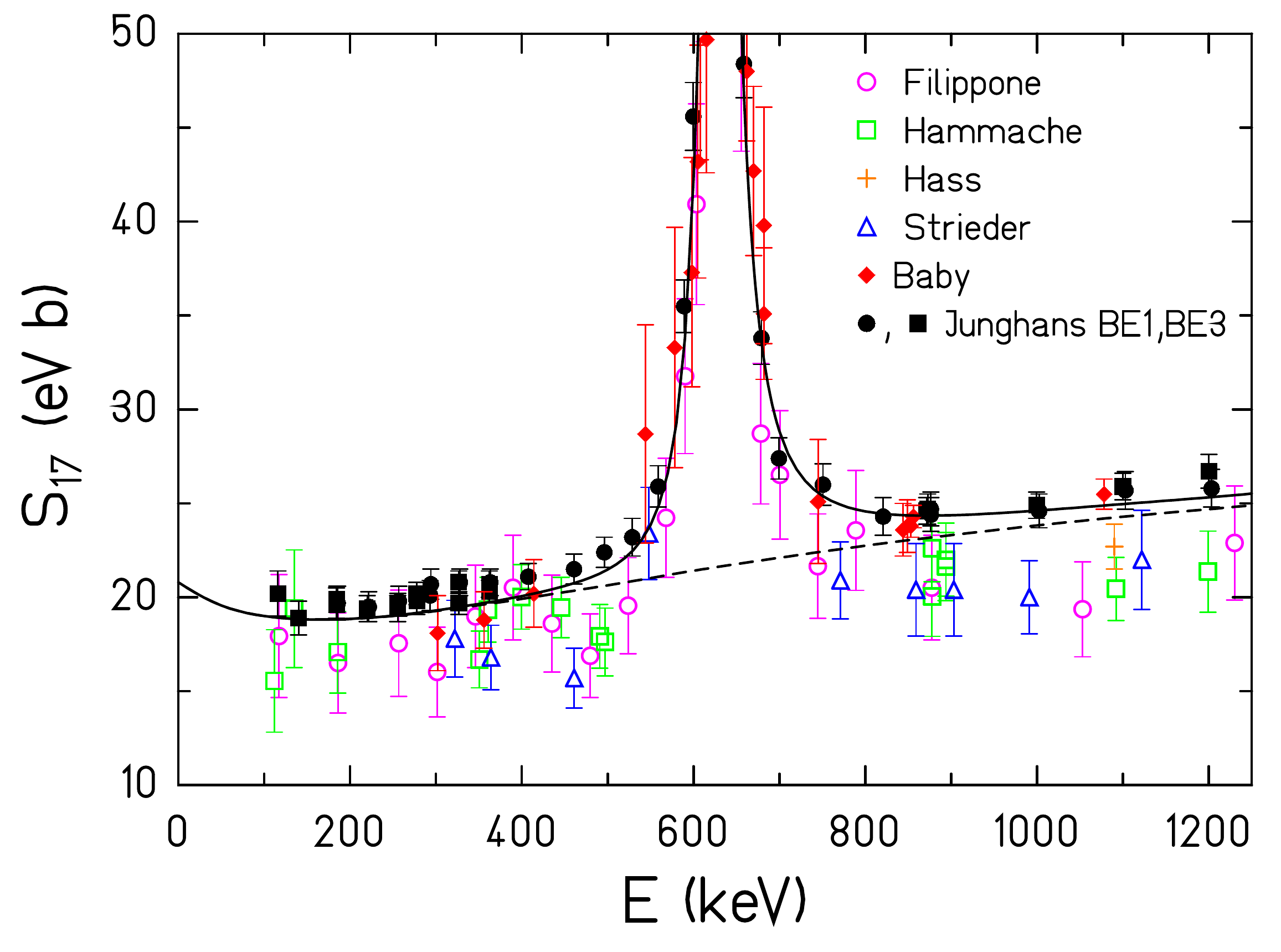}
\caption{(Color online) S$_{17}(E)$ vs. center-of-mass energy $E$, for $E \leq$ 1250 keV. 
 Data points are shown with total errors, including systematic errors.  Dashed line: 
 scaled Descouvemont (2004) curve with S$_{17}$(0) = 20.8 eV b;
solid line: including a fitted 1$^+$ resonance shape. }
\label{s17figure}
\end{figure}

Figure~\ref{s17figure} shows the modern $^7$Be(p,$\gamma$)$^8$B data with center-of-mass 
energy $E \leq$ 1250 keV.  We analyzed the \citet{filippone} data using the $^7$Li(d,p) 
cross section given in Solar Fusion I.   Total errors, including systematic errors, are shown on 
each data point, to facilitate a meaningful comparison of different data sets.   All data sets 
exhibit a similar S$_{17}(E)$ energy dependence, indicating that they differ mainly in 
absolute normalization.  

Following the discussion in Sec.~\ref{s17theory}, we determine our best estimate of S$_{17}(0)$ by 
extrapolating the data using the scaled theory of \citet{d04} (MN calculation). We performed two sets of 
fits, one to data below the resonance, with $E \leq$ 475 keV, where we felt the resonance contribution 
could be neglected.  In this region, all the individual S$_{17}(0)$ error bars overlap, except for 
the Bochum result, which lies low. 

We also made a fit to data with $E \leq$ 1250 keV,  
where the $1^+$ resonance tail contributions had to be subtracted.  We did this using the resonance 
parameters of \citet{junghans2} ($E_\mathrm{p}$=720
keV, $\Gamma_\mathrm{p}=$35.7 keV and $\Gamma_\gamma=$ 25.3 meV), adding in quadrature 
to data errors 
an error of 20\% of the resonance subtraction.  In order to minimize the error induced 
by variations in energy-averaging between experiments, we excluded data close to the resonance, from 
490 to 805 keV, where the S-factor is strongly varying and the induced error is larger than 1.0 eV b. 
Above the resonance, the data have smaller errors.  Only the \citet{filippone} and 
Weizmann group
error bars overlap the UW-Seattle/TRIUMF error bars.  

Figure~\ref{s17figure} shows the best-fit \citet{d04} (MN interaction) curve from the $E \leq$ 475 keV fit (together 
with the 1$^+$ resonance shape determined in \citet{junghans2}, shown here for display purposes).  Our fit 
results are shown in Table~\ref{s17table}.  The errors quoted include the inflation factors, calculated as 
described in the Errors Appendix.  The main effect of including the inflation factors is to increase the error 
on the combined result by the factor 1.7 for $E \leq$ 475 keV, and by 2.0 for $E \leq$ 1250 keV.     
Both the S$_{17}(0)$ central values and uncertainties from the combined fits for these two energy ranges 
agree well,  the latter because the added statistical precision in the $E \leq$ 1250 keV fit is mostly offset 
by the larger inflation factor.

We also did fits in which the low energy cutoff was varied from 375 to 475 keV and 
the high energy exclusion region was varied 
from 425-530 to 805-850 keV.  The central value of S$_{17}(0)$ changed by at most 0.1 eV b.  
On this basis we assigned an additional systematic error of $\pm$ 0.1 eV b to the results for each fit region.  

\begin{table}
\caption{Experimental S$_{17}(0)$ values and (inflated) uncertainties in eV b, and $\chi^2/dof$ determined 
by fitting the \protect\citet{d04} MN calculation to data with $E \leq 475$ keV and with $E \leq 1250$ keV, 
omitting data near the resonance in the latter case. }
\label{s17table}
\begin{ruledtabular}
\begin{tabular}{lcccccc}
Fit Range    &\multicolumn{3}{c}{$E \leq475$ keV } &\multicolumn{3}{c}{$E \leq1250$ keV}\\
\hline 
Experiment   & S$_{17}(0)$ & $\sigma$ & $\chi^2/dof$ & S$_{17}(0)$ & $\sigma$ & $\chi^2/dof$ \\
\hline
Baby              & 20.2  & 1.4\footnotemark[1]   & 0.5/2 & 20.6 & 0.5\footnotemark[1] & 5.2/7 \\
Filippone        & 19.4 & 2.4 & 4.7/6 & 18.0 & 2.2 & 15.8/10\\
Hammache     & 19.3 & 1.1 &  4.8/6 & 18.2 & 1.0 & 12.5/12\\
Hass          &      &  &   & 18.9 & 1.0 & 0/0\\
Junghans BE3    & 21.6  & 0.5   & 7.4/12 & 21.5 & 0.5 & 12.3/17\\ 
Strieder      & 17.2  & 1.7   & 3.5/2 & 17.1	& 1.5 & 5.1/6\\
\hline 
Mean  &  20.8  & 0.7   & 9.1/4     &20.3 & 0.7 & 18.1/5\\

\end{tabular}
\end{ruledtabular}
\footnotetext[1]{We include an additional 5\% target damage error on the lowest 3 points, consistent with 
the total error given in the text of \protect\citet{weiz2} (M. Hass, private communication, 2009).}
\end{table}

\begin{table}
\caption{Experimental S$_{17}(0)$ values and (inflated) uncertainties in eV b, and $\chi^2$ determined by 
fitting nine calculations to the data sets of Table~\ref{s17table}.  The $E \leq475$ keV fits have $dof$ = 4 
and the $E \leq1250$ keV fits have $dof$=5. D04 is \citet{d04}, DB94 is \citet{descouvemont94}, and
NBC06 is \citet{navratil06b}.}
\label{s17table2}
\begin{ruledtabular}
\begin{tabular}{lcccccc}
Fit Range    &\multicolumn{3}{c}{$E \leq475$ keV } &\multicolumn{3}{c}{$E \leq1250$ keV}\\
\hline 	 
Model&	S$_{17}(0)$&	$\sigma$& $\chi^2$&	S$_{17}(0)$&	$\sigma$& $\chi^2$  \\
\hline		 
D04 (central)	&20.8&	0.7&	9.1&	20.3&	0.7&	18.1	\\ 
D04 (upper)	&20.1&	0.7&	10.0&	19.7&	0.7&	18.5	 \\
D04 (lower)	&21.5&	0.7&	8.1&	21.0&	0.7&	17.3	 \\
DB94		&21.4&	0.7&	8.4&	21.5&	0.7&	16.7 	\\ 
NBC06		&22.1&	0.7&	7.4&	21.8&	0.8&	18.5	\\	 
$^7$Be+p (central)	&21.2&	0.7&	8.7&	20.2&	0.7&	19.7	 \\
$^7$Be+p (upper)	&19.4&	0.8&	11.7&	17.3&	0.7&	21.6 	 \\
$^7$Be+p (lower)	&21.7&	0.7&	8.2&	21.0&	0.7&	19.4 	 \\
$^7$Li+n		&20.5&	0.7&	9.7&	19.1&	0.7&	20.9 	 \\
\end{tabular}
\end{ruledtabular}
\end{table}

To estimate the theoretical uncertainty arising from our choice of the nuclear model, we also performed fits 
using the shapes from other plausible models: \citet{d04} plus and minus the theoretical uncertainty 
shown in Fig.~8 of that paper;  \citet{descouvemont94}; the CD-Bonn 2000 
calculation shown in Fig.~15 of \citet{navratil06b}; and four potential model 
calculations fixed alternately to reproduce the $^7\mathrm{Li}+\mathrm{n}$ scattering lengths, the 
best-fit $^7\mathrm{Be}+\mathrm{p}$ scattering lengths, and their
upper and lower limits \cite{davids03}.  The combined-fit results for all these curves, 
including \citet{d04}, are shown in Table~\ref{s17table2}.  

We estimate the theoretical uncertainty on S$_{17}(0)$ from the spread of results in Table~\ref{s17table2}: 
$\pm$ 1.4 eV b for the $E \leq$ 475 keV fits, and 
$^{+1.5}_{-0.6}$ eV b from the $E \leq$ 1250 keV fits (the smaller error estimate in the latter case 
reflects the exclusion of the poorer potential-model fits).  We note that the estimated uncertainties are 
substantially larger than those given in \citet{junghans2} and in \citet{d04}.

We expect the model dependence\footnote{Recently \citet{yam09} discussed a contribution 
of a possible higher energy (3.2 MeV) 2$^-$ resonance
to $^7$Be(p,$\gamma$).  They estimate its contribution by taking the transition strength 
to be a Weisskopf unit. As low-lying E1 transitions are typically strongly inhibited, this estimate is
unlikely to be realistic.  Our $S$-factor estimate
is based on a fit to low-energy data that would be free from any significant influence of this distant resonance,
regardless of such assumptions.} of the fit to be greater above the resonance because of the demonstrated
dependence of the S-factor in this range on the less-constrained short-range part of the wave 
functions \cite{csoto97,jen1998b,d04}.   We base our S$_{17}(0)$ recommendation on the $E \leq$ 475 keV fit, 
\begin{equation}
\mbox{S}_{17}(0) = 20.8 \pm 0.7 \mbox{(expt)} \pm 1.4 \mbox{(theor)}
\hspace{0.2cm} \mbox{eV b.}
\label{combineds0}
\end{equation}  
This value is in agreement with, but substantially more precise than, the 
Solar Fusion I recommendation, S$_{17}$(0) = 19$^{+4}_{-2}$ eV b.

\section{THE SPECTRUM OF $^8$B NEUTRINOS}
\label{sec:spectrum}
The $^8$B neutrino spectrum differs from an allowed shape primarily because the principal state
populated in the decay is a broad resonance.  A precise determination of the neutrino spectrum is 
important to the analyses of the $^8$B neutrino data obtained by the Super-Kamiokande and
SNO collaborations.  Uncertainties in the spectrum are a source of systematic error
in these experiments, potentially affecting conclusions about the hep flux, MSW spectral
distortions, etc.
The neutrino spectrum can be determined from
laboratory measurements of $^8$B $\beta^+$ decay in which the decays of final-state
$^8$Be resonances are observed.

\begin{figure}
\includegraphics[width=8.2cm]{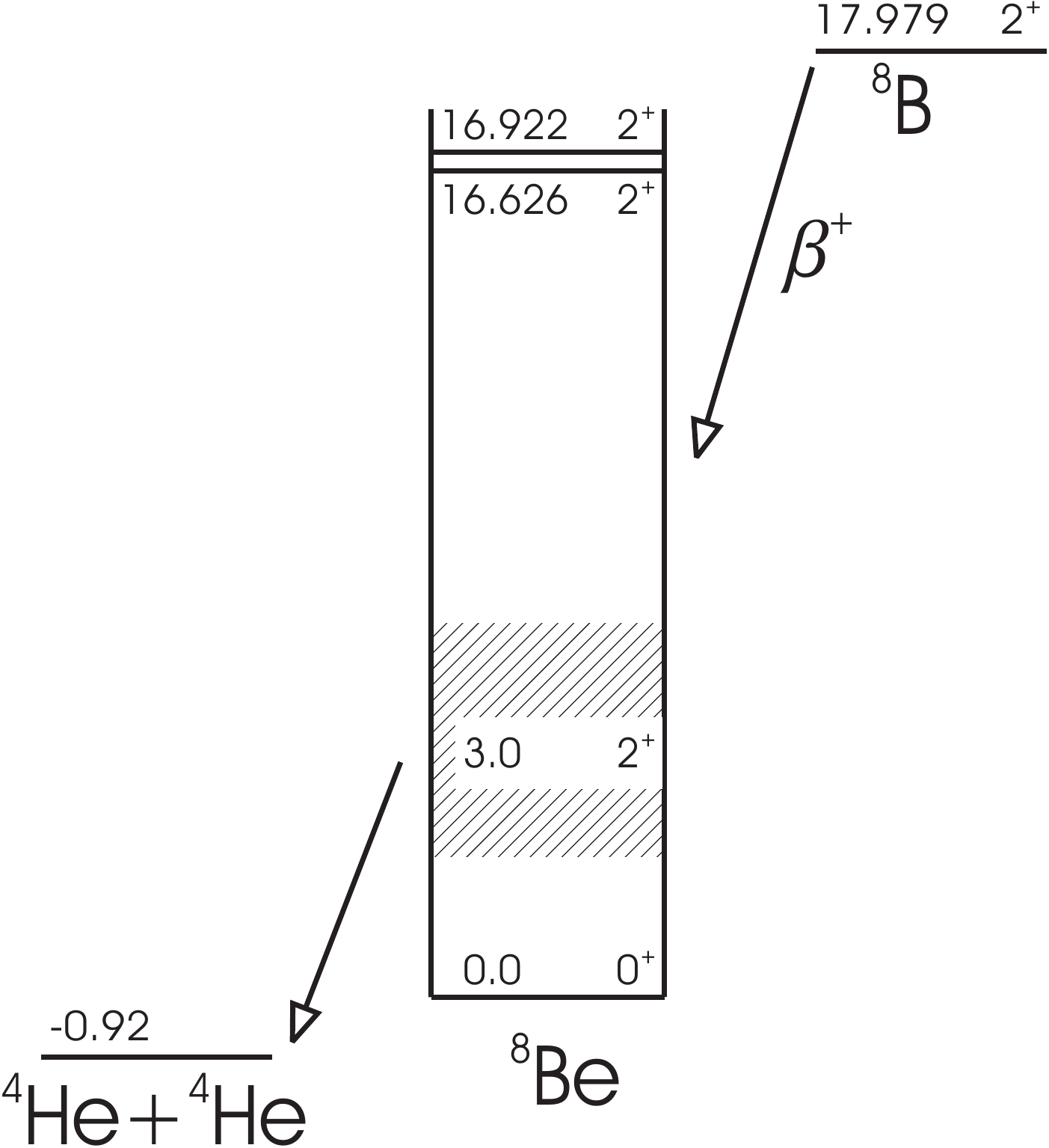}
\caption{Energy levels from the $^8$B($\beta^+$)$^8$Be(2$\alpha$) decay chain.}
\label{fig:b8}
\end{figure}

The $^{8}$B $\beta^+$ decay from the $J^\pi=2^+$ ground state
 is followed by the emission of two $\alpha$ particles from excited $2^+$ states 
of $^{8}$Be (see Fig. \ref{fig:b8}).  Although the region of interest is dominated by a single 
state in $^{8}$Be with $E_x \sim 3$ MeV, the width of this resonance is quite large, 
$\Gamma \sim 1.5$ MeV.
Consequently  the $\alpha$ spectrum yields a continuum, so that other $2^+$ states need 
to be considered.  
The $\alpha$ spectrum was first measured by \citet{fa:60}, and 
later by \citet{wi:71}. R-matrix analyses were presented by 
\citet{ba:89} and \citet{wa:86} (but see the caveat of \citet{ba:02}). 
\citet{ba:96} used the existing data to produce a recommended neutrino spectrum that was 
widely used in subsequent analyses of neutrino experiments. 

\citet{or:00} claimed a discrepancy with previous determinations of the $\alpha$ spectrum. Subsequently \citet{wi:03} and \citet{ba:06} studied the spectrum via experiments with very different systematic uncertainties, finding excellent agreement with each other but disagreement with the claim of \citet{or:00}. It was reported (A. Garc\'{\i}a, 
private communication, 2009))
that \citet{or:00} now recognize that they underestimated uncertainties related to the energy loss generated by carbon buildup in their targets, so that a claim of a disagreement
with earlier measurements no longer should be made.
We recommend using the $\alpha$ spectrum of \citet{wi:06} 
and the consistent and higher precision spectrum of \citet{ba:06}. These experiments 
do not suffer from the energy calibration problems that affected earlier experiments,
as discussed by \citet{ba:96}. Finally we recommend the neutrino spectrum tabulated in
\citet{wi:06}\footnote{The strength function
and the neutrino and positron spectra are in electronic repositories available online
through Phys. Rev. C.}. [The neutrino spectrum was not calculated by \citet{ba:06}.]    

The positron spectrum can be deduced from the $\alpha$ spectrum in a similar fashion,  
and is useful as a test of data consistency. The measurements of
\citet{na:87} have been shown by \citet{wi:06} to be in good agreement with the results 
from the $\alpha$ spectrum.

Forbidden corrections are at the level of a few percent. Many measurements have been 
performed to determine needed matrix elements \cite{{tr:74}, {tr:75},{na:75},{pa:77}, 
{bo:78}, {mk:80},{dB:95}}. Radiative corrections are smaller at a fraction of one percent 
and have been calculated by \citet{si:67} and by \citet{ba:95}. Both sets of corrections 
are described 
by \citet{wi:06}, and 
incorporated into the spectrum given there.
\citet{ba:91} showed that red-shift distortions associated with the
Sun's gravitational potential are insignificant, affecting the spectrum at the fractional level 
of $\sim 10^{-5}$.  \citet{Bacrania} have placed a 90\% confidence-level bound on the
branching ratio for $^8$B $\beta$ decay to the $0^+$ ground state of $^8$Be (a
second-forbidden transition)
of 7.3 $\times$ 10$^{-5}$ (see Fig. \ref{fig:b8}), limiting uncertainties in the high energy
portion of the $^8$B neutrino spectrum.

\section{THE CNO CYCLES}
\label{sec:cn}
The need for two mechanisms to account for the stellar burning of
hydrogen to helium was recognized in the pioneering
work  of  Bethe  and  collaborators.   The pp chain,  which  dominates  energy
production in low-mass  main-sequence  stars,  can operate in metal-free stars,
synthesizing $^4$He from H, while creating equilibrium abundances of deuterium,
$^3$He, and  $^7$Be/$^7$Li, the  elements participating in  intermediate steps
of Fig. \ref{fig:cycles}.

Heavier main-sequence stars produce their energy dominantly through the CNO cycles,
where reactions are characterized by larger Coulomb barriers.  Hence, the energy
production rises more steeply with increasing temperature ($\epsilon_\mathrm{CNO} \propto T^{18}$
compared to $\epsilon_\mathrm{pp} \propto T^4$ at solar core temperature, as illustrated in Fig. \ref{fig:cnopp}).
The CNO cycle was proposed by Bethe and Weizs\"{a}cker to account for the
evolutionary tracks of massive stars.   Unlike the pp-chain,  the CNO
bi-cycle of Fig. \ref{fig:cycles} requires pre-existing metals to process H into $^4$He.
Thus the contribution  to energy
generation is  directly proportional to  the solar-core number  abundance of
the primordial metals.   The CN-cycle, denoted by I  in Fig. \ref{fig:cycles}, is
an important SSM  neutrino source.  It also accounts for about 1\% of solar
energy generation.  The cycle conserves  the number abundance,
but alters the distribution of solar metals as it burns into equilibrium, eventually
achieving equilibrium abundances proportional to the inverse of the respective
rates.   In the Sun this leads to the conversion of almost all of the core's primordial
$^{12}$C into $^{14}$N.  This change in the chemical composition alters the 
core's opacity and, at the 3\% level, the heavy element mass fraction $Z$,
SSM effects first explored by \citet{bu88}.

The  $^{14}$N(p,$\gamma$)  reaction -- the slowest reaction in the CN cycle at
low temperatures and thus the rate-controlling step -- determines  whether  equilibrium has  been
achieved.  The $^{14}$N  lifetime is shorter than the age of  the Sun for temperatures
$\gsim 1.33 \times 10^7$ K.  Therefore equilibrium for the CN cycle has been reached only for
$R \lsim  0.1 R_\odot$, corresponding to the  central 7\% of the  Sun by mass.
Consequently, over a significant portion  of the outer core, $^{12}$C has been
converted   to  $^{14}$N,  but   further  reactions   are  inhibited   by  the
$^{14}$N(p,$\gamma$) bottleneck. 

\subsection{The reaction $^{14}$N(p,$\gamma$)$^{15}$O}
\label{sec:N114}
\subsubsection{Current status and results}
\begin{figure}
\begin{center}
\includegraphics[width=8.5cm]{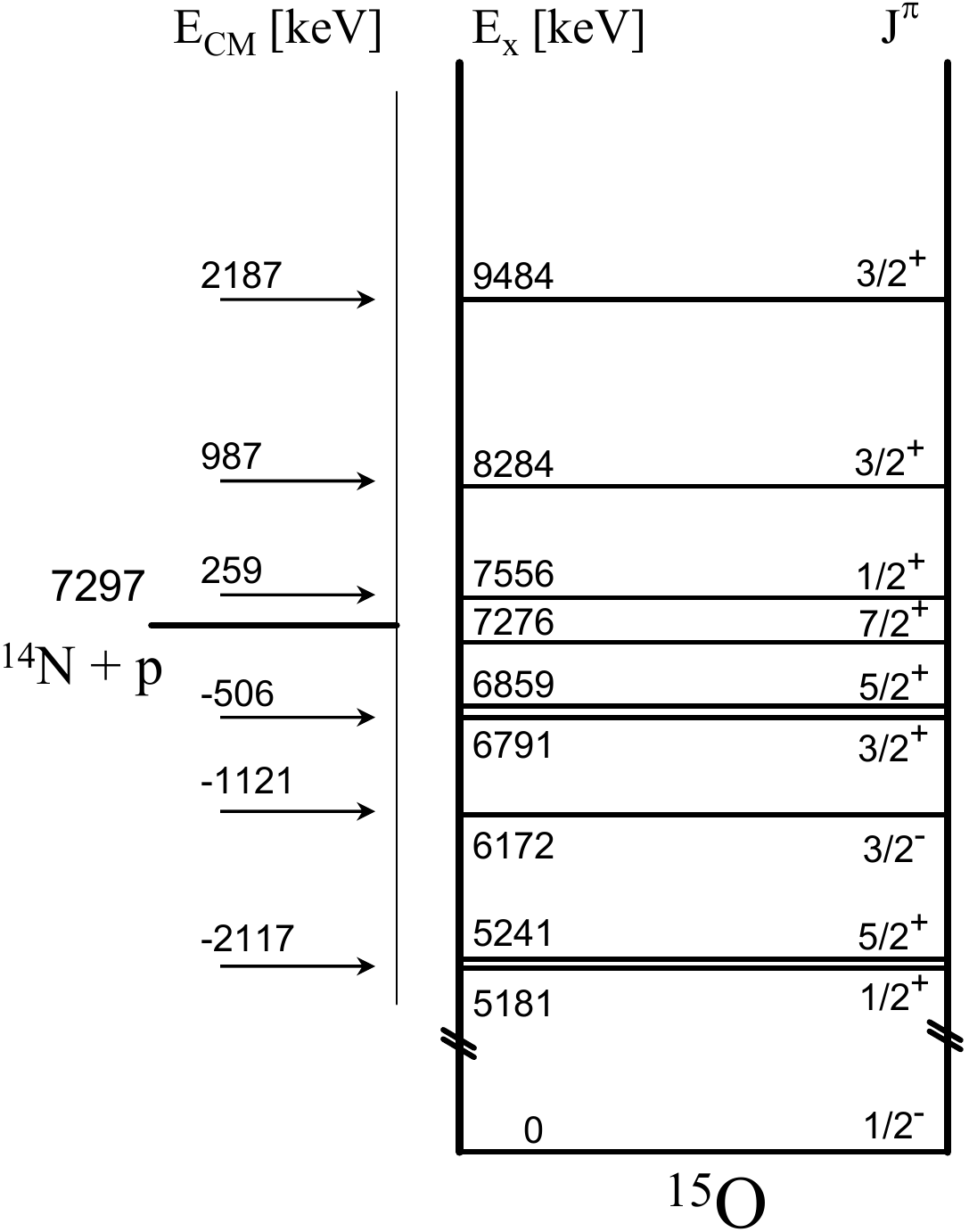}
\caption{The energy levels of $^{15}$O and their relationship to the threshold energy for
$^{14}$N(p,$\gamma$).} 
\label{fig:14Nlevel}
\end{center}
\end{figure}

Figure \ref{fig:14Nlevel} shows the level structure of $^{15}$O, relative to the threshold
energy for $^{14}$N(p,$\gamma$).

Solar Fusion I  gave $3.5^{+0.4}_{-1.6}$ keV b as the recommended
total S-factor for the $^{14}$N(p,$\gamma$)$^{15}$O reaction. This
was based on the energy dependence determined by \citet{Schr87}.
In the \citet{Schr87} analysis the ground state transition accounted for half of the total S-factor
at zero energy, primarily because of the contribution of a subthreshold resonance at
$E$= $-$504 keV (corresponding to the 6.79 MeV state in $^{15}$O). However, 
a reanalysis based on an
R-matrix calculation by \citet{Angulo01} indicated that the strength of the ground state
transition in \citet{Schr87}, S$_{1\,14}^\mathrm{gs}$(0)=1.55 keV b, had been significantly overestimated,
and should be reduced  to 0.08 keV b. 

This finding prompted a series of new experiments using direct \cite{Bemm06,
Formi04,Imb05,Lemut06,Marta08,Runkle05}
and indirect approaches \cite{Bertone01,Bertone02,Muk03,Nel03,
Sch08,Yamada04}. The prompt-capture
$\gamma$-radiation was measured in experiments by the TUNL group \cite{Runkle05}
in a surface laboratory and by the LUNA group \cite{Formi04,Imb05,Marta08}
in Gran Sasso.  From these experiments -- carried out with
Ge detectors -- the contributions of each transition could be extracted. 
In an
additional measurement by the LUNA Collaboration \cite{Bemm06,Lemut06}
the total cross section was determined. These recent experiments cover an energy range
from 70 to 480 keV, still far from the solar Gamow window at $E_0$ = 27 keV. Additional
information is provided by experiments that probe the width of the subthreshold state at $E$ =
$-$506 keV by the Doppler shift attenuation method \cite{Bertone01,Sch08}
and by Coulomb excitation \cite{Yamada04}. 
Asymptotic
normalization coefficients  (ANC) for the ground state
and selected excited states were determined from
transfer reaction measurements for $^{14}$N($^3$He,d)$^{15}$O by \citet{Bertone02} and
\citet{Muk03}.  All experiments and subsequent analyses confirmed that the
value for the ground-state contribution determined in the extrapolations
of \citet{Schr87} had been too high.
Current estimates of S$_{1\,14}^\mathrm{gs}(0)$ range from 0.08 keV b \cite{Angulo01} to
0.45 keV b \cite{Runkle05}. Hence, the S-factor for $^{14}$N(p,$\gamma$)$^{15}$O is now
determined largely by the transition to the 6.79 MeV state. 
Minor contributions arise from
transitions to the 5.18, 5.24, 6.17, 6.86 and 7.28 MeV states in $^{15}$O.

\subsubsection{R-matrix analysis and normalization}
\label{sec:renorm}
We have performed an R-matrix fit to the three strongest transitions using the
data of \citet{Imb05}, \citet{Marta08}, \citet{Runkle05}, and \citet{Schr87} and the
code of Descouvemont \cite{descouvemont-2010}.
In this way we obtain the most robust weighted mean. The recent direct experiments
\cite{Bemm06,Formi04,Imb05,Lemut06,Marta08,Runkle05} 
cover only a relatively narrow energy window. Thus, as no new information is available
for the higher lying resonances, a reliable extrapolation
to zero energy requires the high-energy data of \citet{Schr87}.
However, systematic differences are apparent in the data 
sets of \citet{Imb05}, \citet{Runkle05}, and
\citet{Schr87}.
In order to minimize systematic uncertainties,
all data sets were renormalized to the weighted mean of the strength of the
259 keV resonance in $^{14}$N(p,$\gamma$)$^{15}$O. Table \ref{tab:N1141} summarizes the available 
absolute determinations
of the resonance strength with a weighted mean of $\omega \gamma_{259}$ = 13.1 $\pm$ 0.6 meV. 
The uncertainty was obtained by calculating the error on the weighted mean,
excluding the common systematic uncertainty on the stopping power of protons in nitrogen \cite{Zieg08}.
The latter was summed in quadrature with the weighted mean error to obtain the
final uncertainty.

\begin{table}[here!]
\caption{Summary of the published values for $\omega \gamma_{259}$, along with their estimated
statistical, systematic, and total uncertainties. All quantities are in units of meV.  The 
last row gives the recommended value.}
\label{tab:N1141}
\begin{tabular}{|l c c c c|}
\hline
\hline
			& $\omega \gamma_{259}$       &   ~stat.~  &  ~syst.~ & ~total~ \\
		
\hline
\textcite{Becker82}\footnotemark[1]   	&  14 			     & 		      & 		    & 1.0    \\
\textcite{Runkle05}           &  13.5 		     & 		      & 	1.2	    & 1.2    \\
 \textcite{Imb05}        &  12.9		   	     &   0.4	      & 	0.8	    & 0.9    \\
\textcite{Bemm06}         &  12.8		   	     &   0.3	      & 	0.5	    & 0.6    \\
\hline
recommended value	 &  13.1		     &   	      & 		    & 0.6    \\
\hline
\hline
\end{tabular}
\footnotetext[1]{used in \textcite{Schr87}}
\end{table}

In \citet{Schr87} the data were normalized to an absolute cross section determination at $E$ = 760 keV,
$\sigma (E =760~\mathrm{keV})$ = 620 $\pm$ 80 nb.  
This value is an adopted mean
based on several experimental methods, while the measurement relative to $\omega \gamma_{259}$
gives $\sigma(E=760~\mathrm{keV})$ = 609 nb \cite{Schr87}.   Thus, based on the difference between the
value for $\omega \gamma_{259}$ used by \citet{Schr87}, 14 meV \cite{Becker82},
and the new determination, 13.1 $\pm$ 0.6 meV, a
precise renormalization of $\sigma (E =760~\mathrm{keV})$ can be made,
relative to this resonance.   One finds $\sigma (E =760~\mathrm{keV})$ = 570 nb.  Moreover, we note
that the energy dependence of \citet{Schr87} was corrected for
summing contributions, as discussed by \citet{Imb05}. 
The renormalizations for \citet{Runkle05}
and \citet{Imb05} are 3\% and 2\%, respectively.

The ANCs for the ground, 6.79 MeV, and 6.17 MeV states as well as $\Gamma_\gamma$ of the 6.79
MeV state are important parameters in the R-matrix analysis determining S(0).  Parameter values determined
in the analysis will reflect the quality of the input data.
Thus the R-matrix results can be validated by
comparing these values with those determined independently by transfer reactions and
other indirect measurements (see Table \ref{tab:N1142}).

\begin{table*}
\caption{Published ANC values and $\Gamma_{\gamma}$ for the 6.79 MeV transition. All ANC values are given in the coupling scheme of \textcite{Angulo01}. The recommended values in the last row were obtained as a weighted mean considering as weights the experimental errors only.  Finally, the recommended uncertainty was
obtained by summing in quadrature the weighted mean error and an average theoretical
uncertainty.  The latter is according to information provided by the authors.  As existing measurements of
$\Gamma_{\gamma}$(6.79 MeV) are discrepant, no recommended value is given.}
\label{tab:N1142}
\begin{tabular}{|l c c c c c|}
\hline
\hline
			&~C$_{\mathrm{gs}^{3/2}}$~ (fm$^{-1/2}$)\footnotemark[1]       &~ C$_{6.79}$~(fm$^{-1/2}$)&~C$_{6.17^{1/2}}$~(fm$^{-1/2}$)\footnotemark[2] &~ C$_{6.17^{3/2}}$~ (fm$^{-1/2}$)\footnotemark[1]  & $~\Gamma_{\gamma}$(6.79)~(eV) \\
\hline
\textcite{Muk03}   &  $7.4\pm0.4$ & $4.9\pm0.5$	&  $0.47\pm0.03$  &  $0.53\pm0.03$	&     \\
\textcite{Bertone02}     &   $7.9\pm0.9$ & $4.6\pm0.5$	&  $0.45\pm0.05$  &  $0.51\pm0.06$	&	  \\
 \textcite{Bertone01}    &		&		&	&		& 0.41$^{+0.34}_{-0.13}$\footnotemark[3]         \\
\textcite{Yamada04}           &		&		&	&		& 0.95$^{+0.6}_{-0.95}$       \\
\textcite{Sch08}        &		&		&	&		& $>0.85$       \\
\hline
recommended value	&  $7.4\pm0.5$  & $4.8\pm0.5$   & $0.47\pm0.03$    &  $0.53\pm0.04$     &                     \\
\hline
\hline
\end{tabular}
\footnotetext[1]{channel spin ${\rm I}=3/2$}
\footnotetext[2]{channel spin ${\rm I}=1/2$}
\footnotetext[3]{the quoted uncertainty represents a 90\% confidence limit}
\end{table*}

\subsubsection{Transition to the ground state and 6.79 MeV in $^{15}$O}
\label{sec:gs}
The transitions to the ground and  6.79 MeV states in $^{15}$O are connected
through the reduced proton width of the $-$0.506 MeV subthreshold state.  This width can also
be expressed in terms of the subthreshold state ANC via the 
Whittaker function at the R-matrix radius $a$ that appears in Eq.~(3.60) of \citet{descouvemont-2010} (see
references therein).
Both transitions are discussed together here.

\begin{figure}
\begin{center}
\includegraphics[width=8.5cm]{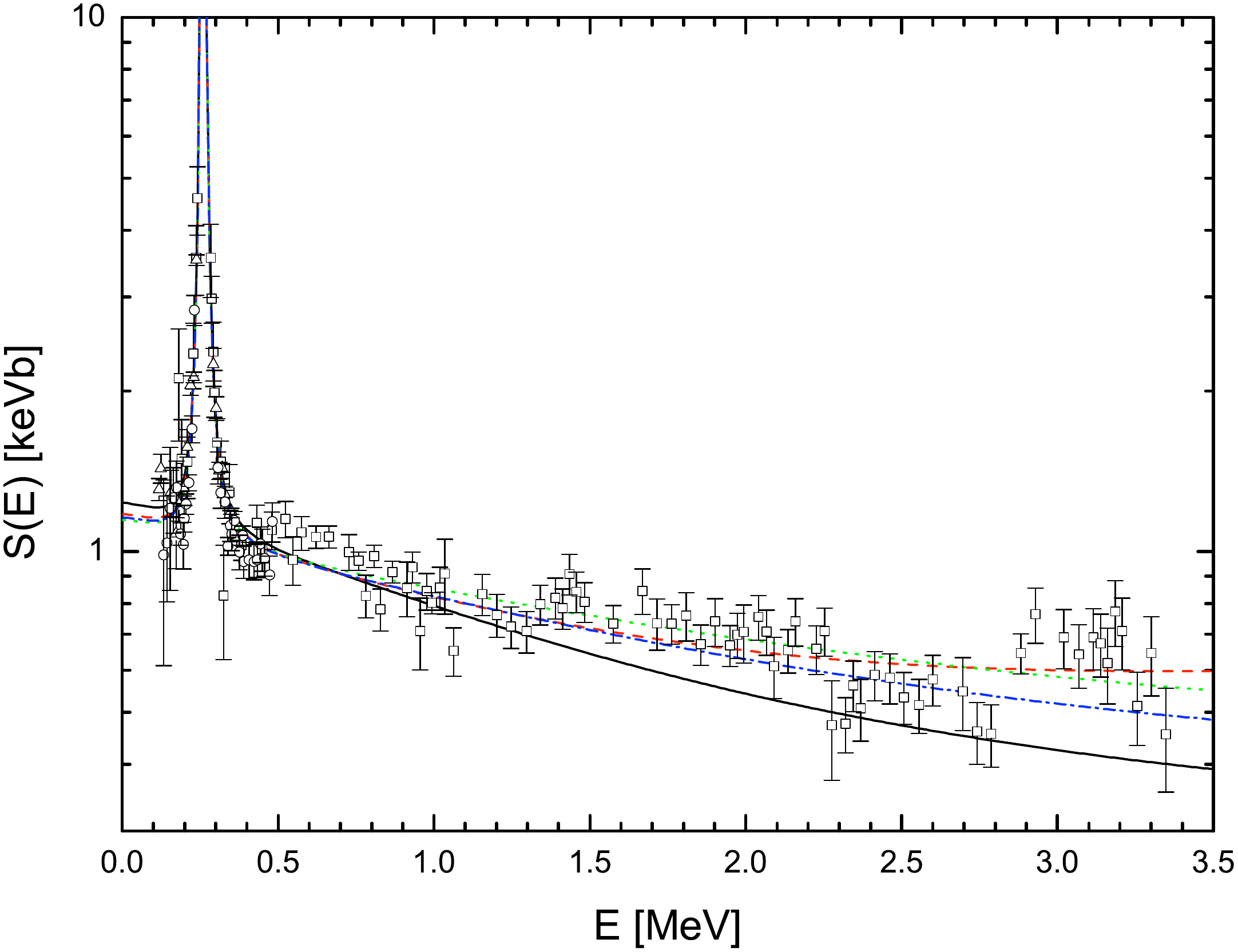}
\caption{(Color online) R-matrix fits to the 
$^{14}$N(p,$\gamma$)$^{15}$O 6.79~MeV transition together with the data of \textcite{Schr87} (open squares), \textcite{Imb05} (open triangles), and \textcite{Runkle05} (open circles). 
The cases i, ii, and iii (see text) are represented by the dotted green, dash-dotted blue, and dashed red lines, respectively.  The black line is a calculation similar to iii), but without the unidentified J$^\pi$ = 5/2$^-$ pole 
at $E$=6 MeV, comparable to fits in past work.} 
\label{fig:N1142}
\end{center}
\end{figure}

{\it Transition to the 6.79 MeV state:}~
The reaction mechanism for the transition to the 6.79 MeV state appears rather simple,
primarily an external capture process whose magnitude is determined by the value of the ANC.
Hence S$_{1\,14}^{6.79}(0)$ is dominated by the external capture process. In the
present analysis the data of \citet{Runkle05}, \citet{Imb05}, and \citet{Schr87}
are included after renormalization, as described above. As the recent low-energy data
do not strongly constrain the R-matrix radius, high-energy data are needed. The
resulting S$_{1\,14}^{6.79}(E)$ fails to reproduce the high-energy data for radii
5.5 fm $<$ $a$ $<$ 6.5 fm, as in
Fig. 4 of \citet{Angulo01}. A better fit can be obtained by choosing smaller radii.  However, this
choice also impacts fits for the ground state transition, which
favor larger radii.  Consequently, we have not used the transition to
the 6.79 MeV state to determine the R-matrix radius in this way.  Instead, 
R-matrix fits were done 
\begin{enumerate}
\renewcommand{\labelenumi}{\roman{enumi})}
\item taking all renormalized data \cite{Imb05,Runkle05,Schr87} into account;
\item limiting the data sets to $E$  $<$ 1.2 MeV;  and
\item same as i), but introducing an unidentified J$^\pi$ = 5/2$^-$ pole at $E$ = 6 MeV.  
\end{enumerate}
In each case the ANC values and the radii were determined.   The results for the three
cases are
\begin{enumerate}
\renewcommand{\labelenumi}{\roman{enumi})}
\item C$_{6.79}$ = 4.61 $\pm$ 0.02 fm$^{-1/2}$ for
$a$ = 4.14 fm and S$_{1\,14}^{6.79}(0)$=1.11 keV b. This solution has the lowest $\chi^2$ but was rejected 
for the reasons given above.
\item C$_{6.79}$ = 4.65 $\pm$ 0.02 fm$^{1/2}$ for $a$ = 4.6 fm and S$_{1\,14}^{6.79}(0)$ = 1.15 keV b.
\item C$_{6.79}$ = 4.69 $\pm$ 0.02 fm$^{1/2}$ for $a$ = 5.4 fm and S$_{1\,14}^{6.79}(0)$ = 1.18 keV b. 
\end{enumerate}
The latter two fits are
in very good agreement with \citet{Runkle05} and about 5\% lower than \citet{Imb05}.
All three fits are shown in Fig. \ref{fig:N1142}.

In summary, the dominant systematic uncertainty for S$_{1\,14}^{6.79}$(0) arises from the interpretation of the 
high-energy data. This uncertainty is estimated from cases i) to iii) to be about 4\%. One could
speculate that the deviation of the higher energy data from the R-matrix fit is due to broad
unidentified structures in this transition (Fig. \ref{fig:N1142}). We recommend S$_{1\,14}^{6.79}$(0) = 1.18 $\pm$ 0.05
keV b. The error includes both systematic and statistical uncertainties, though the former
are much larger.

The weighted mean of the ANC for the 6.79 MeV state from indirect measurements, 
C$_{6.79}$ = 4.8 $\pm$ 0.5 fm$^{-1/2}$  (Table \ref{tab:N1142}), is in excellent agreement
with the results of the R-matrix analysis.

{\it Ground state transition:}~
Three data sets \cite{Imb05,Runkle05,Schr87}, normalized to $\omega \gamma_{259}$ as
discussed above, were used in the ground-state analysis.
The results from \citet{Marta08} -- three data points with high precision above the 259
keV resonance and essentially free from summing effects -- are relative to the yield of the
transition to the 6.79 MeV state. These data were normalized to the weighted mean
of the renormalized S-factor (see Sec. \ref{sec:renorm}) from
\citet{Schr87}, \citet{Runkle05}, and \citet{Imb05} in the energy region 311 keV $<~E~<$ 360 keV.

\begin{figure}
\begin{center}
\includegraphics[width=8.5cm]{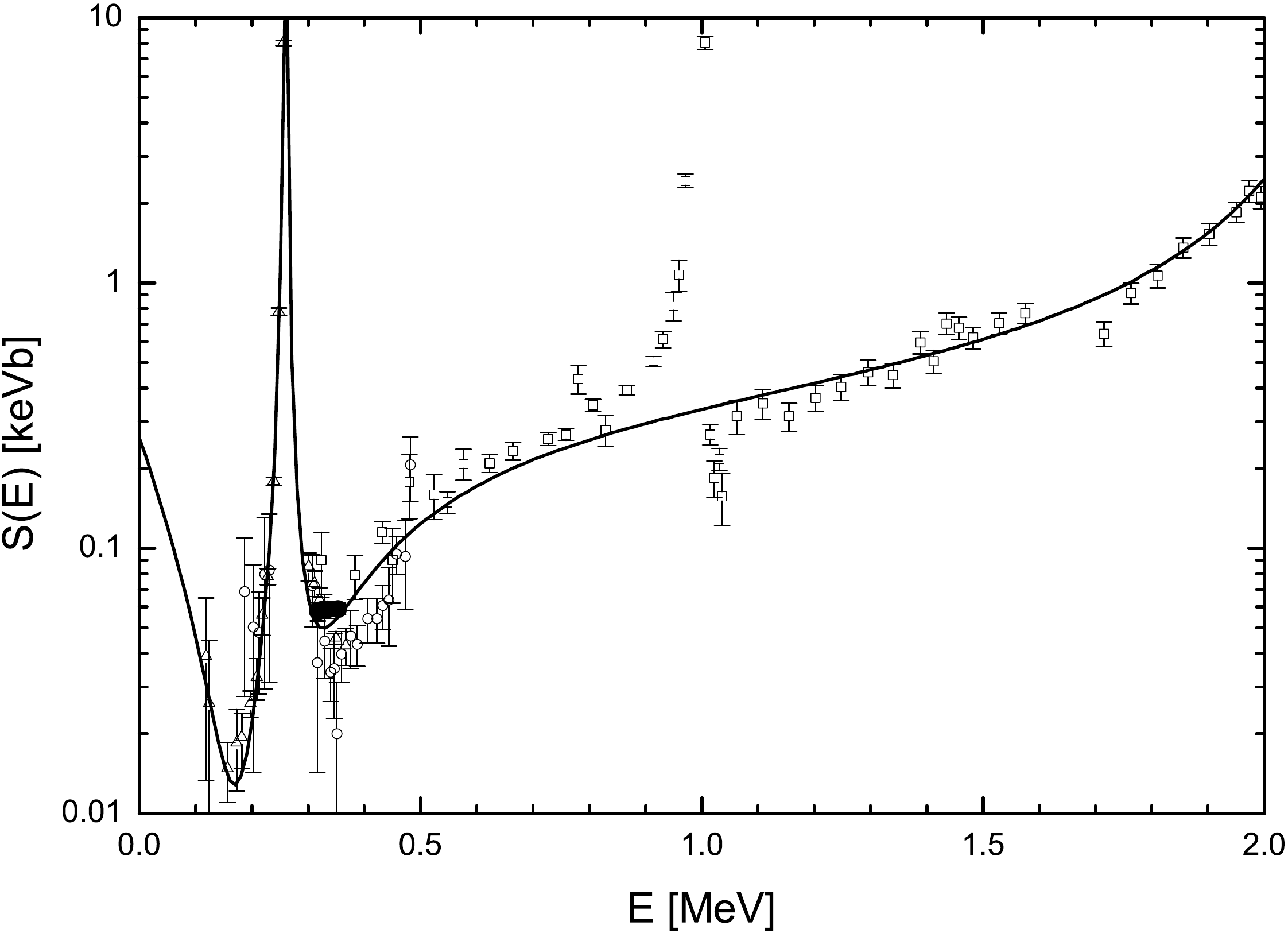}
\caption{R-matrix fit to the $^{14}$N(p,$\gamma$)$^{15}$O
ground state transition.  The filled circles are from \citet{Marta08}.  All
other data are labeled as in Fig. \protect\ref{fig:N1142}.}
\label{fig:N1141}
\end{center}
\end{figure}

The R-matrix fit was based on the same poles as in \citet{Angulo01}
with starting parameters as given in \citet{Ajz91}. The sensitivity to radius was
tested for a broad range of ANC values, 6 fm$^{-1/2}$  $<$ C$_{\mathrm{gs}^{3/2}}$ $<$ 9 fm$^{-1/2}$. 
The minimum $\chi^2$ was obtained for $a$ = 5.6 $\pm$ 0.1 fm.  
Thus, we selected
$a$ = 5.5 fm as an appropriate average for the ground and 6.79 MeV states, employing
this value for all subsequent R-matrix fits. This
value was used previously in \citet{Runkle05}, \citet{Imb05}, \citet{Marta08}, and \citet{Muk03}.
The reduced width for the subthreshold state
was fixed through C$_{6.79}$ (see above) to $\gamma^2$=0.37 MeV. The narrow 
resonances at 0.987 MeV ($\Gamma_p$ = 3.6 keV, see Fig. \ref{fig:N1141}) and 2.191 MeV
 (J$^\pi$ = 5/2$^-$, $\Gamma_p$ = 10 keV) are not
relevant for S$_{1\,14}^\mathrm{gs}(0)$ and thus were excluded from the fit. 
In order to optimize the fit off-resonance, contributions to $\chi^2$ from points near the 
2.191 MeV and 0.259 MeV ($\Gamma_p \sim$ 1 keV) resonances were omitted.
As slopes are steep
and counting rates peak near the resonances, the inclusion of near-resonance data forces the
fit in arbitrary ways.  The region excluded depends on resonance width
and on target thickness, which can spread the effects of a resonance over
a larger energy interval.  We omitted data in the interval between $E_R- 20\Gamma$
and $E_R + 1.5 \Delta$,
where $\Delta$ is the target thickness.  Target thickness effects are especially prominent in the 
data of \citet{Schr87}, representing the integral over the target thickness of $\sim$ 30 keV.

In the fit the $\chi^2$ decreases with increasing ANC, reaching a
minimum at C$_{\mathrm{gs}^{3/2}}$ $\sim$ 11 fm$^{-1/2}$, a value outside the ranges
determined by \citet{Muk03} and
\citet{Bertone02}. At the  9 fm$^{-1/2}$ upper bound for C$_{\mathrm{gs}^{3/2}}$, we
obtain S$_{1\,14}^\mathrm{gs}(0)$ = 0.29 keV, while at  the 6 fm$^{-1/2}$ lower bound, S$_{1\,14}^\mathrm{gs}(0)$= 0.24 keV b. 
These fits do not include the possibility of a small contribution from C$_{\mathrm{gs}^{1/2}}$, 
interfering with the 259 keV resonance.  We expand the uncertainty to account for
such a possibility, recommending S$_{1\,14}^\mathrm{gs}(0)$ = 0.27 $\pm$ 0.05 keV b 
with $\Gamma_\gamma$(int) =
1.1 eV. The latter value is the internal part of the $-$0.504 MeV subthreshold state radiative 
width (at E = 0), a fit parameter in the R-matrix calculation. The total radiative 
width, which can be compared to experimental values obtained from, e.g., lifetime measurements, is 
derived following the approach of \citet{Holt78} and \citet{BarkerKajino91}, giving
$\Gamma_\gamma(6.79)=|\Gamma_\gamma(\mathrm{int})^{1/2}\pm\Gamma_\gamma(\mathrm{ch})^{1/2}|^2$,
where the relative sign of the two amplitudes is unknown. The channel (external) radiative width  
$\Gamma_{\gamma}(\mathrm{ch})= 0.57$ eV can be directly calculated from the adopted 
value of C$_{\mathrm{gs}^{3/2}}$. If the minus sign is chosen in the relationship
for $\Gamma_\gamma(6.79)$, one obtains a
lifetime in excess of 4 fs,  in disagreement with \citet{Bertone01} and \citet{Sch08}. If the
plus sign is chosen, a lifetime shorter than 0.2 fs is obtained.   Such a lifetime
is presently beyond the reach of Doppler shift lifetime 
measurements, but still in agreement with \citet{Sch08}. However, the Coulomb excitation work 
of \citet{Yamada04} 
gives a lower limit of 0.4 fs, apparently ruling out such a short lifetime. We conclude
that the current experimental
situation is unsatisfactory and calls for further work.
Lifetimes larger than 0.4 fs require C$_{\mathrm{gs}^{3/2}}<6$ fm$^{-1/2}$, again in 
disagreement with \citet{Bertone02} and \citet{Muk03}. The somewhat larger range in 
C$_{\mathrm{gs}^{3/2}}$ used in the present analysis, compared to the uncertainty recommended 
in Table \ref{tab:N1142}, takes account of this dilemma. Most recent 
treatments of  $^{14}$N(p,$\gamma$)$^{15}$O 
direct measurements have failed to address issues connected with the total radiative width.
 
\begin{table*}
\caption{S$_{1\,14}(0)$ and the fractional uncertainty $\Delta$S$_{1\,14}(0)$ for the different transitions. Note that tr(5.24)$\to$0 includes contributions from the transition tr$\to$6.86$\to$5.24 and tr$\to$7.28$\to$5.24 with S$_{1\,14}(0)=0.037\pm0.011$ and $0.019\pm0.006$ keV~b, respectively (from \textcite{Schr87} with a 30\% uncertainty). The contribution of tr(7.28)$\to$0 observed by \textcite{Schr87} is negligible.}
\label{tab:N1143}
\begin{tabular}{|l c c c|}
\hline
\hline
transition	& ~~S$_{1\,14}(0)$~(keV~b)~~       &  ~~$\Delta$S$_{1\,14}(0)$~~     & ~~reference~~ \\		
\hline
tr$\to$0  			& $0.27\pm0.05$ &    19\% 	&   present \\
tr$\to$6.79     & $1.18\pm0.05$ &  4\%    &present \\
tr$\to$6.17     & $0.13\pm0.06$  & 38\%    &present \\
tr$\to$5.18     & $0.010\pm0.003$   &  30\%  & \textcite{Imb05} \\	
tr(5.24)$\to$0\footnotemark[1]		& $0.070\pm0.021$   &  30\%  & \textcite{Imb05}\\
%tr$\to$6.86		& $0.037\pm0.011$   &  30\%  & \textcite{Schroeder}\\
%tr$\to$7.28		& $0.019\pm0.006$   &  30\%  & \textcite{Schroeder}\\
\hline
R-Matrix sum    & $1.66\pm0.08$\footnotemark[2]    &  5\% &  \\
additional systematic uncertainty\footnotemark[3] & &5\% & \\
 total & $1.66\pm0.12$    &  7\% &  \\
 \hline
 \hline
\end{tabular}
\footnotetext[1]{value from the analysis of the secondary transition}
\footnotetext[2]{uncertainty from the R-matrix analysis only}
\footnotetext[3]{from normalization to $\omega \gamma_{259}$} 
\end{table*}

\subsubsection{Transition to the 6.17 MeV state}
This transition was analyzed with the poles given by \citet{Angulo01}
except that we also allowed for an external capture contribution (channel spin I = 3/2), improving the fit substantially.
The primary uncertainty in predicting S$_{1\,14}^{6.17}(0)$ arises from the choice
of the poles, i.e., more poles at higher energies and their interference pattern, respectively,
could be included in the fit.  However, a full study of all possible minor contributions
is far beyond the scope of the present work and would be hampered by the lack of precise data.
The best fit yields
S$_{1\,14}^{6.17}(0)$ = 0.13 keV b with C$_{6.17^{1/2}}$ = 0.43 $\pm$ 0.02 fm$^{-1/2}$ and 
C$_{6.17^{3/2}}$ = 0.49 $\pm$ 0.02 fm$^{-1/2}$.
These ANCs are in good agreement with those deduced by \citet{Muk03} and \citet{Bertone02}
(see Table \ref{tab:N1142}). Previous results without the contribution from channel spin 3/2 external capture
led to S$_{1\,14}^{6.17}(0)$ = 0.08 keV b \cite{Imb05} and 0.04 keV b \cite{Runkle05}. Thus, we
have adopted S$_{1\,14}^{6.17}(0)$ = 0.13 $\pm$ 0.06 keV b where the error reflects the uncertainty in the R-
matrix input as well as the spread of this value in the literature \cite{Angulo01,
Imb05,Nel03,Runkle05}. In \citet{Nel03} a
$M1$ contribution was inferred from an analyzing power experiment. The fit only extends
to $E$ $\sim$ 327 keV and trends above the data for higher energies. \citet{Runkle05}
showed that there is no significant difference in S$_{1\,14}^{6.17}(0)$ results from including
the $M1$ contribution specified by \citet{Nel03}.

\subsubsection{Total S$_{1\,14}(0)$ and conclusions}
We have obtained S$_{1\,14}^\mathrm{tot}(0)$ from the data sets of \citet{Imb05}, \citet{Marta08}, and \citet{Schr87},
normalized to the 259 keV resonance, and supported by an R-matrix analysis that defines
the extrapolation to astrophysical energies.  The R-matrix analysis focused on the systematic 
uncertainties associated with fitting and extrapolating the data, and made use
of indirect measurements \cite{Bertone01,Bertone02,Muk03,
Sch08,Yamada04} to constrain parameters in the fitting.   Systematic uncertainties
in this analysis dominate the errors:  statistical uncertainties have minor consequences
for the resulting S$_{1\,14}^\mathrm{tot}(0)$.  The
R-matrix radius $a$ is a key parameter, fixed in the present analysis to the best-choice
value of 5.5 fm (Sec. \ref{sec:gs}). The extrapolation for the strongest transition to the 6.79
MeV state is robust within 4\%, while the extrapolations for transitions to the
ground and 6.17 MeV states are less constrained.  The transitions to the 5.18, 5.24, 6.86, and 7.28
MeV states combine to contribute 0.08 keV b to S$_{1\,14}^\mathrm{tot}(0)$, $\sim$ 5\% of the total.  These contributions
were obtained
from literature \cite{Imb05,Schr87}, scaled to the weighted mean of $\omega \gamma_{259}$.
The errors on the individual transitions were enlarged to a more realistic uncertainty
of 30\%. Note that some of the weak transitions often have been neglected in past work. Finally, an additional
systematic error of 5\% due to the normalization of $\omega \gamma_{259}$
 (see Table \ref{tab:N1141}) is included.  Table \ref{tab:N1143} summarizes the various contributions.

\begin{figure}
\begin{center}
\includegraphics[width=8.5cm]{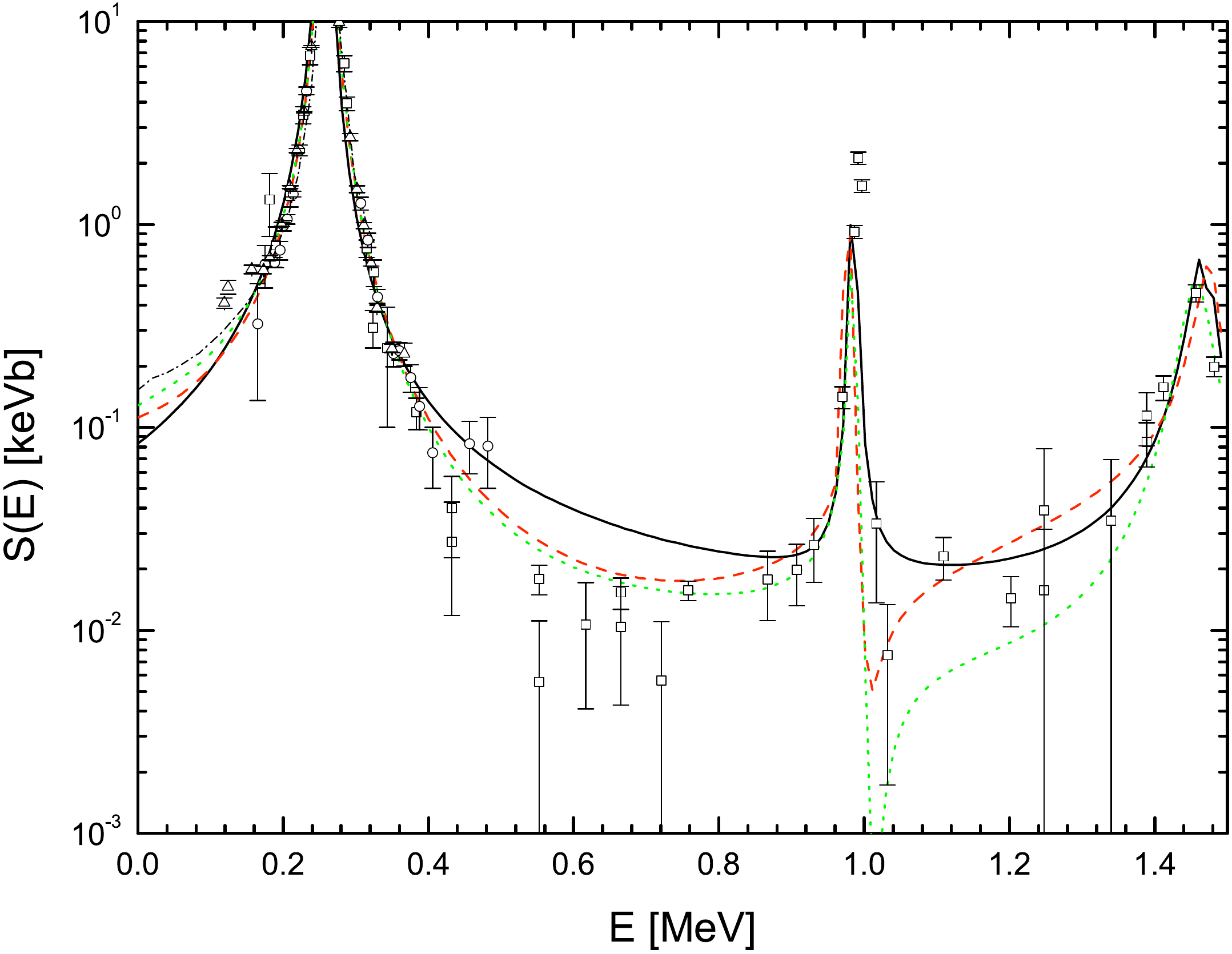}
\caption{(Color online) R-matrix fit to the 
$^{14}$N(p,$\gamma$)$^{15}$O 6.17 MeV transition. The symbols are the same 
as in Fig. \protect\ref{fig:N1142}. The dotted green line corresponds to the present analysis. The solid black, 
dashed red, and dash-dotted black lines are the R-matrix fits of \textcite{Imb05}, 
\textcite{Runkle05}, and \textcite{Nel03}, respectively.}
\label{fig:N1143}
\end{center}
\end{figure}

\begin{figure}
\begin{center}
\includegraphics[width=8.5cm]{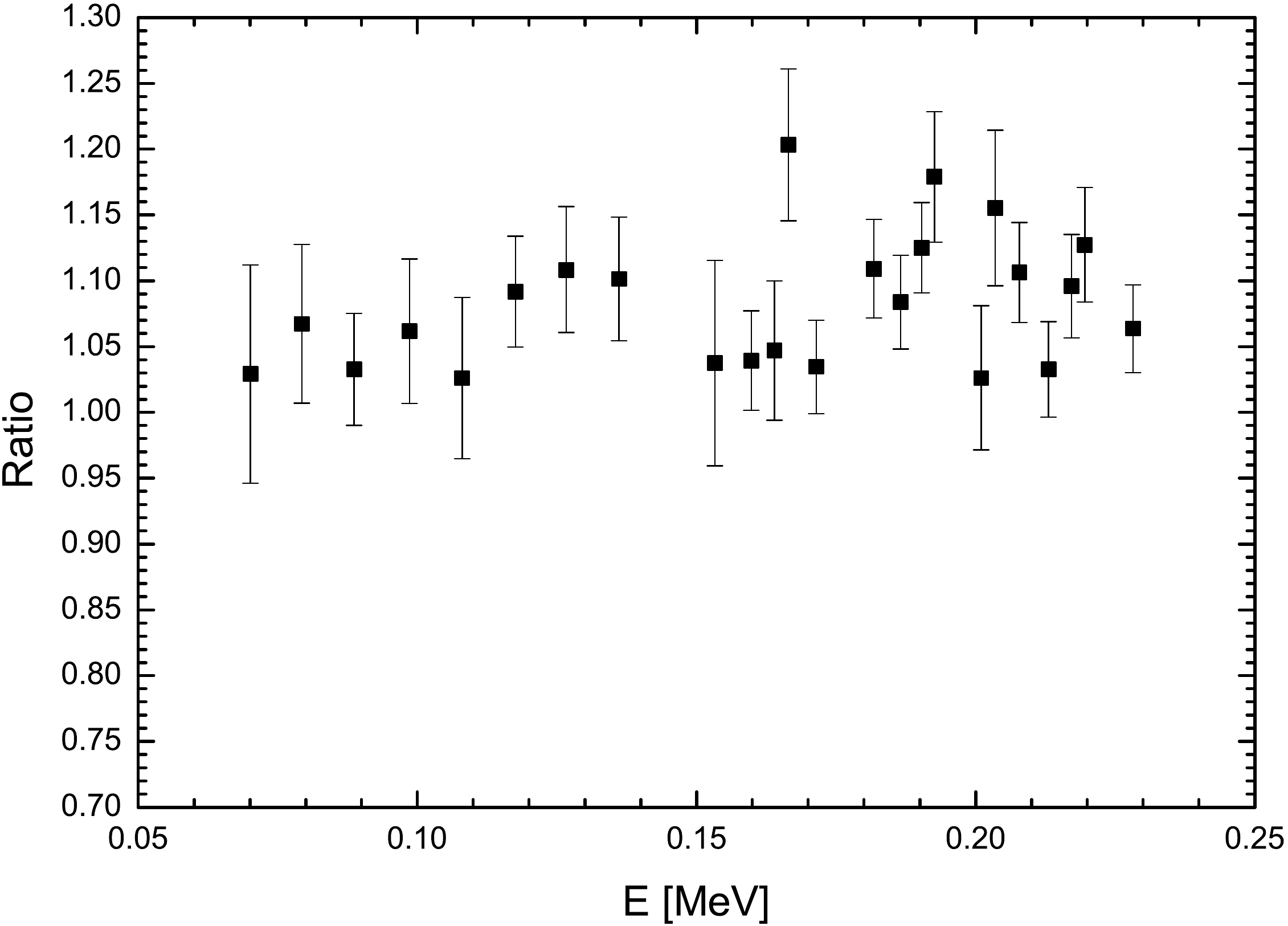}
\caption{Comparison of the S$_{1\,14}^\mathrm{tot}$ obtained from the present R-matrix fit and gas target data. Note that the gas target data are corrected for electron screening (see Table 2 in \citet{Bemm06})
according to calculations of \citet{Ass87}.}
\label{fig:N1144}
\end{center}
\end{figure}

We find, after summing all contributions, S$_{1\,14}^\mathrm{tot}(0)$ = 1.66 $\pm$ 0.12 keV b.
The S-factor fits derived in the present study are shown in Figs. \ref{fig:N1142}, \ref{fig:N1141}, and
\ref{fig:N1143}
together with the renormalized data of \citet{Imb05}, \citet{Marta08}, \citet{Runkle05}, and
\citet{Schr87}. Figure \ref{fig:N1144} compares our results for the  total S$_{1\,14}^\mathrm{tot}(E)$ with the data from
\citet{Lemut06} and \citet{Bemm06}.  Below $E$ = 108 keV
the gas-target results and the R-matrix fit are not inconsistent, given 
uncertainties;  at higher
energies,  $E$ $\sim$ 200 keV, the average deviation is $\sim$ 8\%.
These data are an
absolute determination of the S-factor and thus do not depend on the normalization 
of $\omega \gamma_{259}$.

S$_{1\,14}^\mathrm{tot}(E)$ below $E ~\sim$ 130 keV can be approximated to better than 1\% by a second
order polynomial
\begin{eqnarray}
\mathrm{S}_{1\,14}^\mathrm{tot}(0) &=& 1.66~\mathrm{keV~b} \nonumber \\
\mathrm{S}_{1\,14}^{\mathrm{tot}\,\prime}(0) &=& -0.0033~\mathrm{b} \nonumber \\
\mathrm{S}_{1\,14}^{\mathrm{tot}\,\prime \prime}(0) &=& 4.4 \times 10^{-5}~\mathrm{b/keV}.
\end{eqnarray}
The absolute scale of this energy dependence has an uncertainty of $\pm$ 7\%. Recently, a coupled
channel analysis of the data for $^{14}$N(p,$\gamma$)$^{15}$O has been reported 
\cite{Grin08} which gives S$_{1\,14}^\mathrm{tot}(0)$ = 1.68 keV b, in excellent agreement with the results
presented here.

Further work on $^{14}$N(p,$\gamma$)$^{15}$O is needed.
A better understanding of the reaction mechanism governing the transition
to the 6.79 MeV state at high energies would help reduce
systematic uncertainties.  Moreover, additional experimental and
theoretical work on the transition to the 6.17 MeV state is needed, as the existing database
is lacking. A new determination of $\Gamma_\gamma$ for the 6.79 MeV state with an alternative method
would be desirable to constrain the R-matrix fit and to resolve slight discrepancies in existing
data. Elastic scattering experiments could give an additional constraint.  Finally,
a high-precision measurement of $\omega \gamma_{259}$ with significant improvements in
the accuracy of stopping power data would reduce the systematic uncertainty in the normalization.\footnote{Note added in proof: A new R-matrix analysis of 
$^{14}$N(p,$\gamma$)$^{15}$O reaction appeared \cite{Azuma2010} after 
submission of the present work. This analysis, which 
served as a validity test for the AZURE code, yielded S$_{1\,14}^\mathrm{tot}(0)$ = 1.81 keV b,
9\% larger than the central value recommended here.
No uncertainty was provided.  The differences between \citet{Azuma2010} and Solar Fusion II
are connected with
the 6.79 MeV transition. In the present work
(i) a normalization procedure is employed to address needed corrections in the high-energy 
data and (ii) a background pole is introduced to achieve a better representation of that data.
Without such adjustments, the procedure of \citet{Azuma2010}  
produces a fit that underestimates the 
high-energy data and consequently yields a larger S$_{1\,14}^{6.79}(0)$.
Nevertheless, the present and \citet{Azuma2010} results are consistent
if one assigns a reasonable uncertainty to the latter.}

\subsection{Other CNO-cycle reactions}
\label{sec:N114other}
While the $^{14}$N(p,$\gamma)^{15}$O reaction controls the cycling rate and the 
energy production by CN reactions at solar temperatures, other reactions in the cycle 
determine the extent to which the reaction flow moves out of the CN cycle toward heavier metals, 
oxygen in particular.  These trends in turn affect the opacity evolution and temperature 
profiles as a function of solar age.  There has been significant recent progress in determining the rates of 
many of these other reactions.  The reader is referred to Solar Fusion I for summaries of other 
reactions for which there has not been new work reported since 1998.  More recent reviews 
have been given by \citet{Angulo:1999zz} (the ``NACRE'' compilation) and by \citet{Wiescher10}.

% DB insert begins
\subsubsection{$^{12}$C$(\mathrm{p},\gamma)^{13}$N}
In the starting phase of the CN cycle, before it has reached its equilibrium, this reaction controls the buildup of $^{14}$N \cite{hax08}. A recent study using the ANC method by \citet{Burt08} yields a reaction rate consistent with that of  \citet{Angulo:1999zz}, the rate recommended here.
% DB insert ends. 

\subsubsection{$^{15}$N($\mathrm{p},\alpha$)$^{12}$C}
\label{sec:XIB2}
\begin{figure}
\includegraphics[width=8.6cm]{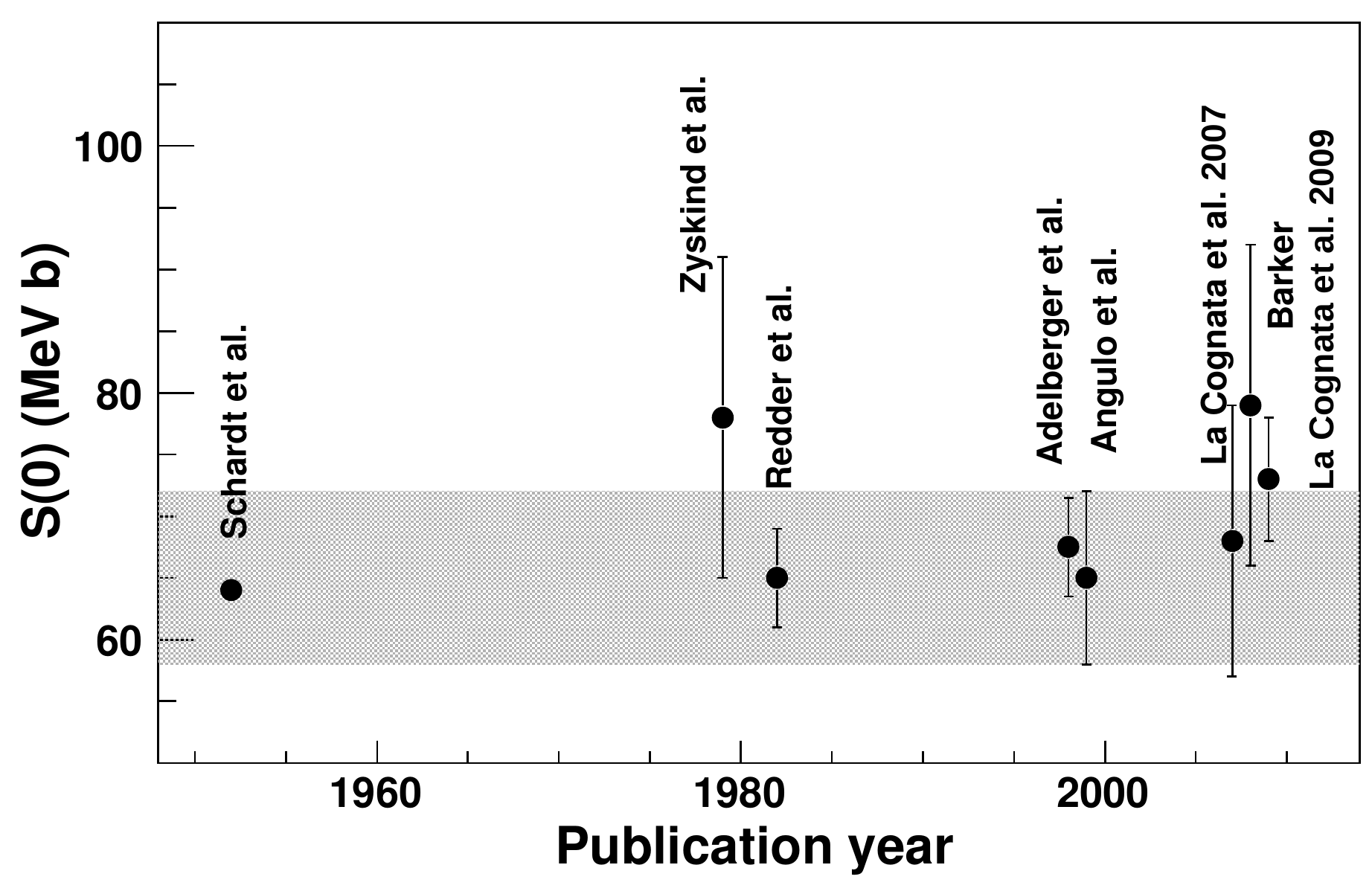}
\caption{Summary of the available measurements of S$_{1\,15}^\alpha(0)$,
showing values as originally reported on the dates indicated.
The shaded band corresponds to the NACRE compilation \cite{Angulo:1999zz}.  From  \citet{Cognata:2009zz}, by permission.}
\label{comparison}
\end{figure}

As the $^{15}{\rm N}(\mathrm{p},\alpha)^{12}{\rm C}$ reaction competes with  
$^{15}{\rm N}(\mathrm{p},\gamma)^{16}{\rm O}$, a parallel study of the two 
is highly desirable. In Solar Fusion I,
a weighted average of S$_{1\,15}^\alpha(0)= 67.5 \pm 4.0$ MeV b was recommended 
using the results of \citet{Redder82}  and \citet{Zyskind79}. Recently 
the $^{15}{\rm N}(\mathrm{p},\alpha)^{12}{\rm C}$ 
reaction has been measured by \citet{LaCognata:2007zz},
using the indirect
Trojan Horse Method (TH method)
(see Sec. \ref{sec:indirect}).  The new data have been analyzed along with 
$^{15}{\rm N}(\mathrm{p},\gamma)^{16}{\rm O}$,
using a common R-matrix approach. The TH method allows one to   
extend the explored energy range down to about
20 keV, without the complication of electron screening enhancements that enter for direct
measurements.  Thus the TH measurements provide complementary information
that can be helpful in checking the overall consistency of S-factor fits.
\citet{LaCognata:2007zz} determined  S$_{1\,15}^\alpha(0)= 68 \pm 11$ 
MeV b from TH measurements.  New 
R-matrix fits to direct data of \citet{Redder82} by \citet{Cognata:2009zz} 
yielded S$_{1\,15}^\alpha(0)= 73 \pm 5$ 
and $74 \pm 9$ MeV b, depending on the respective energy ranges fit (see
\citet{Cognata:2009zz} for details), 
and S$_{1\,15}^\alpha(0)= 70 \pm 13$ MeV b for the indirect TH method data of 
\citet{LaCognata:2007zz}. 
An R-matrix fit by \citet{Barker:2008zza}, which did not include the TH method results, 
gave S$_{1\,15}^\alpha(0)= 80$ MeV b. 
We recommend the  value S$_{1\,15}^\alpha(0)= 73 \pm 5$ MeV b obtained by  \citet{Cognata:2009zz}
by fitting direct data as the new 
best value for the $^{15}{\rm N}(\mathrm{p},\alpha)^{12}{\rm C}$ reaction (see Table \ref{tab:CNO}). It is 
consistent with the two direct measurements,
the indirect TH method data,  and the R-matrix fit by Barker.  A summary given by \citet{Cognata:2009zz} of  S$_{1\,15}^\alpha(0)$ determinations
is shown in Fig. \ref{comparison}.  In Table \ref{tab:CNO} the derivatives shown are those reported by \citet{Zyskind79}, and may therefore not be completely consistent with the R-matrix energy dependence calculated by  \citet{Cognata:2009zz}.

\subsubsection{$^{15}$N(p,$\gamma$)$^{16}$O}
The $^{15}{\rm N}(\mathrm{p},\gamma)^{16}{\rm O}$ reaction provides the path to 
form $^{16}{\rm O}$ in stellar hydrogen burning\footnote{Most of the 
$^{16}{\rm O}$ found in the Sun originates not from hydrogen burning in the Sun 
itself, but instead from the ashes of helium burning in earlier stars.}, thus transforming the CN cycle into 
the CNO bi-cycle and CNO tri-cycle. In stellar environments, the reaction proceeds at very low 
energies, where it is dominated by resonant capture to the ground state through the first two 
interfering $J^{\pi}=1^{-}$ s-wave resonances at $E_{R}=312$ and $964$ keV.  In addition
there is some direct capture to the ground state.
Direct measurements  have been reported by \citet{Hebbard60} for proton energies 
down to  220 keV and by \citet{Rolfs74} down to proton energies of 155 keV.
These measurements disagree significantly below 300 keV.
In order to fit their low-energy data, \citet{Rolfs74} included the interference of the 
two $1^{-}$ resonant capture amplitudes with the nonresonant (direct) component to the 
ground state of ${}^{16}{\rm O}$ calculated in the hard-sphere approximation.  The absolute 
normalization of the direct term is entirely determined by the ANC of the bound state for 
${}^{15}{\rm N} + \mathrm{p} \to {}^{16}{\rm O}$. 
The spectroscopic factor adopted by \citet{Rolfs74}  corresponds to an ANC 
almost an order of magnitude larger than the one determined from $^{15}$N($^3$He,d)$^{16}$O 
by  \citet{Mukhamedzhanov:2008zz}.

A new analysis of the direct data using the two-level, two-channel R-matrix was 
presented by \citet{Mukhamedzhanov:2008zz}. The contribution from the $\alpha- {}^{12}{\rm C}$ channel was 
also taken into account.
The determined astrophysical factor S$_{1\,15}^\gamma(0)= 36 \pm 6$ keV b is about a factor of two lower 
than the previously accepted value
S$_{1\,15}^\gamma(0)= 64 \pm 6$ keV b  from \citet{Rolfs74}.   \citet{Hebbard60} 
reported S$_{1\,15}^\gamma= 32 \pm 6$ keV b at 23.44 keV, which was converted 
by  \citet{Mukhamedzhanov:2008zz} to S$_{1\,15}^\gamma(0)= 29.8 \pm 5.4$ keV b using the 
polynomial extrapolation given by Hebbard.
\citet{Mukhamedzhanov:2008zz} 
conclude that for every $2200 \pm 300$ cycles of the main CN cycle, one CN catalyst is lost 
due to this reaction, rather than $880$ cycles recommended by \citet{Rolfs74} and $1000$ cycles 
recommended by the NACRE compilations \cite{Angulo:1999zz}. Their result coincides with the 
R-matrix analysis by \citet{Barker:2008zz}, which yielded a leak rate of $1/2300$. Barker's analysis 
was completed before the ANC data were available and 
shows a larger spread of S values.

\begin{figure}
\includegraphics[width=8.6cm]{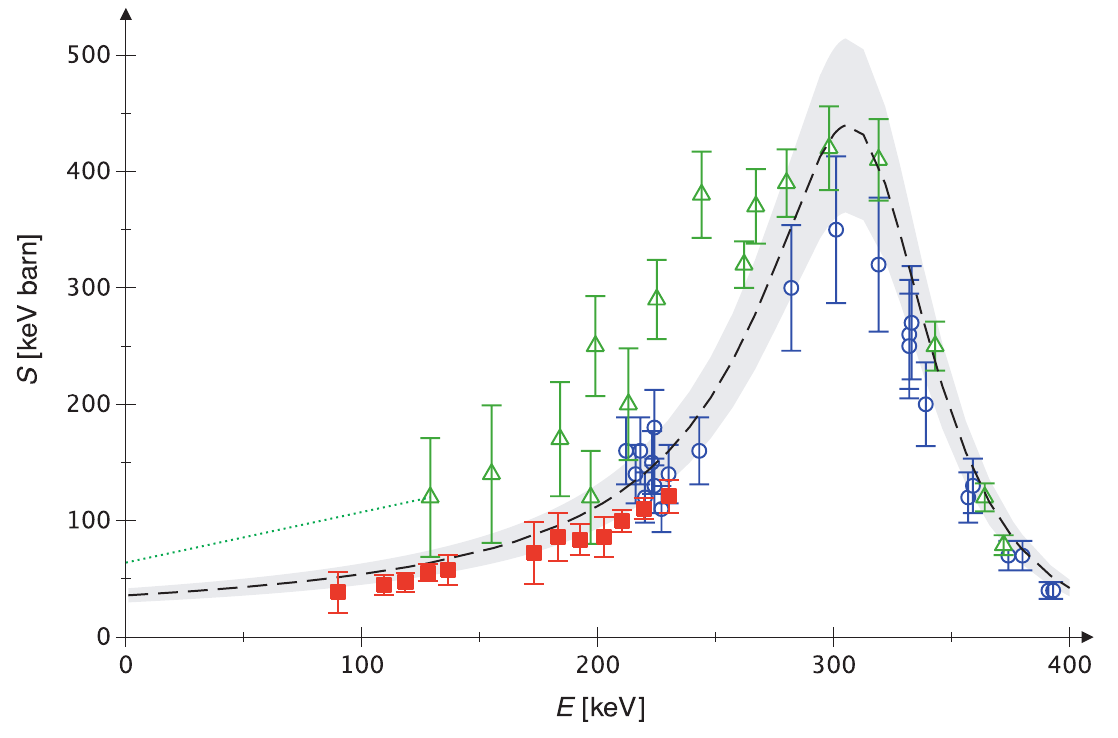}
\caption{(Color online) S$(0)$ for the $^{15}$N(p,$\gamma$)$^{16}$O 
reaction. Data from \citet{Hebbard60} (blue circles,
limited to $E\ge210$ keV), \citet{Rolfs74} (green triangles), and \citet{Bemmerer:2009px}
(red squares). Error
bars reflect statistical and systematic uncertainties summed in quadrature. 
Dashed line, previous R-matrix fit and
shaded area, its quoted 17\% uncertainty, from \citet{Mukhamedzhanov:2008zz}.  
Dotted line: previous extrapolation by \citet{Angulo:1999zz}.  Figure from
\citet{Bemmerer:2009px}, by permission.}
\label{JPG36Bemmerer}
\end{figure}

New measurements of this reaction at LUNA by \citet{Bemmerer:2009px} yielded cross sections 
with improved precision for energies between 90 to 230 keV (Fig.~\ref{JPG36Bemmerer}).
The extent of the agreement between the new LUNA data and the Hebbard data point to a possible 
unidentified systematic error affecting the low-energy data of \citet{Rolfs74}.
The value  S$_{1\,15}^\gamma(0)= 36 \pm 6$ keV b
obtained by  \citet{Mukhamedzhanov:2008zz} 
may be regarded as an interim recommendation pending an updated analysis taking full account of new data
(e.g., completion of the analyses for recent LUNA and Notre Dame experiments).
Further measurements at higher energies are also desirable in order to constrain the R-matrix fits. 

\begin{table*}
\centering
\caption{Summary of updates to S-values and derivatives for  CNO 
reactions.} 
\medskip
\begin{tabular}{|lccccl|}
\hline
\hline
Reaction                &~~~Cycle~~~&~~~~~~~S(0)~~~~~~~&~~~~~~~S$^\prime$(0)~~~~~~~ &~~~~~~~S$^{\prime \prime}$(0)~~~~~~~&References\\ 
                        &        &keV b &b      &keV$^{-1}$ b   & \\
\hline
$^{12}$C$(\mathrm{p}, \gamma)^{13}$N &I &$1.34\pm0.21$ &2.6$\times 10^{-3}$ &8.3$\times 10^{-5}$ & Recommended: Solar Fusion I\\ 
 \hline
$^{13}$C$(\mathrm{p}, \gamma)^{14}$N & I&7.6 $\pm$ 1.0 &-7.83$\times 10^{-3}$ &7.29$\times 10^{-4}$ &Recommended: Solar Fusion I \\
			& & $7.0\pm1.5$ & & & NACRE: \citet{Angulo:1999zz} \\ 
 \hline
$^{14}$N$(\mathrm{p}, \gamma)^{15}$O &I& $1.66\pm0.12$& -3.3$\times 10^{-3}$ & 4.4$\times 10^{-5}$ &   Recommended: this paper    \\
\hline
$^{15}$N$(\mathrm{p}, \alpha_{0})^{12}$C & I & ($7.3\pm0.5$)$\times 10^4$ &351 &11 & Recommended: this paper \\
\hline
$^{15}$N$(\mathrm{p}, \gamma)^{16}$O 	& II & $36\pm6$  & & &  \citet{Mukhamedzhanov:2008zz} \\
			&  &$64\pm6$ &  &  & \citet{Rolfs74}\\
                        & &$29.8\pm5.4$ & &  &  \citet{Hebbard60}\\
\hline
$^{16}$O$(\mathrm{p}, \gamma)^{17}$F & II &$10.6\pm0.8$ & -0.054 & &    Recommended: this paper \\
\hline
$^{17}$O$(\mathrm{p}, \alpha)^{14}$N &II   & & Resonances & &  \citet{Chafa:2007rx}     \\
\hline
$^{17}$O$(\mathrm{p}, \gamma)^{18}$F &III  & $6.2\pm3.1$ & 1.6$\times 10^{-3}$ & -3.4$\times 10^{-7}$  & \citet{Chafa:2007rx} \\ 
\hline
$^{18}$O$(\mathrm{p}, \alpha)^{15}$N & III & & Resonances & &  See text \\
\hline
$^{18}$O$(\mathrm{p}, \gamma)^{19}$F &IV & $15.7\pm2.1$ & 3.4$\times 10^{-4}$ & -2.4$\times 10^{-6}$  & Recommended: Solar Fusion I\ \\ 
\hline
\hline
\end{tabular}
\label{tab:CNO}
\end{table*}

\subsubsection{$^{16}$O(p,$\gamma)^{17}$F}
The cross section is dominated by direct capture to the ground and first excited states of $^{17}$F.  
Because the latter is weakly bound, its S-factor rises rapidly at low energies and the 
ground-state transition plays a minor role.  Calculations of 
the direct capture process by \citet{Rolfs73}, \citet{Morlock:1997zz}, \citet{baye98}, and \citet{bb2000} give a quantitative 
account of the energy dependence of both transitions.  \citet{baye98} calculate  
S$_{1\,16}^\gamma(0)$ with 
two choices for the nuclear force, obtaining S$_{1\,16}^\gamma(0)=10.2$ and $11.0$ keV b when normalized to 
the data of  \citet{Rolfs73} and \citet{Morlock:1997zz}.  The value adopted 
here is  S$_{1\,16}^\gamma(0)=10.6\pm0.8$ keV b and the derivative  is 
S$_{1\,16}^{\gamma~\prime}(0)=-0.054 {\rm \ b}$.
A recent reevaluation by \citet{Iliadis08} using both R-matrix theory and a potential model 
yielded reaction rates at temperatures $\geq 10^7$K that are consistent with these values \cite{Angulo:1999zz}, but with a lower assigned uncertainty.

\subsubsection{$^{17}$O(p,$\alpha)^{14}$N}
The  $^{17}$O(p,$\alpha)^{14}$N reaction  closes
branch II of the CNO bi-cycle.
The reaction rate at solar energies is dominated by a subthreshold resonance at $E_R=-3.1$ keV and 
a resonance at $E_R= 65.1$ keV. Several recent experiments have clarified the strength and location 
of a 2$^-$ resonance at 183.3 keV that plays a significant role at the higher temperatures
characteristic of novae and asymptotic giant branch
stars  \cite{Chafa:2007rx,Chafa:2005uj,Moazen:2007bx}.  \citet{Chafa:2007rx} find a 
low-energy cross section about a factor of three smaller than that given by \citet{Angulo:1999zz},
reflecting a re-evaluation of the proton width of the subthreshold resonance.  No 
calculated value for S$_{1\,17}^\alpha(0)$ has been published.

\subsubsection{$^{17}$O(p,$\gamma)^{18}$F}
The cross section shows a number of resonances in the range relevant to the hot CNO cycle in novae.
Effort has been recently devoted by  \citet{Chafa:2007rx,
Chafa:2005uj} and \citet{Fox:2004zz} to measuring the resonance 
parameters in both $^{17}$O$(\mathrm{p},\gamma)^{18}$F 
and $^{17}$O$(\mathrm{p},\alpha)^{14}$N.
While the higher-lying resonances are not directly 
relevant to solar CNO processing, they do have a significant influence in modern 
interpretations of the work of \citet{Rolfs73}, who measured the direct capture cross section 
that dominates at solar energies. \citet{Fox2005} and  \citet{Chafa:2007rx} both concluded 
that significant corrections 
are required.  The recommended S$_{1\,17}^\gamma(0)$  in Table \ref{tab:CNO} is taken from
\citet{Chafa:2007rx}.
The large uncertainty ($\sim50$\%) makes a new round of measurements of the direct 
capture cross section desirable.\footnote{Note in proof: The direct capture cross section was 
recently extracted from new measurements between lab energies of 193 and 519
keV \cite{Newton2010}.}

\subsubsection{$^{18}$O(p,$\alpha$)$^{15}$N}
The $^{18}{\rm O}$+p interaction represents a branching point in the CNO cycle: the
$^{18}{\rm O}(\mathrm{p},\alpha)^{15}{\rm N}$ reaction leads to a recycling of CN catalytic material,
while $^{18}{\rm O}(\mathrm{p},\gamma)^{19}{\rm F}$ may lead to a loss of this material, depending
on the fate of the produced $^{19}$F.
Nine resonances below 1 MeV influence the astrophysical rate for
$^{18}{\rm O}(\mathrm{p},\alpha)^{15}{\rm N}$, with those at  20, 144, and 656 
keV dominating \cite{Angulo:1999zz}. The presence of strong resonances in the astrophysical 
regime makes extraction of a value for S$_{1\,18}^\alpha(0)$ inappropriate. 
 
The strength of the 20-keV resonance had been known only from spectroscopic measurements 
performed by \citet{Champagne86} through the transfer 
reaction ${}^{18}{\rm O}({}^{3}{\rm He},\mathrm{d})^{19}{\rm F}$ and through the 
direct capture reaction ${}^{18}{\rm O}(\mathrm{p},\gamma)^{19}{\rm F}$ measured by \citet{Wiescher80}. 
The cross section at 20 keV is a factor $\sim$ 10$^{11}$ smaller 
than the one at 70 keV owing to the Coulomb barrier penetration factor. This makes a direct
measurement of the cross section impossible with present-day nuclear physics facilities.
Furthermore the spin and parity of the 8.084 MeV level in ${}^{19}{\rm F}$ (corresponding to a 
90 keV resonance in the $^{18}{\rm O}(\mathrm{p},\alpha)^{15}{\rm N}$ cross section) was
not known. 
In order to reduce the nuclear uncertainties affecting the reaction rate, which
\citet{LaCognata:2008wc} estimated at about an
order of magnitude, a new round of measurements has been made with the TH method
by \citet{LaCognata:2009pq,LaCognata:2008wc,LaCognata:2010zz}.  The deduced strength of the 20 keV resonance 
${\omega \gamma}= 8.3^{+3.8}_{-2.6} \times 10^{-19}$ eV  
eliminates much of the broad range given by NACRE \cite{Angulo:1999zz}, 
${\omega \gamma}= 6^{+ 17}_{-5} \times 10^{-19}$ eV, and
decreases the uncertainty of the reaction rate by about a factor  8.5 \cite{LaCognata:2008wc,LaCognata:2009pq,LaCognata:2010zz}.  In addition, 
the spin ($3/2^{-}$) and strength of the 90-keV resonance, 
which was seen in the work of 
\citet{Lorentz79}, were determined. The \citet{LaCognata:2008wc,LaCognata:2009pq,LaCognata:2010zz} and \citet{Lorentz79} strengths agree.

\section{INDIRECT METHODS AND THEIR VALIDATION}
\label{sec:indirect}
Three classes of experiments contribute to our understanding of solar fusion reactions,
direct cross section measurements, indirect methods, and ancillary nuclear structure
techniques for determining the properties of resonances (energies, $\gamma$ and particle widths,
and spins and parities).   Indirect methods involve the use of nuclear 
reactions related to, but not identical to, the solar reactions under study, as tools to 
probe properties of the solar reactions.  References have been made in this review to three indirect
methods, asymptotic normalization coefficients, Coulomb dissociation, and the Trojan horse method.  
As the connection between the indirect 
observable and the solar reaction of interest must be established through reaction
theory, such methods entail a greater
degree of model dependence, impacting systematic uncertainties.
But indirect methods also have many virtues: they can be applied
when direct measurements are difficult or impossible, have systematic uncertainties that are different
from those of direct measurements, and provide supplementary information that can 
constrain R-matrix and other models used in the extrapolation of data from direct measurements.
The role of indirect measurements in validating and constraining models is apparent from the
discussions, for example, of Sec. \ref{sec:N114}.

\subsection{The asymptotic normalization coefficient method}
The asymptotic normalization coefficient method 
constrains S(0) by exploiting the peripheral nature of many radiative capture
reactions in nuclear astrophysics.  Because of Coulomb and/or centrifugal barriers, most
(p,$\gamma)$ and $(\alpha,\gamma)$ reactions are peripheral 
at solar energies.
The cross section for a nonresonant radiative capture reaction A(p$,\gamma)$B 
at zero relative energy depends only on the  long-distance behavior of the p+A wave function
(and on the overlap of that extended wave function with B).  The detailed short-range 
behavior of the scattering state p+A or bound state B, governed by the strong interaction
and nuclear length scales, are not relevant to the reaction mechanism.   The bound-state
wave function at long distances will contain a component corresponding
to two separated clusters, p and A, with the cluster relative radial motion
given by a Whittaker function.
% This is a third version -- I retained the long-distance modifier, as otherwise the radial
% wave function will not have the Whittaker form
The asymptotic normalization coefficient (ANC) is defined as the amplitude of this component
(apart from an overall phase) \cite{mukhamedzhanov90,xu94}.  A distinct ANC will govern
the nonresonant capture into each final state, i.e., the ground or  bound excited states of B.
Therefore, if one can identify another nuclear reaction that includes the
vertex A + p $\leftrightarrow$ B  and is sensitive only to the tail of the radial overlap function,
the needed ANC can be determined from that reaction.  
This measurement in a different system then determines the radiative capture
cross section at zero relative energy \cite{mukhamedzhanov01},
up to small corrections determined by
the scattering wave function and the potential in the continuum \cite{typ2005,capel06}. 
While the method is limited to S(0), providing a data point below the Gamow peak, this often
complements the data from direct measurements, which are frequently limited to energies above
the Gamow peak.

In most applications, the ANC is deduced from transfer reactions.
The extraction relies on the distorted wave Born approximation
(DWBA) and the direct proportionality between the transfer cross
section and the square of the ANC. 
Provided that the transfer reaction is completely peripheral and the measured 
angular distributions are well described within the single-step DWBA, 
the ANC can be extracted.  The main source of uncertainty 
comes from the optical model description, typically $ \gsim10$\% for reactions above the 
Coulomb barrier.  For this reason, it is often important  to also measure the elastic
channel of the corresponding transfer reaction over a wide angular range,
to help constrain optical model parameters.
Investigations of effects beyond the single-step DWBA arising from target excitation
suggest that deformed targets with strong couplings to low-lying
excited states are not good candidates for the ANC method \cite{azhari01}.
Some of the applications of the method involve loosely bound nuclei, opening
up the possibility of multi-step processes through
continuum states as viable alternatives
to the direct reaction mechanism. So far there has only been one
reaction for which the magnitude of this effect has been evaluated; in this case it was found to be negligible \cite{moro03}, but a more systematic study should be done.

In Solar Fusion I the $^{16}$O(p$,\gamma$)$^{17}$F reaction was identified
as a good test for the method. As a consequence, the 
$^{16}$O($^3$He,d)$^{17}$F reaction was measured at 30 MeV. The angular distributions of the ground state and the first excited state were well described within the DWBA and the inferred S factors agreed with the radiative capture data to better than 9\% \cite{gagliardi99}. 

There have been many subsequent applications of this method,
mostly involving peripheral transfer reactions on intermediate mass targets.
Here we focus on those relevant to validating the method for solar fusion reactions.
Two transfer reactions, $^{10}$B($^7$Be,$^8$B)$^9$Be and $^{14}$N($^7$Be,$^8$B)$^{13}$C,
 were used to extract the ANC for S$_{17}$(0) 
\cite{azhari99a,azhari99b}.  For both targets, the peripheral nature of the transfer reactions 
were checked carefully by evaluating the sensitivity of the extracted 
ANC to the single particle parameters of the binding potential in the DWBA analysis.  
Similar analyses have been done by invoking a radial cutoff in the distorted wave calculation \cite{muk1997,
fer2000}.  
In \citet{tabacaru06} a joint analysis was performed, 
yielding S$_{17}$(0)=18.0 $\pm$ 1.9 eV b, which can be compared to the best value from
direct measurements, 20.8 $\pm$ 0.7 $\pm$ 1.4 eV b.
% Changed from original - number above taken from the paper
In addition, the low-energy reaction $^{7}$Be(d,n)$^{8}$B at $E_\mathrm{lab}$=7.5 MeV 
\cite{liu96,ogata03} was studied, but difficulties were encountered in the analysis.
The (d,n) reaction model
depends on the poorly constrained exit-channel neutron optical potential.
In addition, the use of low energies, necessary to satisfy the  peripherality condition
given the low Z of the deuteron, leads to significant compound nuclear contributions,
introducing additional uncertainties.

 This review includes several illustrations of the use of ANC determinations
  to validate R-matrix descriptions of direct reaction data.  In Sec. \ref{sec:N114} 
  the example of the subthreshold-state
 (6.79 MeV) contribution to
 $^{14}$N(p,$\gamma$)$^{15}$O is described in some detail: the ANC
 determined from the R-matrix 
 fit is in good agreement with that extracted by \citet{Bertone02} and \citet{Muk03}
 from $^{14}$N($^3$He,d)$^{15}$O.
  Analogous work
  using $^{15}$N($^3$He,d)$^{16}$O to study $^{15}$N(p,$\gamma$)$^{16}$O
  is discussed in Sec. \ref{sec:N114other}.
  % Thought this should be shortened, to avoid repetition, with references back to
  % early parts of the review.
  
As ANCs can be related to spectroscopic factors, the latter can also be used to parameterize
cross sections.  However, spectroscopic factors have an additional dependence on the
single-particle bound state orbitals assumed in their extraction.  Consequently radiative
capture reactions parameterized through ANCs and spectroscopic factors have somewhat
different uncertainties.  Further discussion can be found in \citet{mukhamedzhanov01} and
\citet{Bertone02}.

Finally, it should be mentioned that breakup reactions $B+T \rightarrow A+\mathrm{p}+T$
can also be used to extract ANCs when they meet the peripherality condition \cite{trache04}. However a 
detailed study of the uncertainties involved in the reaction theory
has not yet been completed.

\subsection{The Coulomb dissociation method}
Coulomb dissociation (CD), originally proposed as 
a method for extracting information on astrophysical 
fusion cross sections by Rebel, was developed theoretically shortly thereafter \cite{baur86}. 
The process occurs when a beam of fast
projectiles interacts with a heavy target such as Pb. An energetic virtual photon 
from the target can then dissociate the projectile, liberating a nucleon or $\alpha$ particle. 
To the extent that the experimentalist can exploit the kinematics of this process to enhance the
contributions from the long-distance exchange of single photons, this
process can then be related by detailed balance to the corresponding radiative capture reaction.
% Discussed this section with working group leads, and tried for some simplification;
% original discussion seemed to have some contradictions
But several effects complicate this simple picture.
Whereas nonresonant radiative captures generally proceed almost exclusively
by $E1$ transitions, the strong $E2$ field in CD can be important. 
Moreover, the simple radiative capture/CD correspondence is 
complicated by multiple photon exchange and by the strong interaction, which can 
lead to nuclear diffraction dissociation and Coulomb-nuclear interference.
Strong interaction effects can be reduced by restricting
measurements to small angles, where long-range electromagnetic transitions 
dominate nuclear interactions. 
Multiple photon exchange (also known as post-acceleration) can be reduced by
increasing the beam energy, shortening the time the projectile spends in the target's
field.

In Solar Fusion I a proposal was made to test the validity of the CD method quantitatively
through comparison with a corresponding radiative capture measurement.  The radiative capture
reaction was to have suitable properties, including 
a low Q value, a nonresonant $E1$ reaction mechanism, reactants with similar mass/charge ratios, 
and a final nuclear state with relatively simple structure.  Although no perfect reaction was identified, 
$^7$Be(p$,\gamma)^8$B appears to be a good choice.  Several new measurements
were made, and a great deal of theoretical effort was invested in their interpretation and in
extracting the S factor.  This work is summarized in Sec. \ref{coulbreakup} and will not be
discussed further here, except to repeat the conclusion that, while in several
cases agreement between
the CD method and direct measurements has been demonstrated at the 10-20\% level,
remaining uncertainties in the magnitude of S($0$), in independently determining the shape of S($E$),
and in the theory argue that the inclusion of CD data in the current
S$_{17}$ evaluation would be premature.

Efforts also have been made to validate the CD method 
for the $^{14}$C(n,$\gamma)^{15}$C reaction. Although this reaction is not directly relevant to solar fusion,
the radiative capture rate is now known to a precision of $\sim$ 10\% \cite{reifarth08}.
The corresponding CD of $^{15}$C on $^{208}$Pb has recently been 
remeasured at RIKEN \cite{nakamura09}. 
Reaction models predict that the $^{15}$C breakup has an insignificant 
nuclear contribution and is dominated by $E1$ transitions, provided the analysis is limited 
to events in which the $^{15}$C center-of-mass 
scattering angle and the relative energy of the breakup fragments are small.
Independent analyses of these 
data \cite{summers08,esbensen09} find that the neutron capture cross section extracted from 
CD agrees very well with the direct measurement and has 
comparable precision. This appears to be a favorable case for the theoretical 
treatment due to the dominant nonresonant $E1$ reaction mechanism, small $E2$ and 
nuclear contributions, and relative simplicity of $^{15}$C, which can be described reasonably
in a single-particle $^{14}$C+n potential model. While the agreement in this
case is promising, some caution is warranted because the 
radiative capture measurement has not been confirmed by an independent measurement.

The ANC and CD 
methods are both well suited to measurements with low intensity radioactive 
beams because the transfer reaction and CD cross sections 
are much larger than the corresponding radiative capture reactions. Moreover, 
they are both applicable to radiative capture reactions.

\subsection{The Trojan Horse method}
The Trojan Horse (TH) method \cite{baur86th,spitaleri04} is an indirect technique to determine the astrophysical 
S factor for rearrangement reactions. It allows inference of the cross section of the binary  process
\begin{equation}
x+ A \to b+B
\label{binaryreaction1}
\end{equation}
at astrophysical energies through measurement of the TH reaction
\begin{equation}
a + A \to y + b+B.
\label{THreaction1}
\end{equation}
The measurement is done with quasi-free kinematics, 
in which a TH $a$ having a strong $x+y$ cluster structure is accelerated to energies above the Coulomb barrier. 
After penetrating the Coulomb barrier, the nucleus $a$ breaks up,
leaving $x$ to interact with the target $A$ while the projectile fragment $y$ flies away. 
From the measured cross
section of reaction (\ref{THreaction1}), the energy dependence of the binary subprocess 
(\ref{binaryreaction1}) is determined. 
While the reaction (\ref{THreaction1}) can occur in a variety of ways, 
the TH reaction mechanism should dominate
in a restricted region of three-body phase space in which the
momentum transfer to the spectator nucleus $y$ is small, i.e., quasi-elastic scattering 
conditions apply. Since the transferred particle $x$ in the TH reaction (\ref{THreaction1}) 
is virtual, its energy and momentum are not 
related by the on-shell equation $E_{x}= p_{x}^{2}/(2\,m_{x})$. 

The main advantage of the TH method is that the low-energy cross sections can be deduced
from a reaction that is not strongly suppressed by Coulomb barriers or 
strongly altered by electron screening
\cite{Ass87,spitaleri01}.  The TH cross 
section can be used to determine the energy dependence of the bare nuclear S factor for the binary process (\ref{binaryreaction1}) down to zero relative kinetic energy of $x$ and $A$. The absolute value of S$(E)$,
however,  must be determined by normalizing to direct measurements at higher energies. 
To ensure quasi-free kinematics one should measure the momentum 
distribution of the spectator fragment $y$ and the angular distributions of the 
fragments of the binary sub-reaction to check for contributions from 
non-TH mechanisms.  As a check on distortions due to final state interactions, the 
momentum distribution of the spectator can be measured and compared with that 
of the spectator in the free TH nucleus \cite{Pizzone09}.
Final state distortions can be treated in DWBA calculations \cite{LaCognata:2010zz}.
% Cut a bit of material not critical to our focus, solar reactions

The uncertainty of the S$(E)$  
extracted from the TH method includes contributions from statistics, uncertainties due to the 
need to normalize the TH data, finite experimental energy resolution, 
and backgrounds due to other reaction mechanisms. The first successful test of the TH 
method was conducted for the ${}^{7}{\rm Li}(p,\alpha){}^{4}{\rm He}$ reaction  \cite{lattuada01}. 
The extracted S(0)= 55 $\pm$ 6 keV b includes an uncertainty of 10\% from the normalization of the TH data to the direct data \cite{Eng92} and $5.5\%$ from other sources, mainly statistics. 
In addition, in Sec. \ref{sec:XIB2} we compare results for TH
and direct determinations of the cross section for $^{15}$N(p,$\alpha$)$^{12}$C.
Although promising, the TH method requires further validation by experiment, and its
significant dependence on reaction theory calls for more investigation of the
approximations by which TH reactions are related to their astrophysical analogs.

The TH method also provides an important test of electron screening potentials,
which can be obtained from comparisons of direct and TH cross sections.

\subsection{Summary}

The three indirect techniques discussed here provide alternatives to 
direct measurements of astrophysically important reaction rates.  In some cases
they provide the only practical means for determining stellar reaction
rates.  While their connection
to solar reactions requires an additional level of
reaction theory, experimental tests of their validity have 
often yielded agreement with direct measurements within 10-20\%. 
% 20% is more consistent with the discussions in S17
Significant progress
has been made since Solar Fusion I in benchmarking indirect techniques.
Indirect methods are best applied to cases where there is a supporting body
of experimental data that can be used to constrain the needed nuclear model
input, such as optical potentials and effective interactions.

In actual
practice, the distinction between direct and indirect methods is not sharp,
but rather a matter of degree.
While a measurement may probe a stellar reaction directly, it often does so at a different
energy or in a different screening environment.  Thus
direct methods also depend on reaction theory, to extrapolate data to stellar
energies or, in cases like S$_{33}$ where data in the Gamow peak have been obtained,
to correct for the effects of screening in terrestrial targets.   Still, the connection to
stellar physics is typically much closer.  Models play a less important role, and increasingly
the needed modeling can be done microscopically, as direct measurements involve
light nuclei.

For this reason we maintain a distinction between direct and indirect
methods in this review, basing our recommendations on results from the former.
However, indirect methods have had a significant impact on our analysis: they have been used in
this review to constrain R-matrix fits to direct data and to check the consistency of
conclusions based on analyses and modeling of direct data.

We recommend extending the benchmarking of indirect methods against
direct methods over a wider range of reactions, as more data would be useful in quantifying
the uncertainties in such techniques.

\section{FUTURE FACILITIES AND CURRENT CAPABILITIES}
\label{sec:newf}
We noted in the introduction to this review the crucial role nuclear astrophysics 
experiments have played in the development of a quantitative SSM and in
motivating solar neutrino experiments.   We outlined the important goals that
remain in this field -- tests of weak interactions and of solar properties that make use
of high precision solar neutrino measurements, helioseismology mappings
of $c(r)$, and detailed solar modeling.
There are also a host of related problems -- Big Bang nucleosynthesis, red-giant
evolution, the evolution of supernova progenitors, and a variety of transient
explosive phenomena in astrophysics -- where a quantitative understanding of
the nuclear physics is essential.   This chapter deals with the experimental
facilities that have allowed progress in this field, and discusses the instrumental
developments that will be important if we are to continue a similar rate of
progress over the next decade.

The measurements that support the development of a quantitative theory 
of main-sequence stellar evolution primarily involve 
low energy proton- and $\alpha$-capture reactions that
traditionally have been studied with small accelerators.  The machines
must be able to provide proton or $\alpha$ 
beams of sufficient intensity to allow cross section measurements
near the very low energies of 
the Gamow peak.

Because low energy charged-particle reaction cross sections are small, 
experiments must be designed for signal rates much smaller than background
rates associated with cosmic rays, the natural radioactivity of the laboratory 
environment, and the induced activity arising from beam interactions with
target impurities.  The ambient background can 
be roughly divided into muons and neutrons associated with cosmic rays, 
and $\gamma$ rays and neutrons from natural radioactivity (uranium, thorium, 
potassium, and radon from surrounding geology).   Today most charged-particle
reaction measurements for nuclear astrophysics are being 
performed at above ground facilities, with various techniques then employed
to mitigate backgrounds.   The common technique is 
passive shielding around the detection region. Typically a 
layered combination of lead, copper, and polyethylene is used to reduce 
$\gamma$ and neutron backgrounds within detectors with relatively small volumes. 
But additional strategies are available to further reduce backgrounds and
thus allow measurements at energies nearer those relevant for astrophysics,
including
\begin{enumerate}
\item  use of more sophisticated detector setups with both passive
and active shielding and with
triggers to aid in event identification;
\item  measurements in inverse kinematics using recoil 
separators in facilities above ground; and
\item  measurements with direct kinematics using accelerators that are
sufficiently deep underground to suppress penetrating cosmic-ray muons
and the neutrons and other secondary activities they induce.
\end{enumerate}

Passive shielding, active shielding, and coincidence gating techniques can
enhance event identification and significantly reduce backgrounds in
above-ground laboratory environments.  As most 
resonance levels of astrophysical interest decay via $\gamma$-cascades \cite{Row02}
$\gamma \gamma$-coincidence 
techniques can be used to significantly reduce the single-$\gamma$ background. 
Q-value gating techniques, where only events in coincidence with the 
summing peak of the radiative capture reaction are accepted \cite{Couture08}, can
allow one to extend measurements to lower energies, but at the cost of
a decreased overall counting efficiency due to the coincidence requirement.

Alternative techniques have been developed to reduce backgrounds
without such losses in detection efficiency.  Two ideas that have
demonstrated their promise are measurements in inverse kinematics --
one detects the reaction recoil
particles rather than the light particles or $\gamma$s of the reaction --
and measurements in underground environments.  Below we describe
past and current experience with these two techniques as well as the future
facilities, in progress or planned, that would allow these techniques
to be further advanced.

\subsection{Inverse kinematics measurements using recoil separators}
In an  inverse-kinematics experiment a heavy ion induces (p,$\gamma$) or ($\alpha,\gamma$) 
reactions when it interacts in a hydrogen or 
helium gas target. The projectiles and reaction products move within a narrow 
cone in the forward direction.  A recoil separator is used to reject the primary 
beam while focusing the reaction products for detection. 
The charged recoils can be detected with higher efficiency than the $\gamma$s produced in
conventional proton- or $\alpha$-beam experiments.  By detecting the $\gamma$s in 
coincidence with the reaction products, dramatic reductions in backgrounds
can be achieved.  Existing recoil separator facilities fior nuclear astrophysics
experiments include DRAGON
at ISAC in TRIUMF \cite{Hut03}, the Daresbury separator at HRIBF in Oak Ridge \cite{Fit05}, 
ERNA at the DTL in Bochum \cite{Rog03}, and the RMS at KUTL in Kyushu, Japan \cite{Sag05}.

Recoil separators are not useful for ($\alpha$,n) reactions because separator
acceptance angles are too small, given the momentum transfer in this process.

Recoil separators present several experimental challenges
\cite{Rog03}, particularly for the low energies important in
solar fusion cross section measurements. At such energies, the energy spread 
and the angular aperture are, for most solar fusion reactions, larger than 
the acceptance of any of the recoil separators cited above.

The following conditions 
must be fulfilled in experiments on absolute cross sections:
\begin{itemize}
\item the transmission of the recoils must be exactly known and should ideally be 100\%;
\item the charge-state distribution of the recoil products must be known 
or the reaction must be studied for all charge states produced \cite{DiL08}; and
\item the interaction region must be well defined. 
\end{itemize}
Therefore, experiments coming on-line in the 
near future are all planning to use compact
high-density gas-jet targets instead of extended windowless gas targets. 

Recoils of solar fusion reactions
typically have relatively large emission angles and large energy spreads, 
both of which increase with decreasing reaction energies $E$, when $E$ $<$ Q. 
The angular distribution of recoils following emission of
capture $\gamma$-rays of energy $E_\gamma$ is characterized by an emission 
cone half-angle of
\begin{equation}
\theta = \arctan{{E_\gamma \over p}}
\end{equation}
where $p$ is the momentum of the beam ($c \equiv 1$).  
The total energy spread $\Delta E$ of the recoil accompanying $\gamma$ emission is
\begin{equation}
{\Delta E \over E} = {4 E_\gamma \over p}.
\end{equation}
 
Furthermore  a large spatial separation between the reaction products 
and the beam is required, as the primary beam intensity is typically 
many orders of magnitude larger than that of the recoiling reaction products. 
A clean separation is difficult for recoils with large energy 
spreads, making  low-energy solar fusion reactions particularly challenging.
Recoil separators are therefore more typically 
used for higher energies characteristic of helium- or explosive hydrogen-burning reactions. 
For example, the recoil-separator measurements of S$_{34}$ 
at the ERNA facility in Bochum were limited to data above a
center-of-mass energy of 700 keV \cite{erna}. Below this energy the 
angular divergence of the recoils exceeds the angular acceptance of the 
separator, $\pm$ 25 mrad \cite{DiL08}.

\begin{figure*}
\begin{center}
\includegraphics[width=18cm]{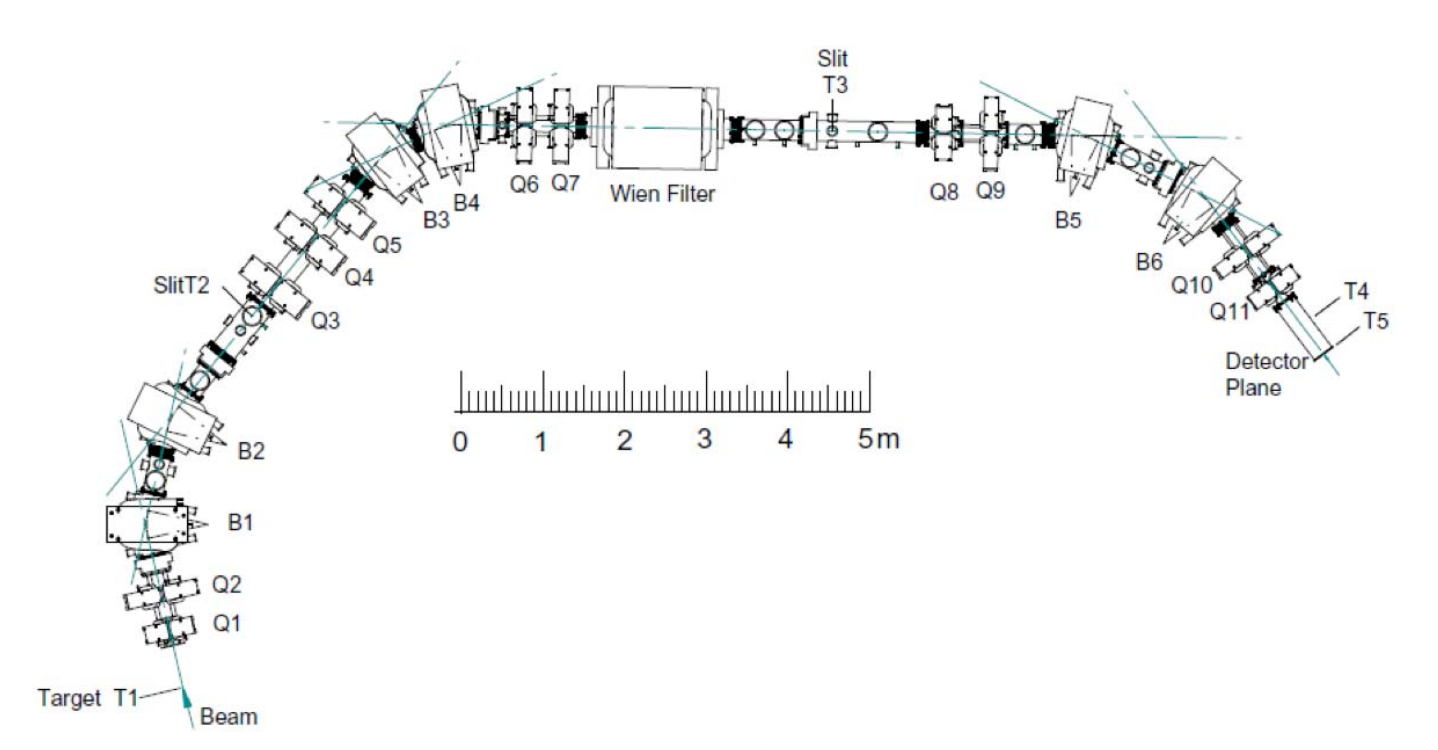}
\caption{Layout of the St. George recoil separator.} 
\label{fig:fac1}
\end{center}
\end{figure*}

Two dedicated next-generation separators for low-energy nuclear 
astrophysics studies with stable ion beams will soon come on line, the
St. George facility at Notre Dame's Nuclear Science Laboratory
 \cite{Cou08} and the ERNA separator at the CIRCE facility in Caserta, 
Italy.  The latter is based on a redesign of the Bochum ERNA separator \cite{Rog03}. 
Both separators feature large acceptances in angle and energy and 
will be equipped with high density gas-jet targets to ensure
well defined interaction regions.  Figure \ref{fig:fac1}
shows the layout of the
St. George recoil separator. The design is optimized for low-energy
radiative
$\alpha$-capture reactions important to stellar helium burning. It has a 
large angular acceptance of $\pm$ 40 mrad, an energy acceptance of 
$\pm$ 7.5\%, and a mass resolving power $M/\Delta M$ $\sim$ 100 \cite{Cou08}. 

\subsection{Underground facilities}

In all direct-kinematics capture-reaction measurements using $\gamma$ or 
neutron spectroscopy,  whether performed above ground or underground, sources of
environmental radioactivity must be controlled. Background sources include
radioactivity 
from intrusions and impurities in the rock and from construction materials, as
well as sources intrinsic
to targets and detectors.  External sources can be  
reduced by careful shielding of the target 
and the detector environment.  In addition, beam-induced backgrounds
(e.g., backgrounds from activation of impurities in the target) must be controlled
through careful ion beam optics and choice of 
vacuum component materials. Active shielding techniques and complex 
event identification can also help.
% Note: Reference 11, Iliadis and Champagne, is not cited in the text.

In surface facilities, however, the most difficult backgrounds are frequently those 
associated with cosmic rays.  This background can be removed by
exploiting the natural shielding provided by the rock overburden in underground
sites.  The improvements possible with this strategy have been demonstrated
by the 50 keV LUNA I and 400 keV LUNA II programs at Gran Sasso.
The laboratory's depth,
$\sim$ 3.0 km.w.e. (kilometers of water equivalent, flat-site equivalent \cite{Mei06}),
reduces the fluxes of muons and secondary neutrons, relative to surface values, by factors
of 10$^6$ and 10$^3$, respectively.
Consequently, the LUNA I collaboration \cite{Bonetti99} was able to map S$_{33}$
throughout the Gamow peak: a counting rate of one event per month
was achieved at the lowest energy, $E$ = 16 keV, with an uncertainty of
20 fb or 2 $\times$ 10$^{-38}$ cm$^{2}$.  
Other critical pp chain and CNO cycle cross sections were made
at energies far lower than 
previously possible \cite{For03,Imb05,Junker98,Greife94}. 

\begin{figure*}
\begin{center}
\includegraphics[width=18cm]{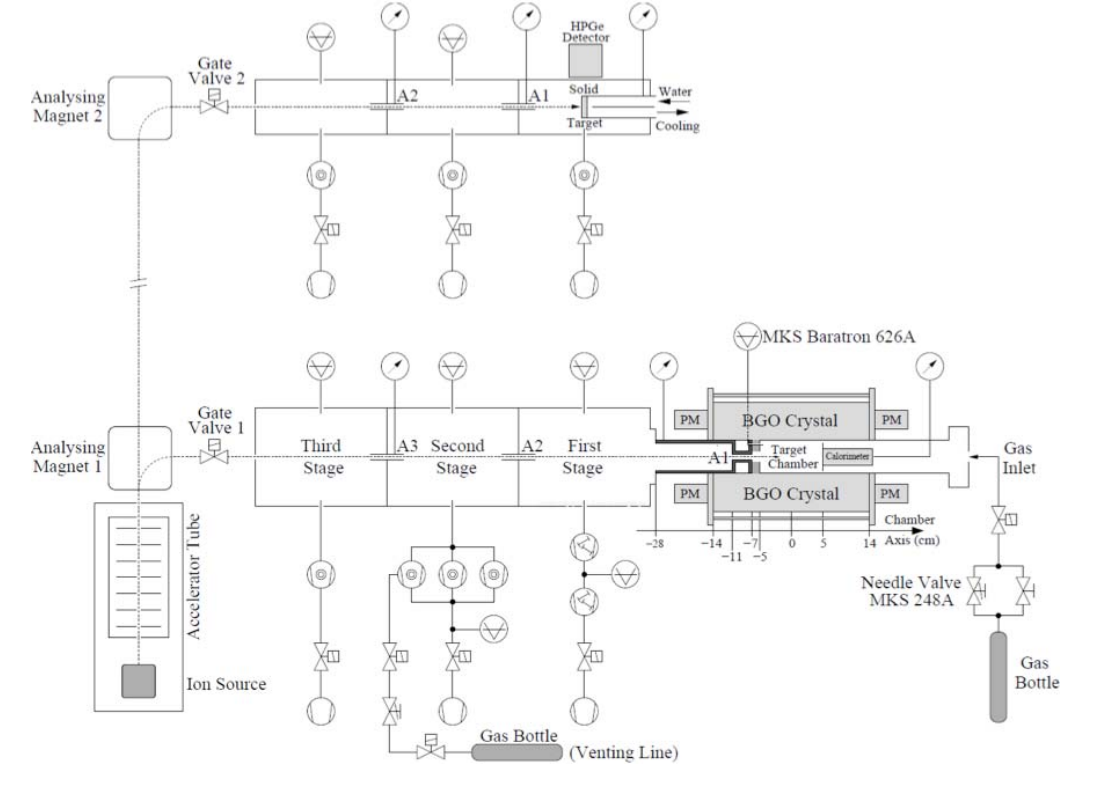}
\caption{A schematic of the present LUNA 400 keV set-up.} 
\label{fig:fac2}
\end{center}
\end{figure*}

The successes of LUNA
have inspired plans for the new underground facilities we discuss in this section.
Figure \ref{fig:fac2} shows a schematic of the present LUNA II set-up in Gran Sasso, 
which consists of a commercial 400 kV accelerator, a windowless gas target, and
a solid target line.

Nuclear astrophysics has rather modest depth requirements.
The hadronic cosmic-ray component
is quickly attenuated, leaving penetrating high-energy muons
as the dominant source of background at depth.  These muons 
interact in the rock to produce neutrons and a continuous spectrum
of high energy $\gamma$s.  Thus the main requirement is an overburden
sufficient to reduce muon-associated activities to a
level well below natural background levels associated with activities
in the laboratory's rock and concrete walls.   The neutron fluxes in Gran Sasso,
$\sim$ 4 $\times$ 10$^{-6}$/cm$^2$/s \cite{Bem05,Lau04}, and in Spain's
underground laboratory Canfranc, 
$(3.80 \pm 0.44) \times 10^{-6}$/cm$^2$/s \cite{Car04}, are almost entirely 
due to local radioactivity.
Taking these deep-laboratory
values as typical of the environmental background component, one can determine
the depth necessary to reduce the cosmic-ray-associated neutron contribution to 1\% of
the total.  The simulations of \citet{Mei06} yield $\sim$ 1.5 km.w.e. (flat site equivalent).

Similar results are found for the $\gamma$-ray flux.  The  
LUNA $^{14}$N(p,$\gamma$) counting goal was 10$^{-4}$ counts/keV/hr.  The
cosmic-ray muon-induced rate at 1.5 km.w.e. would be approximately an order
of magnitude lower \cite{Hax07}.   As almost all deep physics laboratories now
operating are at depths in excess of 1.5 km.w.e, one concludes that many
locations are suitable for nuclear astrophysics -- at least until order-of-magnitude
reductions in the
laboratory environmental neutron and $\gamma$-ray background are made.

Based on the success of the LUNA collaboration, several underground 
accelerator facilities are now being proposed. Table \ref{tab:fac} shows the 
parameters of these facilities.  The plans reflect design improvements from
fifteen years of experience
with LUNA.

The present 
LUNA facility is small and limited to the measurement of proton- and $\alpha$-capture 
reactions below 400 keV,  with typical beam currents between 100 
and 200 $\mu$A.  The available beam current has limited the statistical accuracy
of data taken at the lowest energies.  In addition, many 
reactions have complex resonance structures that must be adequately
mapped, to provide the information needed to extrapolate cross sections
to Gamow energies.   This requires measurements over a broader energy
range than is currently available at LUNA.   
Therefore, the LUNA collaboration has submitted a letter of intent for the 
installation of a higher energy accelerator that would allow the LUNA
 program to grow beyond solar fusion physics.
This upgrade proposal is currently under review \cite{Pra08}  . 

Three initiatives for new underground accelerator facilities are also
under discussion:
\begin{itemize}
\item ELENA is a proposed facility for the Boulby salt mine in the UK, a site that has
environmental neutron backgrounds less than half those of Gran Sasso
[(1.72 $\pm$ 0.61 (stat.) $\pm$ 0.38 (syst.)) $\times$ 10$^{-6}$ /cm$^2$/s above 0.5 MeV \cite{Car04}] 
and $\gamma$-ray backgrounds that are 5-30 times lower than Gran Sasso values, 
for $E_\gamma \lsim$ 3 MeV \cite{Ali09}.
This reflects the low U and Th concentrations in salt.
As the site is approximately at the same depth as Gran Sasso (2.8 vs. 3.1 km.w.e.,
taking proper account of the topography \cite{Mei06}), full advantage can be taken
of the reduced environmental background.
%Note: many Gran Sasso depths are badly off because they quote overburden
%directly overhead: the cosmic ray flux has a csec behavior that makes the slant
%depth critical.  Mei and Hime do things properly
% Mei and Hime depths are used in the table, with the exception of Canfranc, where
% the effective depth is computed from the measured muon flux
\item CUNA is a 3 MeV accelerator facility that has been proposed
for Spain's Canfranc Laboratory, located in an 
abandoned train tunnel in the Pyrenees mountains \cite{Bet09}.
\item DIANA, Dakota Ion Accelerators for Nuclear Astrophysics, would be the
nuclear astrophysics facility for DUSEL
(Deep Underground Science and Engineering Laboratory), a laboratory
being planned in the abandoned
Homestake gold mine, South Dakota \cite{ecr}. 
\end{itemize}
%changed reference immediately above -  link was dead -- is this the site you wanted?
As in the case of the proposed LUNA upgrade, these facilities would be capable
of mapping cross sections over broad energy ranges with fixed
configurations for target and detector.

\begin{figure*}
\begin{center}
\includegraphics[width=18cm]{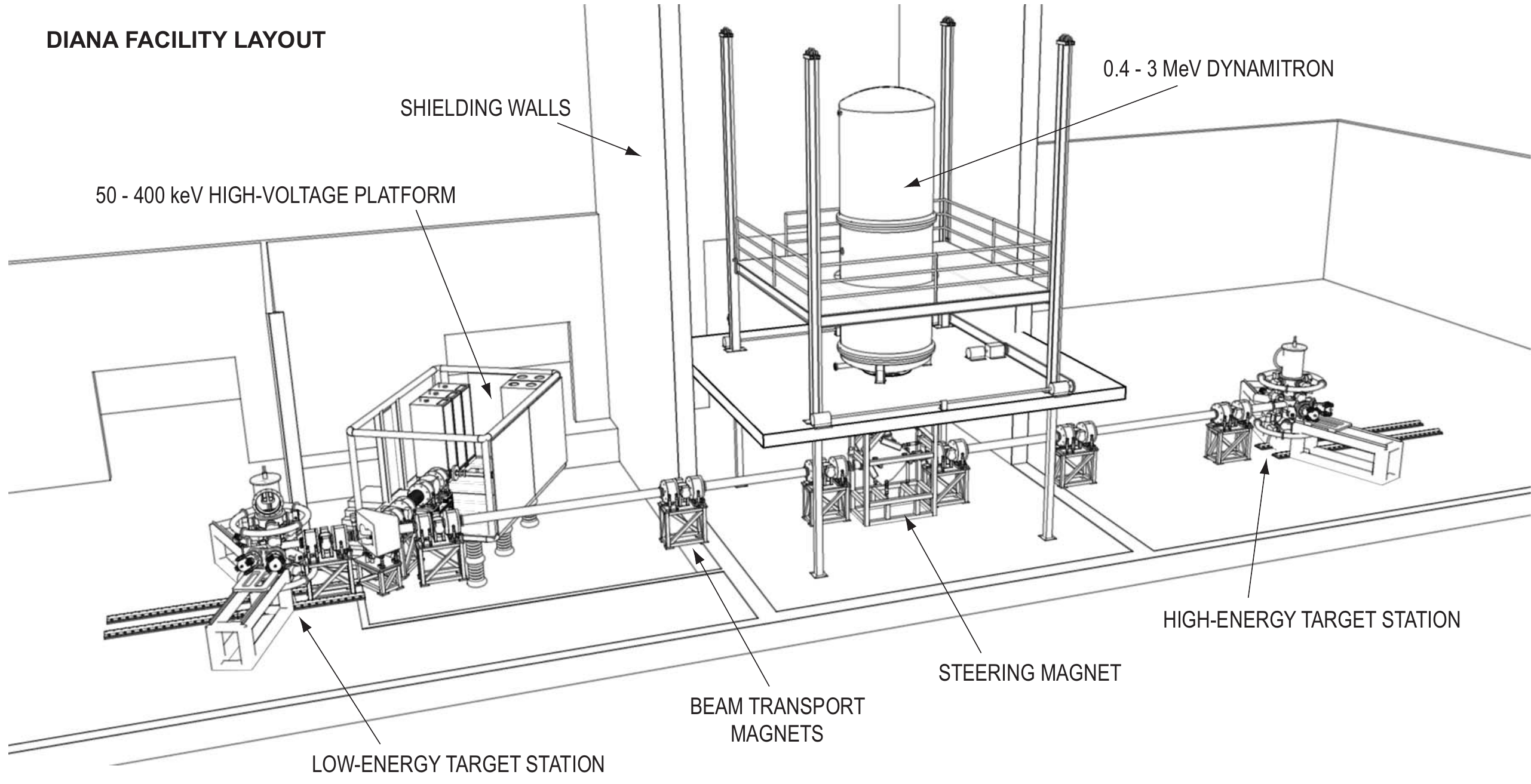}
\caption{Proposed layout of the DIANA facility.} 
\label{fig:fac3}
\end{center}
\end{figure*}

We discuss DIANA is more detail, as an example of the improvements that
would be possible in next generation nuclear astrophysics facilities.  DIANA's proposed
site is the 4850-foot level of Homestake, the same level where
Davis operated his chlorine detector.
The design combines a low-energy 400 kV high-intensity accelerator, 
a high-energy accelerator with a maximum voltage of 3 MV, and
flexibly configured target stations and detector systems.  
Both accelerators will be coupled to a shared target station, in order to
reduce uncertainties that would arise
when cross sections are measured at different facilities,  with different targets 
and detector configurations.  The accelerators will have a substantial overlap
in their energy ranges due to the design of the ion source 
on the high-voltage platform of the low energy accelerator.  This will reduce
uncertainties in combining data sets.  The 
proposed beam current of several mA is at least one order 
of magnitude higher than any presently available.  This enhances counting rates,
but also requires increased attention to beam-induced backgrounds
as well as targets capable of handling the power. 
Figure \ref{fig:fac3} shows DIANA's conceptual
design.
% Some figures are not sharp, may lead to objections from RMP.  Note
% that labels need to be readable when reduced to journal sizes

\begin{widetext}

\begin{table}
\caption{Attributes of proposed second-generation underground facilities for 
nuclear astrophysics.}
\label{tab:fac}
\begin{center}
\begin{tabular}{|l|l|l|l|l|}
\hline\hline
 & LUNA & DIANA & ELENA & CUNA \\
 FACILITY & Laboratory Underground & Dakota Ion Accelerators & Experimental Low-Energy & Canfranc Nuclear \\
  & for Nuclear Astrophysics & for Nuclear Astrophysics & Nuclear Astrophysics & Astrophysics Facility \\
\hline
Location & Gran Sasso, Italy & Homestake Mine, USA & Boulby Mine, UK & Canfranc, Spain \\
\hline
Rock type & hard limestone & metamorphic rock & salt & hard limestone \\
\hline
Depth & 3.1 & 4.3 & 2.8 & $\sim$ 2.0 \\
(km.w.e, flat site) & & & & \\
\hline
Low energy & 50-400 keV & 50-400 kV HV platform & none & none \\
 accelerator& 0.5-1.0 mA & $\gsim$ 10mA & & \\
  & RF ion source (p,He$^+$) & ECR ion sources, & & \\
  & & single, multiply charged & & \\
  \hline
High energy &  0.4-3.5 MeV & 0.35-3.0 MeV & 3.0 MeV accelerator & up to 5.0 MeV \\
 accelerator & electrostatic & electrostatic & electrostatic & electrostatic \\
  & up to 0.3 mA & up to 10 mA & 0.5 mA & \\
  & ECR ion source & ECR ion sources & ECR ion source & \\
   & & single, multiply charged& & \\
\hline\hline
\end{tabular}
\end{center}
\end{table}

\end{widetext}

\section*{Acknowledgments and Dedication}
\label{sec:acknowledgment}
We would like to thank the Institute for Nuclear Theory for hosting and supporting the Solar Fusion II workshop
and for providing technical assistance during the writing of
this review.  We thank A. Champagne, P. Descouvemont, A. Di Leva, and J. Toebbe
for their generous help, including many discussions and advice or assistance with fitting.
The research described in this review was supported by various agencies 
including the U.S. Department of
Energy, the U.S. National Science Foundation, the
Deutschen Forschungsgemeinschaft (cluster of excellence ``Origin and Structure
of the Universe" and grant SFB 634) and the Alliance Program EMMI of the Helmholtz Association.
% Are there major nonUS funding agencies that should be cited?

We dedicate Solar Fusion II to John Bahcall, who proposed and led the effort on Solar Fusion I.
John's advocacy for laboratory astrophysics and his appreciation of its importance to solar neutrinos 
paved the way for many advances in our field.

\section*{Appendix: Treating Uncertainties}
\label{errorsappendix}
\addtocounter{subsection}{-2}
\subsection{Introduction}
\label{errorsintro}
This section describes our method 
for dealing with discrepant data sets that may occur, for example, when deriving recommended S(0)
values from experimental measurements of nuclear reaction cross sections. 

While the conventional $\chi^2$ minimization method is adequate for analysing 
data sets that are
in good agreement, there is no rigorous method for dealing with
discrepant data sets and their underlying unidentified systematics.  But reasonable procedures exist.
In Solar Fusion II we adopt the Scale Factor method, here called the inflation factor method (IFM), 
that is used by the PDG \cite{Amsler:2008zzb}.  
In this method, the fit  errors from a conventional
$\chi^2$ minimization are inflated by a factor that depends on $\sqrt{\chi^2/\nu}$, where $\nu$ is the number of degrees of freedom.
This method is well known, widely used, and straightforward to apply.   

While the IFM is the only one discussed in the PDG Introduction,
alternatives exist.  We discuss some examples at the end of this Appendix.

\subsection{The inflation factor method}
\label{sfmethod}
The IFM addresses systematic uncertainties when combining
results from different
and possibly discrepant data sets.
The method inflates errors in proportion to the quoted errors originally given
by the  experimenters.
% wording changed - verify meaning has been maintained

Discrepant data may be defined by the $P$-value of the fit, where $P \equiv
P(\chi^2,\nu)$ is the probability of obtaining a $\chi^2$ value at least as large
as the observed value.  
The  inflation factor is conventionally chosen to be $\sqrt{\chi^2/\nu}$  
 and is commonly applied in cases
where $\chi^2/\nu > 1$.  We use an alternative inflation factor
$\sqrt{\chi^2/\chi^2(P=0.5)}$ to
account for the fact that, for small $\nu$
and non-discrepant data, the expected value of $\chi^2$ is smaller than unity.
For large $\nu$, the two scaling factors are equivalent.

The IFM scales all experimental errors by
the same fractional amount, resulting in equal internal and external errors on
the mean.  Because one generally cannot identify a specific mechanism accounting
for discrepant data, this procedure (like all other procedures) has no rigorous mathematical justification.
However qualitative arguments support its reasonableness.   As the method maintains
the relative precision of discrepant data sets, it apportions a larger absolute fraction of the 
identified systematic error to the less precise data sets.  This is consistent with naive
expectations that a large, unidentified systematic error is more likely to ``hide" within
a low-precision data set than within a high-precision one, given the advantages a high-precision
data set offers an experimentalist who does ``due-diligence" cross checks to
identify systematic errors.  The IFM is generally considered the most appropriate procedure
in the absence of information that would support alternatives, such
as omitting certain data, or increasing errors on some data but not others.

We employ error inflation whenever $\chi^2
> \chi^2(P=0.5)$, and no error scaling otherwise.  With this general rule,
errors are inflated a bit even when  $\chi^2$ is only slightly in excess of 
$\chi^2(P=0.5)$, despite the lack of compelling
evidence of discrepancy in such a case.  This procedure yields a continuous
formula and avoids the introduction of an arbitrary threshold for inflation.

In extreme cases one may obtain
errors that are deemed too small.  For example, when analyzing data
containing a few results with
small errors and a larger number of results with large errors, the large-error data
will reduce the error on the mean by increasing $\nu$, even though they may
have little effect on the central value.  In such a case, we agree with the
PDG's recommendation that, to mitigate this problem, data be excluded which have
an error larger than some (arbitrary) limit, specifically 3$\delta \sqrt{N}$,
where $N$ is the number of measurements and $\delta$ is the unscaled error on
the mean.  However, applying this exclusion
criterion may not be adequate to resolve this difficulty in all cases. 
%  It appears to me that this argument is not primarily addressing the IFM:
%  would not this issue arise for data sets that are not discrepant, but where a subset
% of the data has much larger errors than the remainder?  If so, then this
% argument should not be directed at the IFM. 

While the IFM makes no assumptions about the reasons for discrepant data, in actual
applications it may be apparent that not all data sets are equally reliable.  In such cases
judgment is necessary, and data selection is appropriate.  Data should be
discarded if the error analysis is poorly documented or inadequate.  Data may be
discarded if the procedure used to generate them involves questionable
assumptions, or if corrections were not made for effects now known to be
important.  Data errors may be modified (e.g. increased) if such new information is available. 

\subsection{Application of the inflation factor method}
\label{sfapplication}
The following is based on the discussion in the
Introduction of the PDG compilation of \citet{Amsler:2008zzb}:
\begin{enumerate}
\item
In general, statistical and systematic data errors should be identified and
specified separately.  Systematic errors should be subdivided into varying (random) and
common-mode (normalization) errors.  For a single data set, normally the
statistical and {\it varying} systematic errors should be combined in quadrature
and used as data errors in a $\chi^2$ minimization to determine unknown
parameters.  The resulting fit error(s) should be multiplied by the inflation factor (see below). 
The common-mode error is then folded in quadrature with the inflated fit
error to determine the overall normalization error.  

For multiple data sets, the
systematic errors should be examined to determine if they are independent among
the different data sets.  Parameters determined from multiple, independent data
sets may be combined in a separate $\chi^2$ minimization in which each parameter
value is characterized by its total error determined by combining statistical
and systematic (normalization) errors in quadrature.  Again, this fit error should be 
multiplied by the inflation factor.  If the systematic errors in different
data sets are correlated, then this correlation must be taken into account in
the fitting. A convenient method for handling correlations is described in the
2008 PDG compilation.
\item
Whenever $\chi^2 > \chi^2(P=0.5)$ the fit errors should be increased by the
multiplicative inflation factor $\sqrt{\chi^2/\chi^2(P=0.5)}$, where $\chi^2(P=0.5)$ is
the $\chi^2$ corresponding to a $P$ value of 0.5 for $\nu$ degrees of freedom.
The $\chi^2$ and
$\nu$ should be stated, along with the inflation factor when it
is larger than unity.  Large reported inflation factors serve to alert the reader to
potential problems. 
\item
Data with uncertainties larger than 3$\sqrt{N}\delta$, where $N$ is the number of
measurements and $\delta$ is the (unscaled) error on the mean should be
excluded.  One should be aware of possible error underestimation in certain
cases as mentioned above.  The resolution of such situations
may require additional judgment.
\end{enumerate}
%We note that error overestimation is just as bad as error underestimation as it
%. Error overestimation 
%devalues 
%experimental results.  
%Error underestimation can lead to confusion. xxx I don't know what this is supposed to mean 
%We must, in every case, respect the tension between these two
%situations, and strive for realistic error estimates, avoiding both extremes. 

\subsection{Other methods}
Other error analysis methods follow somewhat different strategies.
The cost function methods used in CODATA analyses
\cite{cohen86} are designed to reduce the $\chi^2$ by selective
re-weighting of data; i.e. by increasing the errors nonuniformly on the data, in
such a manner as to minimize the ``cost", i.e. the error on the mean. Alternatively,
\citet{dagostini94} has advocated a procedure for fitting multiple data sets
in which one minimizes the sum of a data $\chi^2$ and a 
normalization $\chi^2$.  

One method that has been applied to the analysis of solar fusion cross
section is that of \citet{cyburt04} (see also \citet{cyburt08}).
This approach introduces 
a ``discrepancy error", $\sigma_{\rm disc}$, that is added in quadrature with the 
normalization errors of individual experiments when fitting mixed data sets.  
Effectively this procedure distributes the unexplained discrepancy equally over  the data sets,
regardless of their stated accuracy, 
in contrast to the PDG procedure, which assigns the discrepancy in way that preserves the
relative stated accuracy of data sets.  
The \citet{cyburt04} method leads, in cases where there is excess dispersion, to increased 
de-weighting of the more precise data points, compared to the IFM.  In addition, the contribution 
of $\sigma_{\rm disc}$ to the error of the mean does not
decrease as the number of measurements $N$ increases.

The \citet{cyburt04} and IFM methods reflect two limits in how one apportions an unexplained discrepancy
among data sets: one could construct other models that interpolate between these two
limits (equal vs. proportionate allocation of the discrepancy error).    
The argument for the IFM procedure has been stated previously:  it is easier to miss
a large systematic error within a low-quality data set than within a high-quality one.   In addition,
it avoids a situation where archival data of poor quality, containing an unidentified systematic error,
unduly impact the weight that would otherwise be accorded a new experiment of exceptional
quality -- thereby inappropriately diluting the impact of the best results.  
Alternatives to the IFM tend to produce roughly equivalent results unless the discrepancies
among data sets are large.   We are fortunate in this paper to be dealing with discrepancies
that are modest.

\bibliography{solfus}

\end{document}